\documentclass[]{elsarticle}

\usepackage[graphicx]{realboxes}
\usepackage{amssymb}
\usepackage{amsmath}
\usepackage{lineno}
\usepackage{rotating}
\usepackage{ulem}
\usepackage{hyperref}
\usepackage{paralist}
\usepackage{subfigure}

\newcommand{\e}{\text{e}}

\begin{document}

\begin{frontmatter}

\author[eug]{Eugene d'Eon}
\author[norm]{Norman J. McCormick}
\address[eug]{Autodesk - Level 5, Building C - 11 Talavera Road - North Ryde NSW 2113 - Australia - ejdeon@gmail.com}
\address[norm]{Department of Mechanical Engineering - University of Washington - Seattle, WA 98195-2600, USA - mccor@uw.edu}

\title{Radiative transfer in half spaces of arbitrary dimension}

\begin{abstract}
  We solve the classic albedo and Milne problems of plane-parallel illumination of an isotropically-scattering half-space when generalized to a Euclidean domain $\mathbb{R}^d$ for arbitrary $d \ge 1$.  A continuous family of pseudo-problems and related $H$ functions arises and includes the classical 3D solutions, as well as 2D ``Flatland'' and rod-model solutions, as special cases.  The Case-style eigenmode method is applied to the general problem and the internal scalar densities, emerging distributions, and their respective moments are expressed in closed-form.  Universal properties invariant to dimension $d$ are highlighted and we find that a discrete diffusion mode is not universal for $d > 3$ in absorbing media.  We also find unexpected correspondences between differing dimensions and between anisotropic 3D scattering and isotropic scattering in high dimension.
\end{abstract}


\end{frontmatter}

\section{Introduction}

    Linear transport theory~\cite{SC60,davison57,vanrossum99} has a long history of utility in many fields for modeling the motion of particles and waves in random media and is most often concerned with transport in a spatially three-dimensional (3D) domain (even when symmetries in the solution reduce the equations to a single spatial coordinate, prompting a ``1D'' label).  It is, however, occasionally useful to consider Euclidean domains apart from $\mathbb{R}^3$.  For infinite homogeneous media, monoenergetic problems in $\mathbb{R}^d$ have been mostly solved~\cite{grosjean53}.  In this paper, we solve the classic problem of diffuse reflection from an isotropically-scattering half space in $\mathbb{R}^d$ for general $d \ge 1$.  The familiar $H$ functions, rigorous diffusion lengths, extrapolation constants, and singular eigenfunctions of radiative transfer and neutron transport are given new mathematical context as special cases in a general family of solutions expressed in closed form using hypergeometric functions.  We also derive a number of new analytic results for the ``Flatland'' ($d=2$) domain, which is used in a wide number of applied settings.

  \subsection{Motivation and related work}

    There are both practical and theoretical reasons to consider transport problems in general dimension.  The 1D rod has long been a useful domain for transport research and education~\cite{wing62,kohler1973power,hoogenboom08,davis14}.  The same can be said for the 2D ``Flatland'' domain~\cite{JSKJ12}, where it is possible to visualize the entire lightfield with 2D images~\cite{tantalum}.

    Transport in both the rod and Flatland settings finds numerous application in practice.  The rod model is equivalent to the two-stream approximation in plane-parallel atmospheric scattering~\cite{schuster05,kubelka31}, which is still a common method of solution for radiation budgets~\cite{meador80}.  Transport in Flatland also has many real-world applications, such as sea echo~\cite{jakeman1976model}, seismology~\cite{sato}, animal migration~\cite{jeanson05,XYSD09}, and wave propagation and diffraction in plates and ice~\cite{fock44,Bal00,MHDM06}.  Also, planar waveguides comprised of dielectric plates with controlled or random patterns of holes lead to 2D transport and have proven useful for studying engineered disorder~\cite{KVDSW12}.  Similarly, bundles of aligned dielectric fibers, such as clumps of hair or fur, can also be treated with a Flatland approach~\cite{gorodnichev89,gorodnichev90,mishchenko1992multiple,grzesik2018radiative}, where it is common to employ an approximate separable product of 1D and 2D solutions~\cite{marschner03}.  Reactor design also makes use of such 2D/1D decompositions~\cite{BWKEWL2016}.

    Going beyond 3D, higher-order dimensions occasionally find application in practice.  Exact time-dependent solutions in 2D and 4D have been combined to approximate the unknown 3D solution for the isotropic point source in infinite media~\cite{paasschens97}, and later applied to a time-dependent searchlight problem using the method of images~\cite{martelli07}.  In the study of cosmic microwave background radiation it has even been considered to change dimension over the course of a single random flight \cite{reimberg2015random}.

    The study of transport problem in the case of general dimension can reveal how dimension $d$ impacts various aspects of the solution.  New insights about 3D transport have been found by identifying correspondences between problems of differing configurations and dimensionalities~\cite{ciesielski1962first,majumdar06,comtet2011excursions}, which often appear unexpectedly.  Most investigations of this nature have considered only infinite media.  In this work, we identify some new exact correspondences regarding anisotropic scattering in a 3D half space.

	  The literature on stochastic processes includes a number of fully general studies in $\mathbb{R}^d$ (see \cite{chandrasekhar43,grosjean53,watanabe1970convergence,fournier78,dutka1985problem,zoia11} and references therein) that reveal the influence of dimension.  Such studies often permit $d \ge 1$ to be a free real parameter~\cite{kingman1963random} by using the general surface area $\Omega_d$ of the unit sphere $S^{d-1} \equiv \{x \in \mathbb{R}^d, ||x|| = 1 \}$ in $d$ dimensions,
    \begin{equation}
    	\Omega_d = \frac{d \pi ^{d/2}}{\Gamma \left(\frac{d}{2}+1\right)},
    \end{equation}	
    as expressed using the Gamma function~\cite{abramowitz1965handbook}, and taking the meaning of a non-integral dimension flight abstractly and ``by definition'' from the equations that follow.   Where possible, we consider the same freedom on $d$ in this paper.

    For infinite homogeneous media, Green's functions for monoenergetic linear transport with isotropic~\cite{paasschens97,kolesnik2005planar,kolesnik2008random,zoia11,deon13,deon18ii} and anisotropic~\cite{grosjean53,liemert11,MM16} scattering are known in domains apart from 3D.  The unidirectional point source was also considered in Flatland~\cite{rossetto17}.  In these infinite domains, the non-universal role of diffusion as a ``rigorous asymptote'' of the full solution for general dimension with absorption was observed~\cite{birkhoff70,deon13}.  We expand on these findings for the isotropic scattering case, and find a simple algebraic condition for diffusion asymptotics to arise.

    For the case of bounded domains, isotropic scattering in a Flatland half space has been solved in a number of works~\cite{fock44,Bal00,EdEMMR18,grzesik2018radiative}.  Slab geometry \cite{gorodnichev90} and layered problems~\cite{liemert12a} in Flatland have also been solved.  The study of inverse problems in plane-parallel domains of general dimension has been considered in a number of works (see, for example, \cite{mcdowall2009stability}).  We expand on these solutions by considering general dimension and by producing the singular eigenfunctions, whose orthogonality properties allow derivation of the moments of the internal scalar flux and angular distributions.  These moment derivations complement the mean, variance and general moments previously produced for 3D~\cite{abushumays67,NM15} and are useful for forming approximate searchlight approximations~\cite{deon14dual} and for guiding Monte Carlo estimators towards zero-variance~\cite{krivanek14zero}.

    \subsection{Outline}

    Section~\ref{sec:radiometry} provides a discussion of general dimensional ($d$D) radiometry. The integral and integrodifferential transport equations are given in Section~\ref{sec:transporteqs}, together with the dispersion equation and related eigenvalues.  The albedo problem is solved in Section~\ref{sec:albedo} using Chandrasekhar $H$-functions.  This section also discusses evaluation strategies for the $H$ functions, escape probabilities from within the half space, and a generalized enhancement factor for coherent backscattering.  Case's method is applied to the general problem in Section~\ref{sec:case}, and used to derive equations for spatial and directional moments.  Extrapolation distances for the Milne problem for a half-space and two adjacent half spaces are derived in Section~\ref{sec:Milne}.  We derive some universal properties of half-space transport in Section~\ref{sec:universal} before concluding in Section~\ref{sec:conclusion}.  A closed-form equation for the discrete eigenvalues is given in \ref{appendix:eigen} and additional closed-form equations for the $H$ function are given in \ref{sec:Heval}.  Monte Carlo sampling and connections between anisotropic scattering and nonclassical transport are given in \ref{appendix:MC} and \ref{appendix:connections}, respectively.

\section{Radiometry in $\mathbb{R}^d$}\label{sec:radiometry}

  Previous studies of the transport equation for $d$-space \cite{monin1956statistical,birkhoff69} implicitly assumed the generalization of standard radiometric quantities, which we briefly review here, along with a derivation of the generalized Lambertian bidirectional reflectance distribution function (BRDF).

 \subsection{Radiance}

    Let us consider time-independent mono-energetic specific intensity (radiance) $I(\mathbf{x},\omega)$ at a position $\mathbf{x} \in \mathbb{R}^d$ such that the rate of energy $dE$ flowing across a surface element of area $d\sigma$ in directions comprising a solid angle $d\omega$ about direction $\omega \in S^{d-1}$ is
    \begin{equation}
    	dE = I(\mathbf{x},\omega) \cos \theta d\sigma d\omega
    \end{equation}
    where $\theta$ is the angle between $\omega$ and the outward surface normal of $d\sigma$.  With $\Omega_d$ ``steradians'' in the unit sphere $S^{d-1}$, the integral of an angularly-uniform $(I(\mathbf{x},\omega) = 1)$ field of unit radiance is
    \begin{equation}
    	\int_{S^{d-1}} I(\mathbf{x},\omega) d\omega = \Omega_d
    \end{equation}
    and the isotropic phase function is
    \begin{equation}
    	P(\omega_i \rightarrow \omega_o) = 1 / \Omega_d.
    \end{equation}

  \subsection{Axial symmetry}
    In plane-parallel problems with axial symmetry the radiance is assumed to be symmetric in all but one axis and rotationally invariant about that axis, allowing us to express the transport equations over an integrated radiance that depends on a single position, the optical depth $x$, and a single direction parameter $\mu = \cos \theta$, the cosine with respect to the depth axis.  The parameter $\mu$ indexes a pair of directions in Flatland, a cone of directions in 3D, and higher-dimensional hypercones for $d > 3$.

      We choose a definition of integrated radiance $I(x,\mu)$ that is constant in $\mu \in [-1,1]$ whenever $I(x,\omega)$ is constant in $\omega \in S^{d-1}$ by introducing the appropriate angular measure $G(\mu) d\mu$.  The total rate of radiant energy flowing across $d\sigma$ at depth $x$ confined to directions with cosines in $d\mu$ about $\mu$ is then
    \begin{equation}
    	dE = I(x,\mu) \, \mu  \, G(\mu)\,   d\mu \, d\sigma.
    \end{equation}
    We normalize $G(\mu)$ such that
    \begin{equation}
        \frac{1}{2} \int_{-1}^1 G(\mu) d\mu = 1,
    \end{equation}
    a uniform integral over the sphere $S^{d-1}$.
    The function $G(\mu)$ that satisfies these conditions is (\cite{grosjean53}, Eq.(33))
    \begin{equation}\label{GnDmu}
    G(\mu) = \frac{2(1-\mu^2)^{\frac{d-3}{2}} \Gamma(\frac{d}{2})}{\sqrt{\pi}\Gamma \left(\frac{d-1}{2} \right) }, \; \; \; d > 1.
    \end{equation}
    For integer dimensionalities $d = 2$ to $7,$ $G(\mu)$ is
    \begin{equation}
    \quad \frac{2}{\pi \sqrt{1-\mu^2}};\quad 1;\ \frac{4\sqrt{1-\mu^2}}{\pi};\quad \frac{3}{2}(1-\mu^2);\quad  \frac{16(1-\mu^2)^{3/2}}{3\pi};\quad \frac{15}{8} \left(1-\mu ^2\right)^2 . \nonumber
    \end{equation}
    Scattering in a 1D rod is included in this definition by noting~\cite{paasschens97}
    \begin{equation}
        \lim_{d \rightarrow 1+} G(\mu) = \delta(\mu-1) + \delta(\mu+1).
    \end{equation}
    The probability that a single photon leaves an isotropic collision into $d\mu$ about $\mu$ is $(c/2) G(\mu) d\mu$, where $0 < c \leq 1$ is the single-scattering albedo, and so the integrated radiance about $x$ arising from isotropic collisions happening at a unit rate is $I(x,\mu) = c / 2$.

  \subsection{Uniform diffuse illumination}

    In additional to unidirectional illumination of the half space, we will also consider the case of uniform diffuse illumination, which we define to be uniform radiance in all directions arriving from the hemisphere to a given surface patch.  Such a source produces flux across the patch with an intensity proportional to $I(\mu) \mu$.  To produce a unit flux across a patch of unit area, we require a normalization constant such that a $\mu$-weighted integral over the hemisphere is $1$,
      \begin{equation}\label{eq:Ghatnorm}
        \frac{\sqrt{\pi } \, \Gamma \left(\frac{d+1}{2}\right)}{\Gamma \left(\frac{d}{2}\right)} \int_0^1  \mu G(\mu) d\mu = 1.
      \end{equation}
    The uniform diffuse boundary source condition is thus
    \begin{equation}\label{eq:boundarydiffuse}
        I(0,\mu) = \frac{\sqrt{\pi } \, \Gamma \left(\frac{d+1}{2}\right)}{\Gamma \left(\frac{d}{2}\right)}, \quad -1 \leq \mu \leq 0
    \end{equation}
    or $I(0,\omega) = 1 / \pi_d$ where
     \begin{align}
        &\pi_d \equiv \frac{\Omega_d}{2} \int_0^1 \mu \, G(\mu) \, d\mu = \frac{\pi ^{\frac{d-1}{2}}}{\Gamma \left(\frac{d+1}{2}\right)}, \\
        &\pi_1 = 1, \; \; \; \; \pi_2 = 2, \; \; \; \; \pi_3 = \pi, \; \; \; \; \pi_4 = \frac{4 \pi }{3}, \; \; \; \; ... \nonumber
    \end{align}

  \subsection{Bidirectional reflectance distribution function}

    The BRDF gives the radiance $f_L(\omega_i,\omega_o)$ leaving surface area $d\sigma$ in direction $\omega_o$ due to a unit radiance arriving at $d\sigma$ \textit{from} direction $\omega_i$.  This form of expressing the diffuse reflection law is convenient for image synthesis~\cite{hanrahan} and for comparing the behaviour to other known BRDFs.  Of particular interest is the Lambertian BRDF with total diffuse albedo $0 < k_d \le 1$, whose generalization to arbitrary dimension is
    \begin{equation}
        f_L(\omega_i,\omega_o) = \frac{k_d}{\pi_d}.
    \end{equation}

\section{Transport equations}\label{sec:transporteqs}

  We now review the plane-parallel transport equations for isotropic scattering in a halfspace in $\mathbb{R}^d$.  Energy balance in an infinitesimal slab in plane geometry with axial symmetry yields a transport equation of one spatial and one angular variable.  If the intensity distribution arriving at the slab of thickness $dx$ is given by $I(x,\mu')$, the flux in direction $\mu'$ crossing $dx$ is proportional to $\mu' G(\mu')$, and the track lengths extending through the slab are $dx / \mu'$, so the rate of photons entering collisions within $dx$ is
    \begin{equation}
        C(x)dx = \int_{-1}^1 I(x,\mu') G(\mu') \mu' \frac{dx}{\mu'} d\mu'.
    \end{equation}
    The inscattered contribution to $I(x,\mu)$ is thus
    \begin{equation}
        \frac{c}{2} C(x) = \frac{c}{2} \int_{-1}^1 I(x,\mu') G(\mu') d\mu'
    \end{equation}
  and the full integrodifferential form of the transport equation is then
  \begin{equation}\label{mu-transport}
    \left (\mu \frac{\partial}{\partial x} + 1 \right) I(x,\mu) = \frac{c}{2}\int_{-1}^1  I(x,\mu') G(\mu') \, d\mu',   
  \end{equation}
  which reduces to the familiar 3D form with $G(\mu) = 1$ and the Flatland equation~\cite{gorodnichev89,Bal00,grzesik2018radiative,EdEMMR18} with $G(\mu) = 2/(\pi \sqrt{1-\mu^2})$.

  Equation (\ref{mu-transport}) is a ``pseudo problem'' of the form studied by Chandrasekhar (Section 89 of~\cite{SC60}).  In his notation,
  \begin{equation}\label{eq:chandrapseudo}
    \left (\mu \frac{\partial}{\partial x} + 1 \right) I(x,\mu) = \int_{-1}^1 \Psi(\mu') I(x,\mu') \, d\mu'.   
  \end{equation}
  Chandrasekhar considered pseudo problems in relation to anisotropic scattering in a 3D half space.  Multiple pseudo problems with polynomial characteristic functions $\Psi_i(\mu)$ arise in each case, and their related $H$ functions appear in the exact solution.  No individual pseudo problem on its own corresponds to a complete transport problem, hence the label.  For isotropic scattering in $d$-space, however, we see a single pseudo problem does describe the complete problem.
  Comparing Eqs.(\ref{mu-transport}) and (\ref{eq:chandrapseudo}), we find the characteristic function for our problem to be
  \begin{equation}
    \Psi(\mu) = \frac{c}{2} G(\mu).
  \end{equation}
  For $d > 1$, $\Psi(\mu)$ is an even, non-negative function satisfying
  \begin{equation}
    \int_{0}^1 \Psi(\mu) d\mu = \frac{c}{2} \le \frac{1}{2},
  \end{equation}
  provided $0 \le c \le 1$.  Further, for $d > 1$, $\Psi(\mu)$ is also regular on $(-1,1)$.  At the boundaries,
  \begin{equation}
    \Psi(\pm 1) = \begin{cases}
        \infty, \quad (1 \leq d < 3)\\
        c/2, \quad (d = 3)\\
        0, \quad (d > 3).
    \end{cases}
  \end{equation}
  Busbridge studied a very general class of pseudo problems, relaxing the assumption of polynomial characteristic function.  With the above conditions satisfied, we can apply the findings of Chapter 2 of Busbridge~\cite{Busbridge60}.

  Before solving Eq.(\ref{mu-transport}), we consider the related integral equation for the collision-rate density $C(x)$ at optical depth $x$ in the half space.  This can be formed by integrating the total attenuated intensity at $x$ arriving from collisions at each depth $x'$ inside the half space,
  \begin{equation}\label{eq:WH2}
    C(x) = C_0(x) + c \, \int_0^\infty K(x-x') C(x') dx',
  \end{equation}
  where $C_0(x)$ is the forcing function (the collision-rate density of first collisions from any external source, in this case).  The kernels $K$ are symmetric and account for the total collision rate density at optical depth $x$ arising from energy that leaves a collision from a hyperplane at depth $x'$.

  Eq.(\ref{eq:WH2}) is an integral equation of the Wiener-Hopf (W-H) kind, named after the authors who first solved it for isotropic scattering in 3D.  That original W-H equation was first posed by Chwolson \cite{chwolson1889grundzuge}, who considered the Schwarz-schild-Milne (exponential integral) displacement kernel
    \begin{equation}\label{eq:KE1}
        K(x) = \frac{1}{2} E_1(|x|) = \frac{1}{2} \int_1^\infty \frac{\e^{-|x| t}}{t} dt
    \end{equation}
    in his study of the translucent appearance of milk glass\footnote{112 years later, the computer rendering of a glass of milk with multiple scattering was one of the iconic images in a seminal paper~\cite{jensen01} that sparked the subsurface revolution in film rendering and earned the authors an Academy award.}.

    For 1D and 2D, the kernels are also already known, and their W-H equations have been studied.  The Picard-Lalesco kernel
    \begin{equation}\label{eq:PicardK}
        K(x) = \frac{1}{2} \e^{|x|}
    \end{equation}
    describes exponential flights in a rod.  The MacDonald / Hankel kernel
    \begin{equation}\label{eq:MacDonaldK}
        K(x) = \frac{K_0(|x|)}{\pi}
    \end{equation}
    describes isotropic scattering in Flatland~\cite{EdEMMR18}, and has also appeared in studies of wave diffraction problems~\cite{bouwkamp1954diffraction,daniele2014wiener}.
    It is interesting that Fock, Case and Krein each considered the three kernels (\ref{eq:KE1}), (\ref{eq:PicardK}), and (\ref{eq:MacDonaldK}) in papers~\cite{fock44,case57,Krein62} on general W-H methods, but did not explicitly identify them as pertaining to isotropic scattering in 3D, 1D and 2D, respectively.  To the best of our knowledge, they have not previously been shown as members of the same unified family.  We show this now, expressing the general kernel in terms of the plane-geometry measure for $d$-space,
  \begin{equation}\label{eq:WHK}
    K(x) = \frac{1}{2} \int_0^1   \e^{-|x|/\mu} \frac{1}{\mu} G(\mu) d\mu = \frac{1}{2 \pi } \Gamma \left(\frac{d}{2}\right) G_{1,3}^{3,0}\left(\frac{x^2}{4}\left| \vphantom{\frac{1}{2}} \right.
\begin{array}{c}
 \frac{d-1}{2} \\
 0,0,\frac{1}{2} \\
\end{array}
\right).
  \end{equation}
  Here, $G_{1,3}^{3,0}$ is a Meijer $G$ function.  Again, we have used the assumption of isotropic scattering (which can be lifted, at considerable complexity, and won't be treated here) and also that the free-path distribution between collisions $p_c(x)$ is an exponential, $p_c(|x|/\mu) = \e^{-|x|/\mu}$, which can be easily generalized for the case of complete-frequency redistribution in line formation~\cite{ivanov94} and non-classical media with non-exponential free paths~\cite{deon18ii} (see \ref{A:nonexp}).

  The kernels in Eq.(\ref{eq:WHK}) are positive symmetric normalized displacement/convolution kernels
  \begin{equation}
    \int_{-\infty}^\infty K(x) dx = 1
  \end{equation}
  of the Laplace type~\cite{ivanov94}, expressible as
  \begin{equation}
    K(x) = \int_0^\infty h(s) \e^{-|x| s} ds
  \end{equation}
  where
  \begin{equation}
    h(s) = \frac{1}{2} \Theta(s-1) G(1/s) / s,
  \end{equation}
  using the Heaviside theta function $\Theta(x)$.  The kernels are singular at $x = 0$ for $d > 1$,
  \begin{equation}
  	\lim_{x \rightarrow 0} K(x) = \infty.
  \end{equation}

  The Fourier transform $\tilde{K}(t)$ of the kernels will play a central role in solving the albedo problem and can be expressed for the general case $d \ge 1$ using the hypergeometric function $_2F_1$~\cite{abramowitz1965handbook}, by taking the Fourier transform of Eq.(\ref{eq:WHK}) and exchanging the order of integration,
  \begin{equation}\label{eq:Kfourier}
  	\tilde{K}(t) \equiv \int_{-\infty}^\infty K(x) \e^{i x t} dx = \, _2F_1\left(\frac{1}{2},1;\frac{d}{2};-t^2\right).
  \end{equation}
  The common cases of $d \in \{ 1, 2, 3\}$ reduce to the familiar set
  \begin{equation}
    \left\{ \frac{1}{1+t^2}, \frac{1}{\sqrt{1+t^2}}, \frac{\tan^{-1} t}{t} \right\} \subset \tilde{K}(t),
  \end{equation}
  of Fourier transforms of the Picard, MacDonald and Schwarz-schild-Milne kernels, respectively.

  We see that a change in dimension $d$ amounts to a change of kernel $K(x)$ and related characteristic function $\Psi(\mu)$, and these two functions completely characterize the problem.  The solutions that follow will have much in common with analogous variations of $K$ and $\Psi$ that arise when considering general phase function~\cite{SC60}, reflectance conditions at the boundary~\cite{williams05}, and line-formation and other energy-dependent problems~\cite{JSIKNM66,ivanov73,frisch88}.  As such, we will rely on general studies of W-H equations~\cite{fock44,Busbridge60,Krein62,JSIKNM66,carlstedt66,mullikin1968some,frisch88,ivanov94}.

  \subsection{Dispersion equation and eigenvalues}\label{sec:roots}

    For infinite, half space or slab geometry problems, the solutions all depend on the eigenspectra of the transport kernel $K(x)$, which can include both a continuous and a discrete component.  We review the eigenvalues now, and their conditions for existence, before solving the general albedo problem.

    The discrete eigenvalues, when they exist, are real zeros $\nu_0$ of the dispersion function $\Lambda(z)$~\cite{Busbridge60}, which is related to the Fourier transform of the kernel by
    \begin{equation}\label{eq:lambdait}
      \Lambda(i / t) = 1 - c \int_{-\infty}^\infty K(x) \e^{i t x} dx
    \end{equation}
    or, equivalently, from the characteristic function $\Psi(\mu)$,
    \begin{equation}
         \Lambda(z) = 1- \frac{c \, z}{2} \int_{-1}^1 \frac{G(\mu)}{z-\mu} \, d \mu.
    \end{equation}
    From Eq.(\ref{eq:Kfourier}) we have the generalized dispersion equation in terms of a hypergeometric function
    \begin{equation}\label{eq:Lambdad}
        \Lambda(z) = 1 - c \, \, _2F_1\left(\frac{1}{2},1;\frac{d}{2};\frac{1}{z^2} \right),
    \end{equation}
    in agreement with previous derivations in infinite spherical geometry~\cite{birkhoff70,paasschens97,zoia11}.  By known properties of $_2F_1$~\cite{abramowitz1965handbook}, $\Lambda(z)$ satisfies the differential equation
    \begin{equation}\label{eq:LambdaDE}
         \Lambda ''(z) \left(z^2-z^4\right) + \Lambda '(z) z \left(z^2 (d-3) - 2\right) + 2 \Lambda
       (z)-2=0.
    \end{equation}
    Here we see $d = 3$ as a special case, the unique dimension where the $z^3 \Lambda'(z)$ term vanishes.

    The discrete eigenvalues fall into three cases~\cite{Busbridge60}, based on dimension $d$ and absorption $c$.  For the case of conservative scattering $c = 1$, double zeros at infinity arise for all $d \ge 1$.  This follows immediately from the limit as $z \rightarrow \infty$,
    \begin{eqnarray}
    \Lambda(\infty) &=&  1 - \frac{c}{2} \int_{-1}^1 G(\mu) \,d\mu  \nonumber \\
    &=& 1 - c \, ,\label{Lambda infty}
    \end{eqnarray}
    and from $\Lambda'(\infty) = 0$.  For absorbing media, $0 < c < 1$, given the symmetry and non-negativity of $K$, there will be either $0$ or $1$ pairs of real eigenvalues $\pm \nu_0$, satisfying $\Lambda(\pm \nu_0) = 0$~\cite{Krein62}.  

    For $d \leq 3$, $\Psi(1) \neq 0$ and so $\Lambda(z)$ always has a real root $\nu_0 > 1$.  For $d > 3$, discrete eigenvalues will only exist when $\Lambda(1) < 0$~\cite{Busbridge60}.  After observing that
    \begin{equation}
        \, _2F_1\left(\frac{1}{2},1;\frac{d}{2};1\right) = \frac{d-2}{d-3}, \quad d > 3,
    \end{equation}
    we find that $\Lambda(z)$ admits a finite zero $\nu_0 > 1$ if and only if
    \begin{equation}\label{eq:v0condition}
       (d-3)/(d-2) < c < 1.
    \end{equation}
    This condition simplifies several prior observations and conditions for diffusion modes disappearing in dimensions $d > 3$~\cite{birkhoff70,deon13}\footnote{It was incorrectly reported in~\cite{deon13} that the number of discrete eigenvalues increases past $d = 4$.} and is new, to the best of our knowledge.  
    Closed form expressions are known for $d \in \{ 1, 2, 4, 6\}$~\cite{deon13},
    \begin{align*}
        &\nu_0 = \pm 1/\sqrt{1-c},  &(d = 1) \\
        &\nu_0 = \pm 1/\sqrt{1-c^2},  &(d = 2) \\
        &\nu_0 = \pm 1/\left(2\sqrt{(c-c^2)}\right),  &(d = 4, \; c > 1/2) \\
        &\nu_0 = \pm \frac{3}{2 \sqrt{(9-8 c) c-\sqrt{c (4 c-3)^3}}},  &(d = 6, \; c > 3/4).
    \end{align*}
    Mathematica is able to find the roots in 8D and 10D in closed form but we omit these bulky expressions for space reasons.

    In 3D, the eigenvalues always exist, the dispersion equation being
    \begin{equation}
        0 = 1 - c \, \nu_0 \tanh^{-1}(1/\nu_0) . \nonumber
    \end{equation}
    In odd dimensionalities $d \ge 5$, the discrete eigenvalues are (like in 3D) also solutions of transcendental equations of increasing complexity, such as in 5D,
    \begin{equation}
        6 c \nu _0 \left(\nu _0+\nu _0^2 \left(-\coth ^{-1}\left(\nu _0\right)\right)+\coth
       ^{-1}\left(\nu _0\right)\right)=4.
    \end{equation}
    In~\ref{appendix:eigen} we derive a general closed-form expression for $\nu_0$ for any dimension $d \ge 1$.

    A related function that plays an important role in the solution of half space problems is
    \begin{equation}\label{lambdapvintegral}
    \lambda(\nu) = 1-\frac{c\nu}{2} \mathcal{P}\int_{-1}^1 \frac{G(\mu)}{\nu-\mu} \, d\mu
    \end{equation}
    with $\mathcal{P}$ indicating the Cauchy principal value of any integral over $\nu$ or $\mu$ must be taken.  For $\nu \in [-1,1]$ the principal value integral can be expressed in closed form,
    \begin{equation}\label{eq:lambdad}
        \lambda(\nu) = 1 - c + c \, _2F_1\left(1,1-\frac{d}{2};\frac{1}{2};\nu^2\right),
    \end{equation}
    which simplifies to
    \begin{equation}
         \left\{ 1, 1, 1-\frac{1}{2} c \nu  \ln \left(\frac{1+\nu}{1-\nu }\right), 1-2 c \nu ^2 \right\} \subset \lambda(\nu)
    \end{equation}
    for $d \in \{ 1, 2, 3, 4 \}$, respectively.  Equation (\ref{eq:lambdad}) can also be written
    \begin{multline}
        \lambda(\nu) =  1 + c\frac{ d-2}{d-3} \left[\left(\nu^2-1\right) \,
       _2F_1\left(1,2-\frac{d}{2};-\frac{1}{2};\nu^2\right) + \right. \\ \left. \left((d-6) \nu^2+1\right) \,
       _2F_1\left(1,2-\frac{d}{2};\frac{1}{2};\nu^2\right)\right]
    \end{multline}
    which shows how $c = (d-3)/(d-2)$ is a special value.
    Again, using known properties of $_2F_1$, it is straightforward to find the differential equation satisfied by $\lambda(z)$,
    \begin{equation}\label{eq:lambdaDE}
        \left(z ^2-1\right)\lambda ''(z ) - (d-5) z  \lambda '(z ) + (4-2 d) \lambda (z ) + 2 (1-c) (d-2)   =0,
    \end{equation}
    with conditions
    \begin{equation}
        \lambda(0) = 1, \, \, \, \, \, \lambda'(0) = 0.
    \end{equation}

    From the Plemelj-Sohotski formula\cite{muskhelishvili2008singular} the boundary values of
    $\Lambda(z)$ are
    \begin{equation}
        \Lambda ^{\pm} (\nu)= \lambda(\nu) \pm i \pi c  \nu G(\nu)/2, \, \, \,  \, \, \, \, \nu \in [-1,1],\label{Lambda pm}
    \end{equation}
    from which it follows that
    \begin{equation}
         \Lambda^+(\nu) + \Lambda^-(\nu) = 2\lambda(\nu).
    \end{equation}
    Also, $\Lambda (0) = \Lambda ^\pm(0)  =1$.  It is straightforward to verify that
    Eq.(\ref{Lambda pm}) satisfies the differential equation in Eq.(\ref{eq:LambdaDE}).  

\begin{figure}
      \centering
      \includegraphics[width=1.1\linewidth]{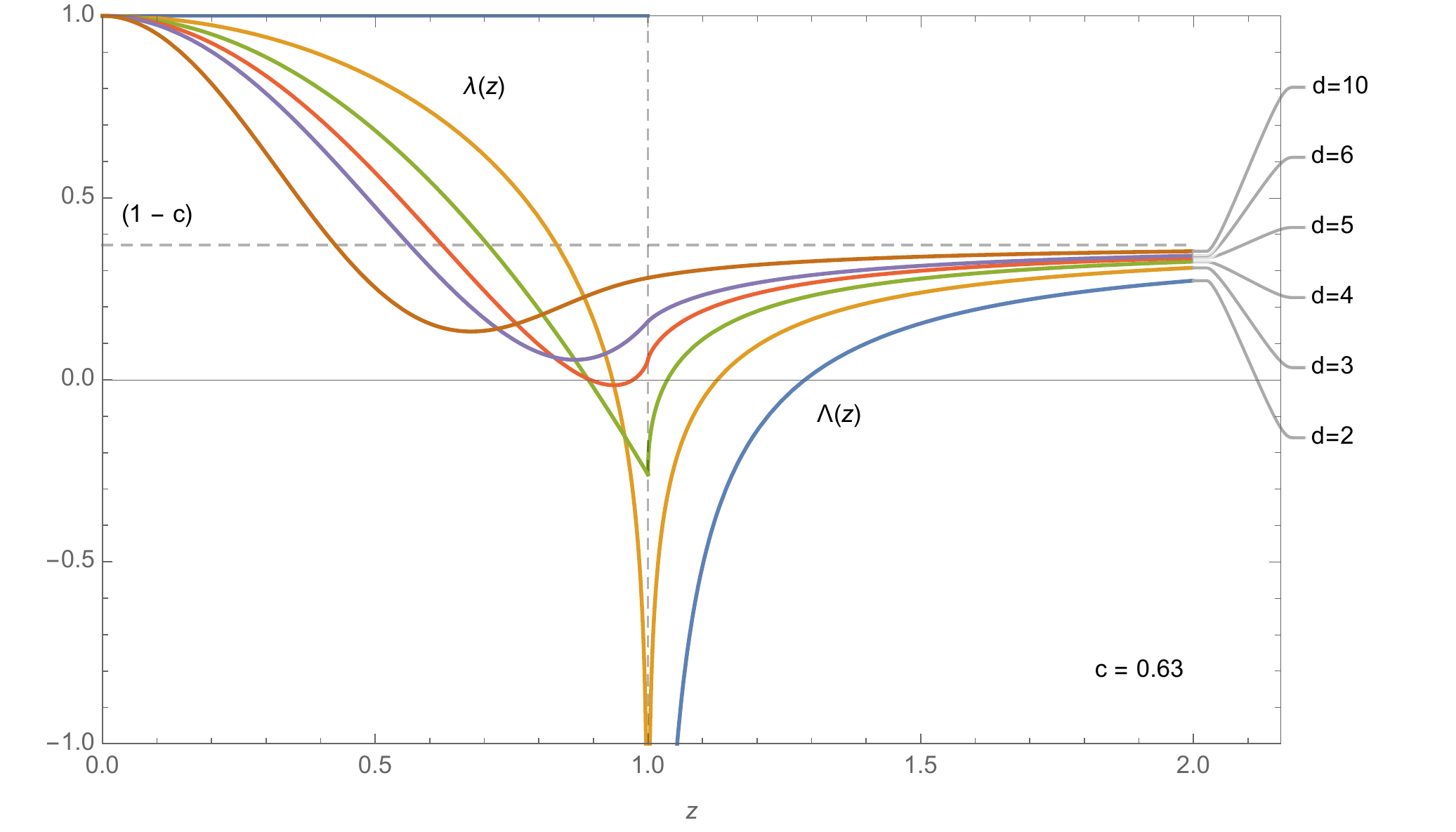}
      \caption{\label{fig:lambdas}The functions $\Lambda(z)$ and $\lambda(z)$ for real $z$ and $c = 0.63$.} 
    \end{figure}

    Both $\Lambda(z)$ and $\lambda(z)$ are shown in Figure~\ref{fig:lambdas} for a variety of dimensions $d$.  For the illustrated case of $c = 0.63$ we see that there are no discrete roots $\nu_0 > 1$ of $\Lambda(\nu_0)$ for $d > 4$ and also that $\lambda(\nu)$ admits either $0$, $1$, or $2$ roots $0 < \nu < 1$.

\section{The albedo problem}\label{sec:albedo}

    Let us now consider a unit current azimuthally-symmetric plane-parallel illumination arriving along direction cosine $\mu_\ell$, given by
    \begin{equation}\label{eq:eugboundary}
      I(0,\mu;-\mu_\ell) = \delta(\mu-\mu_\ell) / ( \mu_\ell G(\mu_\ell) )
    \end{equation}
    for $\mu$ and $\mu_\ell \in [0,1]$.  The density of initial collisions inside the half space at optical depth $x \ge 0$ is then,
    \begin{equation}\label{eq:Salbedo}
        C_0(x) = (1/\mu_\ell) \exp(-x / \mu_\ell),
    \end{equation}
    which is independent of dimension.
    Here, $C_0(x)dx$ is the rate of particles \textit{entering} their first collision within $dx$ about $x$.
    
    The inhomogeneous integral equation~(\ref{eq:WH2}) with the forcing function $C_0(x)$ in Eq.(\ref{eq:Salbedo}) defines the $dD$ albedo problem with a unidirectional source.  Its solution can be found in a number of ways, and directly yields the internal distribution and, indirectly, the emerging distribution.

    Once the collision rate density $C(x)$ is found, the scalar flux of particles in flight inside the medium is known immediately because we are assuming a classical medium with no correlation or memory between scattering events.  Given our parametrization of optical depth $x$, the scalar flux is proportional to $C(x)$ by a unit constant, the mean-free path, which is simply a change of units.

    The integrated radiance at the boundary is
    \begin{equation}\label{eq:I0mu}
        I(0,\mu;-\mu_\ell) =  \int_0^\infty \frac{c}{2} \, C(x) \exp(-x / \mu) \frac{1}{\mu} dx,
    \end{equation}
    and the rate of photons leaving the medium along directions in $d\mu$ about $\mu$ is $I(0,\mu;-\mu_\ell) \mu G(\mu) d\mu$.
    Eq.(\ref{eq:I0mu}) is simply a reduced-intensity calculation of radiance leaving collisions at depth $x$, using the exponential Beer-Lambert law.  The quantity $(c/2) C(x) G(\mu) d \mu$ is the integrated radiance leaving collisions at $x$ into $d\mu$ about $\mu$, $\exp(-x / \mu)$ is the Beer-Lambert calculation, and $(1/\mu) dx$ is the source measure at depth $x$; i.e. the length of the line segment tilted by $\mu$ inside the slab of thickness $dx$ from which the source of collided photons arises.

    Equation (\ref{eq:I0mu}) gives the law of diffuse reflection for the half space and shows that the emerging distribution is related to the internal distribution by a Laplace transform.  The Laplace transform of the internal distribution is given in terms of $H$-functions, which we consider next.

  \subsection{$d$D Chandrasekhar $H$-functions}

\begin{figure*}
      \centering
      \includegraphics[width=1.1\linewidth]{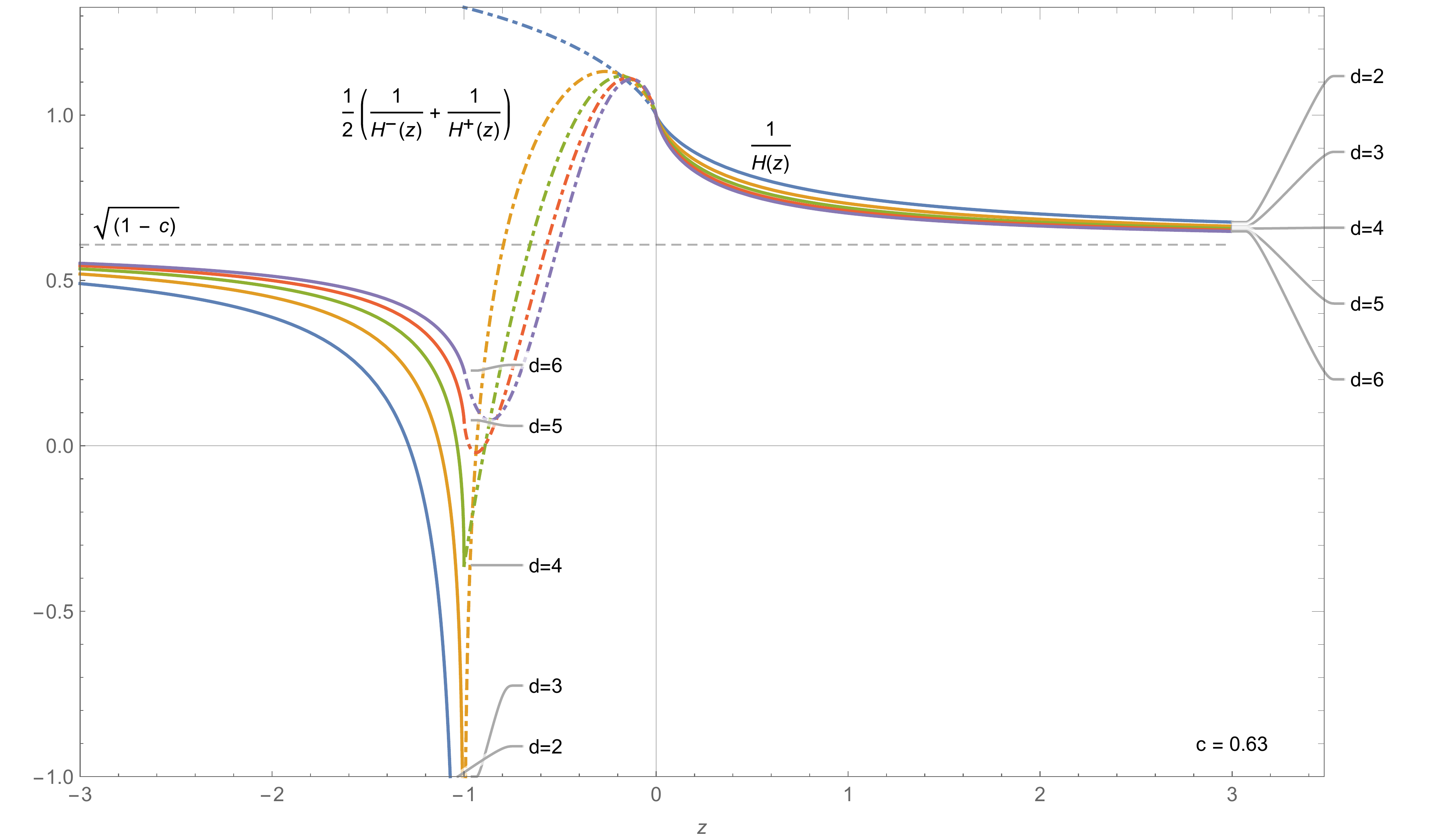}
      \caption{\label{fig:H}The $H$-functions of isotropic scattering in $\mathbb{R}^d$.  The continuous curves show $1/H(z)$ for real $z$ and $c = 0.63$.  The arithmetic mean of the boundary values is plotted (dot-dashed) for $z \in \left(-1,0\right)$.} 
    \end{figure*}

    Given the characteristic function $\Psi(\mu) = (c/2)G(\mu)$ for isotropic scattering in $d$D (Section~\ref{sec:transporteqs}), the related $H$ functions satisfy the integral equation~\cite{Busbridge60},
      \begin{equation}\label{1stH(mu)eq}
        \frac{1}{H(\mu)} = 1 - \frac{c\,\mu}{2}\int_0^1 \frac{H(\mu')}{\mu + \mu'} G(\mu') \,d \mu',
      \end{equation}
    for $0 \leq \mu \leq 1$ or, more generally
    \begin{equation}\label{H(z)eq}
      \frac{1}{H(z)} = 1 - \frac{c\,z}{2}\int_0^1 \frac{H(\mu')}{z + \mu'} G(\mu') \,d \mu', \quad z \notin [-1,0]
    \end{equation}
    Regardless of dimension $d \ge 1$ or absorption $0 < c \leq 1$, the solution of Eq.(\ref{1stH(mu)eq}) is known in closed form by the Fock/Chandrasekhar equation~\cite{fock44,Busbridge60}
      \begin{multline}\label{eq:HsolutionRd}
        H(z) = \exp \left( \frac{z}{2 \pi i} \int_{-i \infty}^{i \infty} \frac{1}{t^2 - z^2} \ln \Lambda(t) dt \right) 
        \\ = \exp \left( \frac{-z}{\pi} \int_0^\infty \frac{1}{1+z^2 k^2} \ln \Lambda(i/k) dk \right), \;\; \text{Re} \, z > 0.
      \end{multline}
    Figure~\ref{fig:H} illustrates values of $1 / H(z)$.

    Let us pause for a moment to consider the significance of the appearance of $\Lambda(i/k)$ in Eq.(\ref{eq:HsolutionRd}).  By Eq.(\ref{eq:lambdait}),
    \begin{equation}
        \frac{1}{\Lambda(i/t)} =  \frac{1}{1 - c \, \tilde{K}(t)} = 1 + c \tilde{K}(t) + c^2 \left(\tilde{K}(t)\right)^2 + ... ,
    \end{equation}
    which is the Fourier-space Neumann-series Green's function for the isotropic plane source in infinite geometry.  So we see an exact infinite-space solution inside the $H$ function expression for the half space,
    \begin{equation*}
        H(z) = \exp \left( \frac{z}{\pi} \int_0^\infty \frac{1}{1+z^2 t^2} \ln \left[ \frac{1}{1 - c \, \tilde{K}(t)} \right] dt \right).
    \end{equation*}
    Ivanov~\cite{ivanov94} noted a similar relationship for general displacement kernels $K$.  From his analysis we also have, for all $d \ge 1$,
    \begin{align}
        &H(0) = 1, \\
        &H(\infty) = (1-c)^{-1/2} \label{eq:sqrtepsilon}.
    \end{align}

  \subsubsection{$H$-function moments}
    The moments of the $H$ functions can form an additional integral equation for $H$, and also arise in later expressions for the internal and emerging distributions of the albedo problem (and related extrapolation distances), and so we look at these now.  

    If Eq.(\ref{H(z)eq}) is expanded around infinity, one obtains
    \begin{equation}
        \frac{1}{H(z)} = 1 - 
        \frac{c}{2}\sum_{n=0}^\infty (-1)^n\frac{\alpha_{\,n}}{z^{n}}\, ,\quad |z| > 1, \label{Hexpansion}
    \end{equation}
    where $\alpha_n$ are moments of the $H$-function defined by
    \begin{equation}
        \alpha_n = \int_0^1 \mu^n  H(\mu) G(\mu) \,d\mu \, . \label{alphadef}
    \end{equation}
    In her general study of pseudo problems, Busbridge~\cite{Busbridge60,Busbridge57} found that the $\alpha_n$ moments of Eq.(\ref{alphadef}) satisfy the recurrence equations
    \begin{equation}
        \alpha_{2n} \sqrt{1-c} = g_{2n} + \frac{c}{4} \sum_{k=1}^{2n-1}(-1)^k \alpha_{2n-k}\alpha_k , \quad n=0,1,2,\dots \,\label{halphamoments}
    \end{equation}
    where moments related to the characteristic function are given by
    \begin{equation}
        g_{2n} = \int_0^1 \mu^{2n} G(\mu) d\mu .\label{gdef}
    \end{equation}
    For our present study of isotropic scattering in $d$-space, we find
    \begin{equation}
        g_{2n} = \frac{\Gamma \left(\frac{d}{2}\right) \Gamma \left(n+\frac{1}{2}\right)}{\sqrt{\pi } \,
   \Gamma \left(\frac{d}{2}+n\right)},
    \end{equation}
    which, for $n \in \{ 0, 1, 2, 3\}$, is
    \begin{equation*}
        1; \quad 1 / d; \quad  \frac{3}{d^2+2 d};  \quad  \frac{15}{d^3+6 d^2+8 d}.
    \end{equation*}
    In Flatland, $g_{2n}$ reduces to
    \begin{equation}
        g_{2n} = \frac{(2n-1)!!}{(2n)!!},
    \end{equation}
    differing from the familiar 3D case, $g_{2n} = (2n+1)^{-1}$.

\begin{figure*}
        \centering
        \subfigure[$H$-function moments $\alpha_i$ for varying $c$.]{\includegraphics[width=0.88 \linewidth]{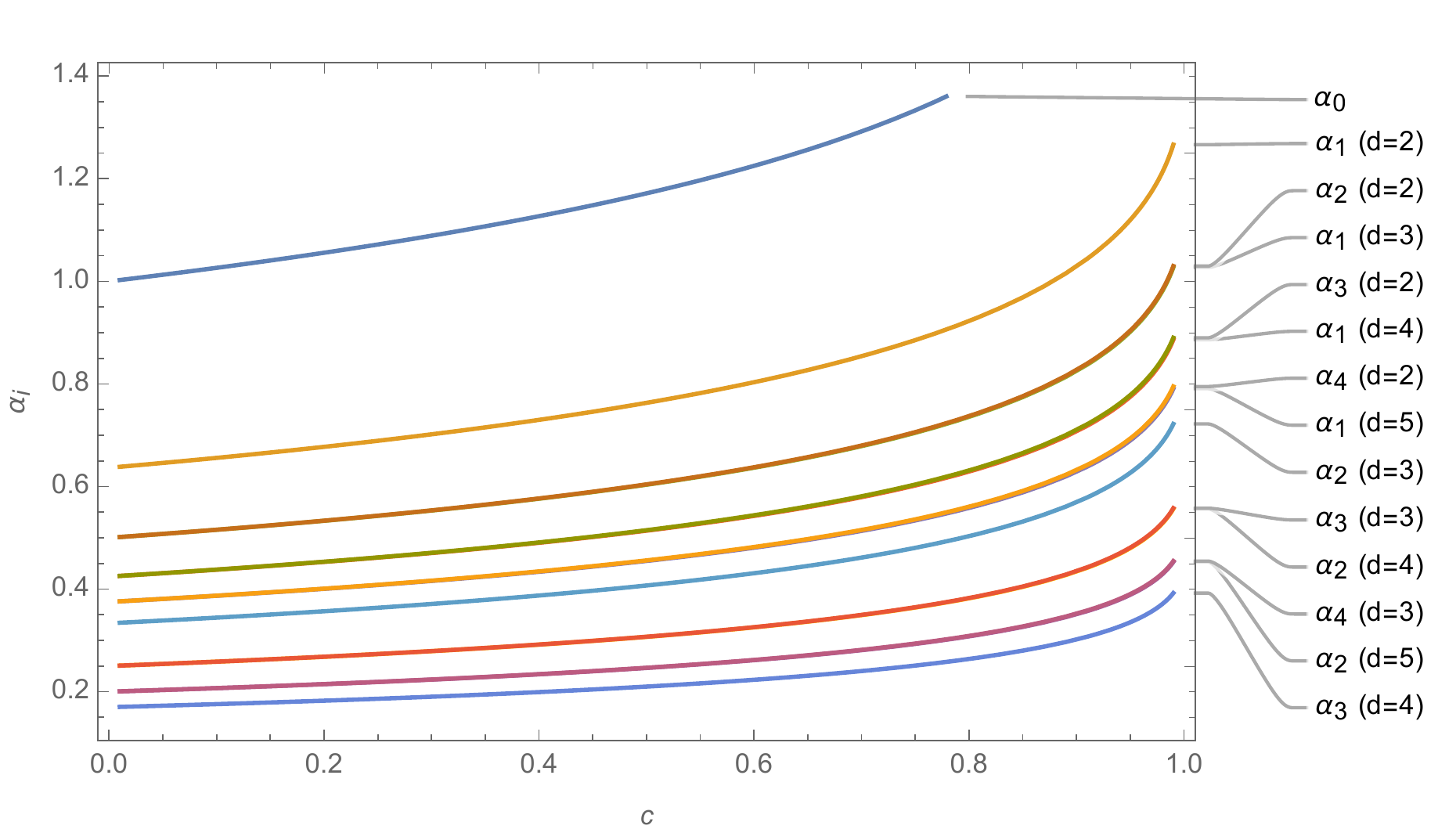}}
        \subfigure[Ratios of close pairs of moments $\alpha_i(c,d)$.]{\includegraphics[width=0.9 \linewidth]{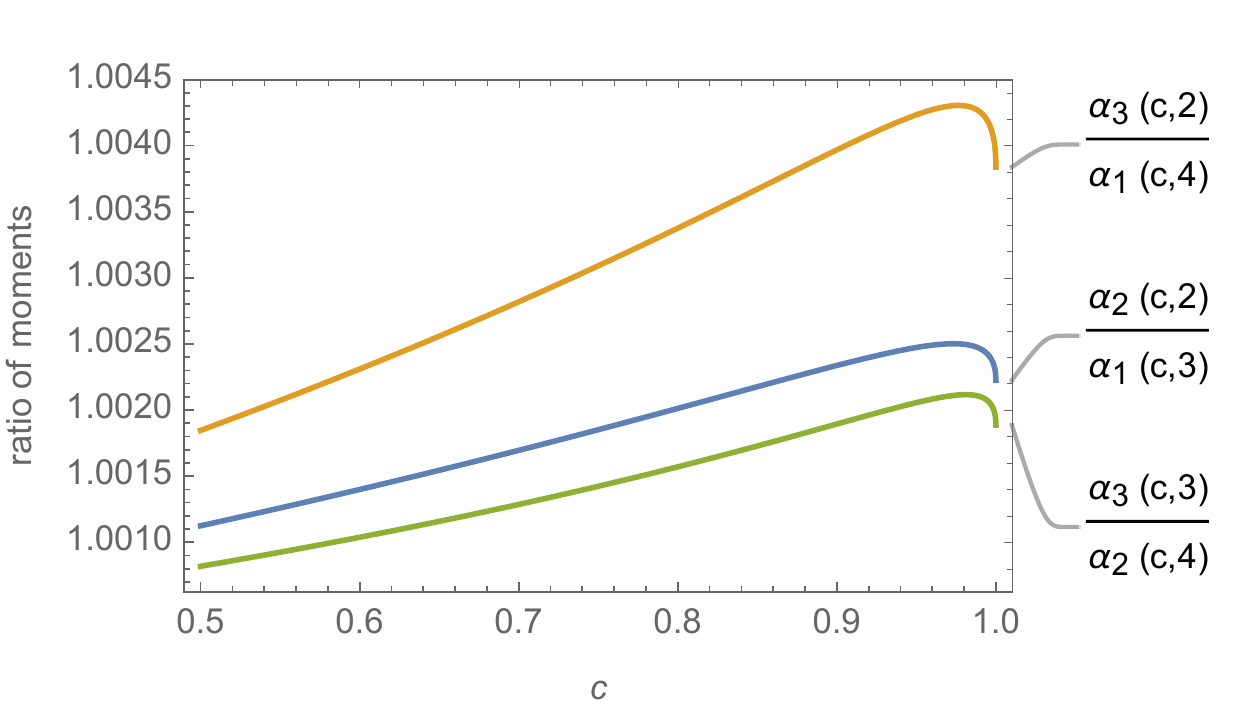}}
        \caption{The $H$ function moments appear to exhibit a peculiar pairing over various orders and dimensions (a).  Upon closer inspection (b), we see that the agreement is not exact.}          
        \label{fig-Hmoments} 
      \end{figure*}

    For nonconservative scattering $0 < c < 1$, the odd moments must be evaluated numerically, and the even moments are given from the recurrence relations.  For example,
    \begin{align}
        &\alpha_0 = \frac{2}{c} \left(1 - \sqrt{1-c}\right) \label{alpha0} \\
        &\alpha_2 = \frac{1}{\sqrt{1-c}} \left( \frac{1}{d}-\frac{c}{4} \alpha_1^2  \right).
    \end{align}
    For conservative ($c = 1$) scattering, it is the even moments which must be numerically evaluated, and the odd moments deduced,
    \begin{align}
        &\alpha_0 = 2 \\
        &\alpha_1 = 2 / \sqrt{d}.
    \end{align}
    Figure~\ref{fig-Hmoments} and Table~\ref{tab:moments} illustrate a peculiar, but not exact, pairing of moments for different dimensions that we cannot explain and leave as an interesting area for future investigation.

    Equation (\ref{alpha0}) can be used to rewrite Eq.(\ref{1stH(mu)eq}) as
    \begin{equation}
        \frac{1}{H(\mu)} = (1 - c)^{1/2} + \frac{c}{2} \int_0^1 \frac{\mu' H(\mu')}{\mu + \mu'} G(\mu') d\mu' \,, \label{2ndH(mu)eq}
    \end{equation}
    which is the form that should be used if evaluating $H$ directly from an integral equation~\cite{SC60}.

    \begin{table*}
      \centering
      \scalebox{0.75}{
      \begin{tabular}{c | llllllll}
        & $\alpha_1$ & $\alpha_2$ & $\alpha_3$ & $\alpha_4$ & $\alpha_5$ & $\alpha_6$ & $\alpha_7$ & $\alpha_8$ \\
         \hline 
         \text{d=2} & 1.26655 & 1.02971 & 0.889917 & 0.794985 & 0.725119 & 0.670931 & 0.627316 & 0.591231 \\
         \text{d=3} & 1.02718 & 0.721955 & 0.557658 & 0.45458 & 0.383773 & 0.332099 & 0.292712 & 0.261689 \\
         \text{d=4} & 0.886147 & 0.556488 & 0.392232 & 0.295955 & 0.233692 & 0.190647 & 0.159405 & 0.135873 \\
         \text{d=5} & 0.790625 & 0.452909 & 0.295462 & 0.208513 & 0.15522 & 0.120126 & 0.0957642 & 0.078151 \\
         \text{d=6} & 0.720483 & 0.381903 & 0.232999 & 0.155012 & 0.109429 & 0.0809076 & 0.0616274 & 0.0487154 \\
         \text{d=7} & 0.666181 & 0.330172 & 0.189885 & 0.119836 & 0.0806849 & 0.057 & 0.041798 & 0.03158 \\
      \end{tabular}}
      \caption{\label{tab:moments}$H$ function moments of various orders $\alpha_i$ and various dimensions $d$, for $c = 0.99$.}
    \end{table*} 

    \subsubsection{$H(z)$ calculation methods}

    Given their central role in the solutions that follow, we now consider important details regarding the uniqueness and evaluation strategies for the $H$ functions.

    Equation (\ref{eq:HsolutionRd}) is the unique solution of Eq.(\ref{H(z)eq}) unless there is a finite discrete eigenvalue $\nu_0$ of the dispersion Eq.(\ref{eq:Lambdad}).  In the latter case, there is one other non-physical solution~\cite{Busbridge60}, which is not relevant to our problem.  Equation (\ref{eq:HsolutionRd}) is the unique solution of Eq.(\ref{2ndH(mu)eq}) in all cases~\cite{Busbridge60}.

    Fox~\cite{fox61} considered the expression of $H$ with general characteristic $\Psi$ as the solution to a Riemann-Hilbert problem involving the function
    \begin{equation}\label{eq:theta}
        \tan \theta(t) = \frac{\pi t \Psi(t)}{\lambda(t)} = \frac{c}{2} \frac{ \pi t G(t)}{\lambda(t)}
    \end{equation}
    where $\theta(0) = 0$ and $0 \leq \theta(t) \leq \pi$.  When $(d-3)/(d-2) < c < 1$ and one real eigenvalue $\nu_0 > 1$ exists, a useful extension of Fox's approach is~\cite{carlstedt66,williams12}
    \begin{equation}\label{eq:HthetaZero}
        H(z) = \frac{1+z}{\nu_0+z} \frac{1}{\sqrt{1-c}} \exp \left( -\frac{1}{\pi} \int_0^1 \frac{\theta(t)}{t+z} dt \right).
    \end{equation}
    For $c \leq (d-3)/(d-2)$, with no eigenvalues $\nu_0 > 1$, the simpler result is~\cite{carlstedt66}
    \begin{equation}\label{eq:HthetaNoZero}
        H(z) = \frac{1}{\sqrt{1-c}} \exp \left( -\frac{1}{\pi} \int_0^1 \frac{\theta(t)}{t+z} dt \right), \; \; \; z \notin [-1,0].
    \end{equation}
    We noted improved numerical stability over Eq.(\ref{eq:HsolutionRd}) for $d > 4$ when using these last two forms of $H(z)$.  In practice, when $\lambda(\nu)$ has zeros, negative $\tan^{-1}$ results need to be detected manually and remapped to ensure $0 < \theta(t) < \pi$ when performing numerical evaluation of Eqs.(\ref{eq:HthetaZero}) and (\ref{eq:HthetaNoZero}).  

    Carlstedt and Mullikin~\cite{carlstedt66} declare that when there are no discrete roots $\nu_0 > 1$ of $\Lambda(z)$ that there are then no roots of $\lambda(\nu)$ for $\nu \in [-1,1]$, which we found to not hold in general (Figure~\ref{fig:lambdas} shows that $\lambda(\nu)$ will admit $0, 1$ or $2$ roots, depending on $d$ and $c$), but we noted no issues in applying Eqs.(\ref{eq:HthetaZero}) and (\ref{eq:HthetaNoZero}), provided $\theta(t)$ was strictly non-negative.

    Additional closed-form expressions for computing $H(z)$ are given in~\ref{sec:Heval}.

    \subsection{Law of diffuse reflection}

    We can derive the law of diffuse reflection for the half space by solving for the internal collision rate density $C(x)$ due to external unidirectional illumination along cosine $\mu_\ell$, the solution of the W-H equation (\ref{eq:WH2}) with source term (\ref{eq:Salbedo}).  This collision rate can be found using the Green's function for an internal isotropic plane source at depth $x_0$ (with $C_0(x) = \delta(x-x_0)$).  

    The Green's function $\mathbb{G}(x,x_0)$ is a source function such that $\mathbb{G}(x,x_0)dx$ is the rate of photons leaving collisions (or the source directly) from depths $dx$ about $x$.
    If we define the Laplace transform
    \begin{equation}
        \mathcal{L}_s\left[ f(x) \right] \equiv \int_0^\infty f(x) \e^{-s x} dx,
    \end{equation}
    then we have, from Ivanov (\cite{ivanov94}, Eqs. (19) and (21)), that the double Laplace transform of the Green's function is
    \begin{equation}
        \bar{\bar{\mathbb{G}}}(s,s_0) = \mathcal{L}_s\left[ \mathcal{L}_{s_0}\left[ \mathbb{G}(x,x_0) \right] \right] = \frac{H(1/s)H(1/s_0)}{s+s_0}.
    \end{equation}
    Before considering the external source, we first note a number of exact properties that relate to the life of a photon in the half space.  From $\mathbb{G}(x,x_0)$, we can find the collision rate density due to isotropic emission (or leaving a collision) at depth $x_0$.
    We convert the source function $\mathbb{G}(x,x_0)$ to a collision rate density by removing the Dirac delta for direct emission (since this is not a collision), which is
    \begin{equation}
        \mathcal{L}_s\left[ \mathcal{L}_{s_0}\left[ \delta(x - x_0) \right] \right] = \frac{1}{s+s_0},
    \end{equation}
    and then apply a factor $1 / c$, to convert densities for leaving collisions into densities for entering collisions,
    \begin{equation}
        \bar{\bar{C}}(s;s_0) = \frac{1}{c} \left( \frac{H(1/s)H(1/s_0)-1}{s+s_0}  \right).
    \end{equation}
    The double Laplace inversion of $\bar{\bar{C}}(s;s_0)$ gives the collision rate density $C(x;x_0)$ at any position $x \ge 0$ inside the half space, due to an isotropic plane source at depth $x_0$.
    We also easily have the mean number of collisions, by taking the Laplace inversion of $\bar{C}(0;s_0)$, which is, by Eq.(\ref{eq:sqrtepsilon}),
    \begin{equation}
        \bar{C}(0;s_0) = \frac{1}{c} \left( \frac{ (1-c)^{-1/2} \, H(1/s_0)-1}{s_0}  \right).
    \end{equation}
    Of those collisions, $1-c$ will absorb the photon, so the escape probability $p(x_0)$ after isotropic emission (or leaving a collision) at $x_0$ is then
    \begin{equation}\label{eq:escape}
        p(x_0) = 1 - (1-c) \mathcal{L}^{-1}_{x_0}\left[ \bar{C}(0;s_0) \right].
    \end{equation}

\begin{figure}
        \centering
        \includegraphics[width=.7 \linewidth]{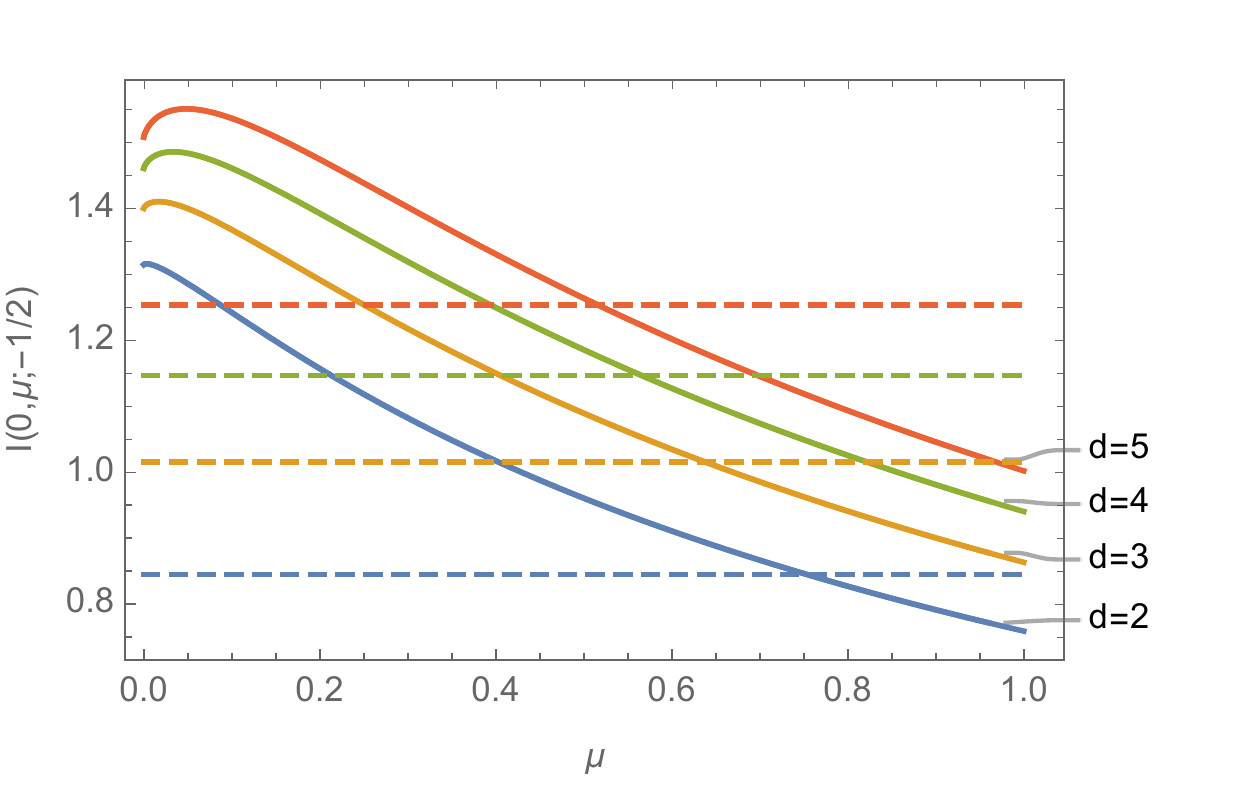}
        \caption{Comparison of integrated emergent intensity $I(0,\mu;-\mu_\ell)$ for various dimensions, $c = 0.9$, $\mu_\ell = 1/2$.  A Lambertian exitance (constant) of matched total albedo $R(\mu_\ell)$ is shown (dashed) for reference.}          
        \label{fig-emergent_compare_d} 
      \end{figure}

\begin{figure*}
        \centering
        \subfigure[Fixed single-scattering albedo ($c=0.9$)]{\includegraphics[width=0.9 \linewidth]{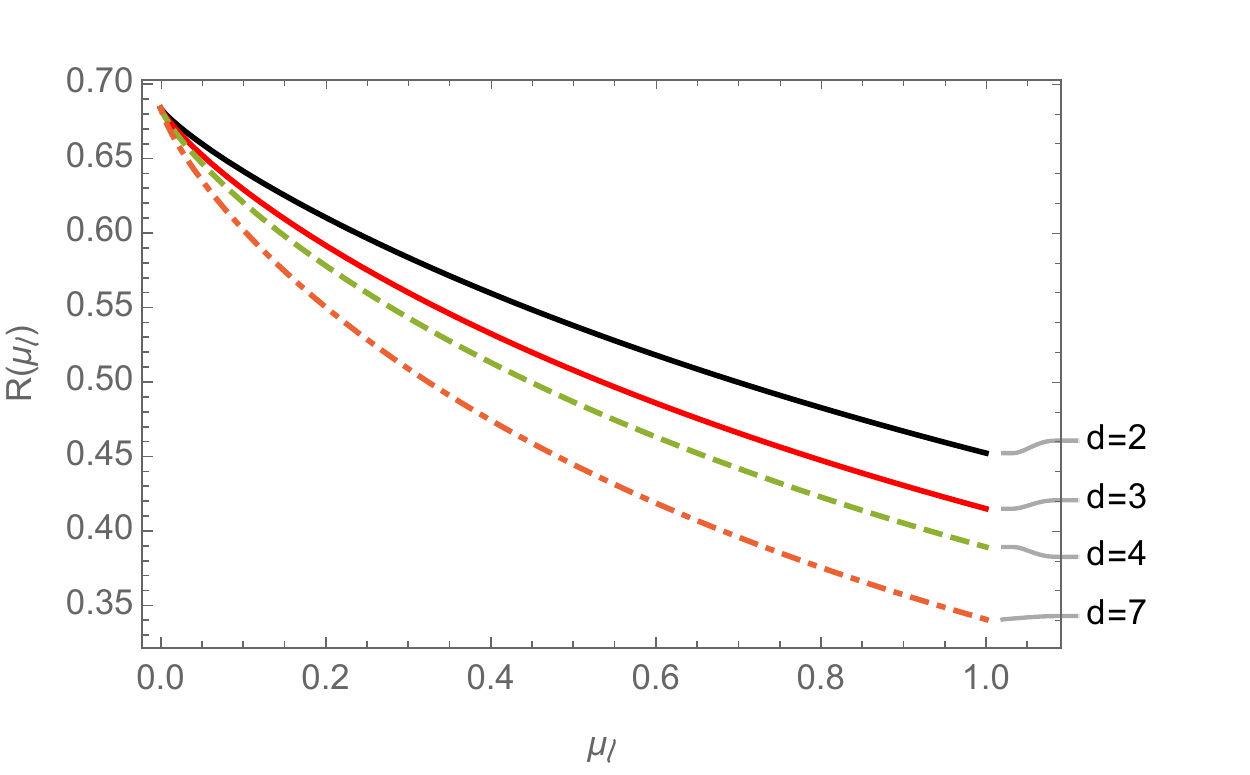}}
        \subfigure[Normal incidence ($\mu_\ell = 1$)]{\includegraphics[width=0.9 \linewidth]{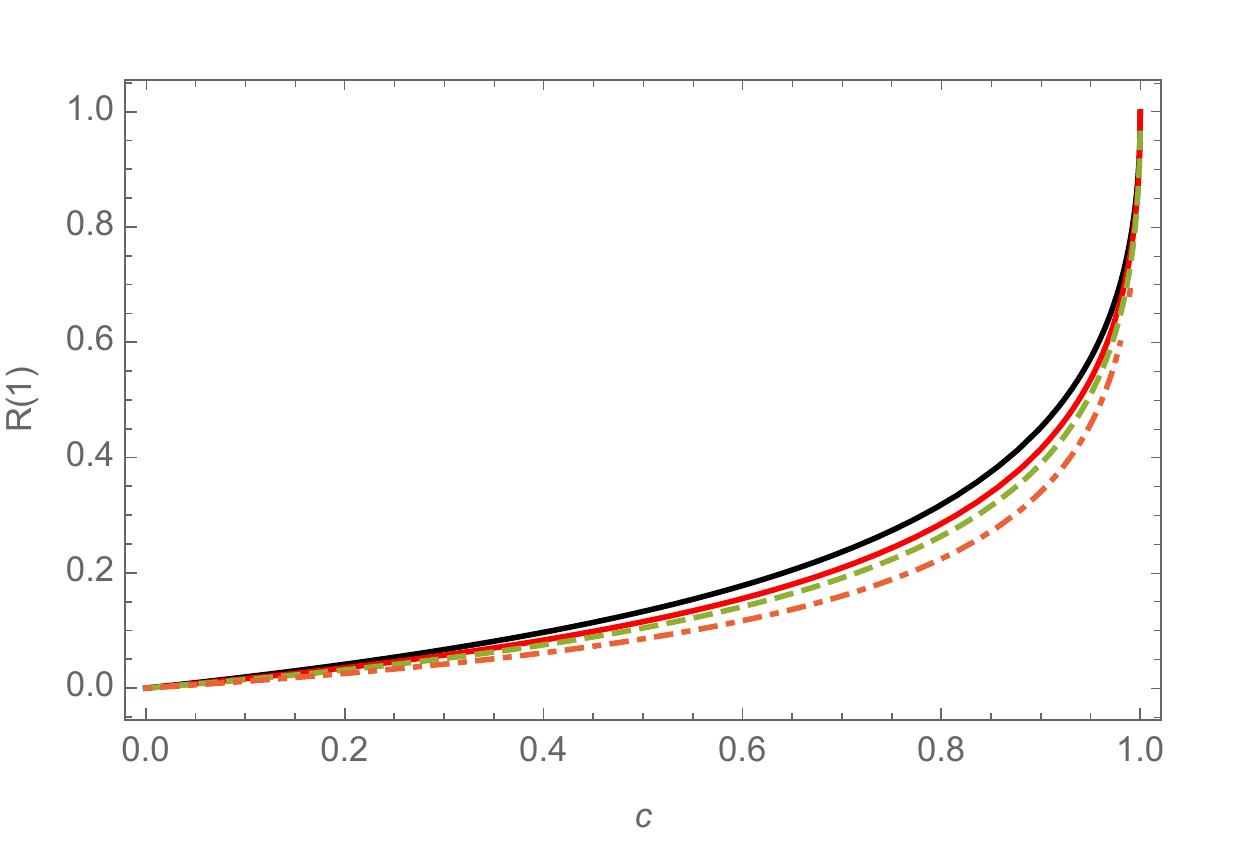}}
        \caption{Comparison of Eq.(\ref{eq:albedo}) for the total diffuse albedo of the half space for 2D, 3D, 4D, and 7D.}          
        \label{fig-albedo_compare_d} 
      \end{figure*}

    For unidirectional illumination along cosine $\mu_\ell$, the collision-rate density inside the half space will be
    \begin{equation}\label{eq:Cdelta}
        C(x) = \int_0^\infty \mathbb{G}(x,x_0) \frac{\e^{-x_0 / \mu_\ell }}{  \mu_\ell } d x_0 = \frac{1}{\mu_\ell} \bar{\mathbb{G}}(x,1/\mu_\ell).
    \end{equation}
    After combination of Eqs.(\ref{eq:I0mu}) and (\ref{eq:Cdelta}), we find
    \begin{equation}\label{eq:diffuselaw}
        I(0,\mu;-\mu_\ell) = \frac{c}{2} \frac{1}{\mu \mu_\ell} \bar{\bar{\mathbb{G}}}(1/\mu,1/\mu_\ell) = \frac{c}{2} \frac{H(\mu) H(\mu_\ell)}{\mu + \mu_\ell},
    \end{equation}
    which is the generalized law of diffuse reflection for a half space of general dimension.  The probability that a photon arriving along cosine $\mu_\ell$ escapes the half space along a direction within $d\mu$ of $\mu$ is $\mu I(0,\mu;-\mu_\ell) G(\mu) d \mu$, and so the total albedo of the half space is
    \begin{equation}\label{eq:albedo}
        R(\mu_\ell) = \int_0^1 \mu I(0,\mu;-\mu_\ell) G(\mu) d\mu = 1 - \sqrt{1-c} H(\mu_\ell),
    \end{equation}
    where we have used Eq.(\ref{2ndH(mu)eq}).  Equations~(\ref{eq:diffuselaw}) and (\ref{eq:albedo}) are in agreement with previous derivations for Flatland~\cite{EdEMMR18} and show how the familiar expressions for 3D are universal over dimension $d \ge 1$, with all variation due to the $H$ function and the measure $G(\mu) d\mu$.  The variation of the emergent distribution and albedo with respect to dimension $d$ is illustrated in Figures~\ref{fig-emergent_compare_d} and \ref{fig-albedo_compare_d}.

    \subsubsection{Low-order scattering}
     The once- and twice-scattered portions of the reflection law and albedo shine further light on the structure of the solutions and can also provide accurate approximations for high absorption.  These are found via Taylor expansions about $c = 0$ of Equations~(\ref{eq:diffuselaw}) and (\ref{eq:albedo}).  The once-scattered reflection law reduces to
    \begin{equation}\label{eq:diffuselaw1}
        I(0,\mu;-\mu_\ell|1) = \frac{c}{2} \frac{1}{\mu + \mu_\ell}.
    \end{equation}

\begin{figure}
        \centering
        \includegraphics[width= .95\linewidth]{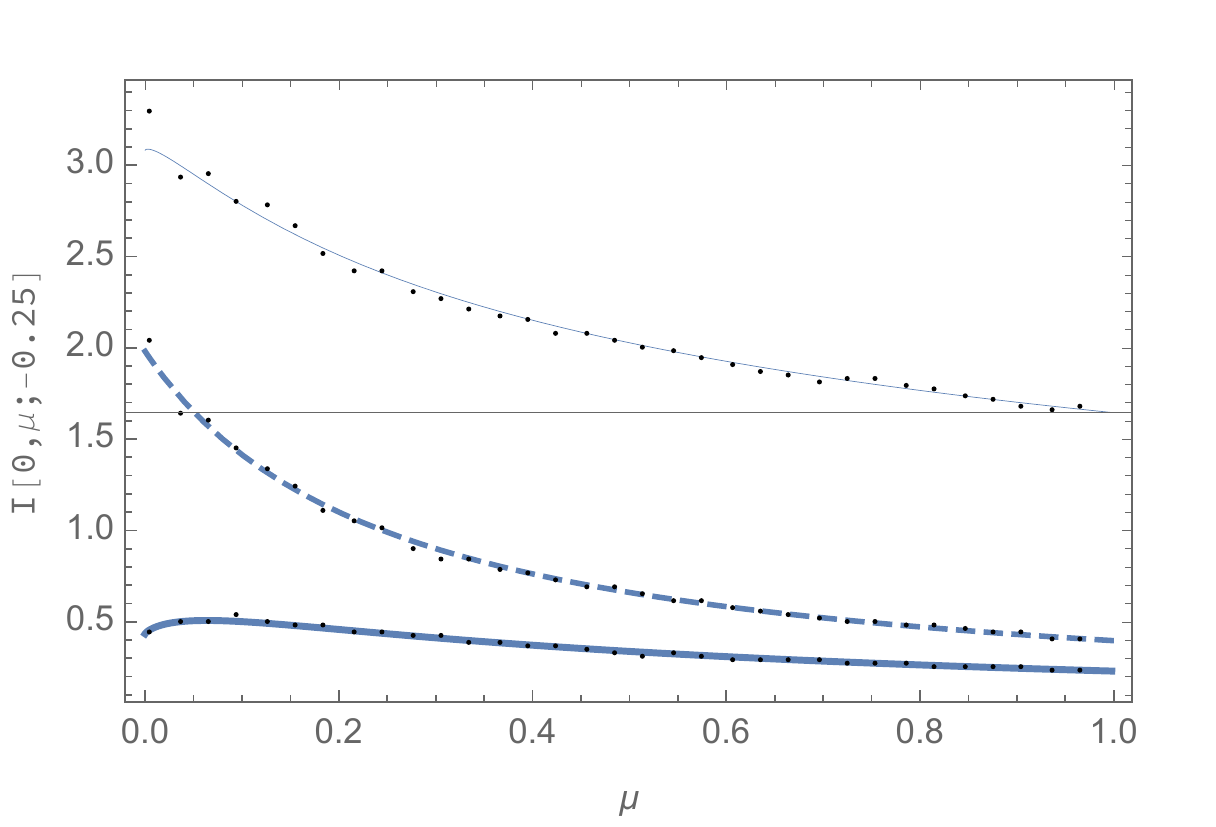}
        \caption{Emergent distribution $I(0,\mu;-\mu_\ell)$ from a 4D half space with isotropic scattering, $c = 0.8$, $\mu_\ell = 0.25$.  Total emergent distribution (Eq.(\ref{eq:diffuselaw}), thin), single-scattered component (Eq.(\ref{eq:diffuselaw1}), dashed), and double-scattered (Eq.(\ref{eq:diffuselaw2}), thick).  MC simulation shown as dots.}          
        \label{fig-brdf4D} 
      \end{figure}

    In the Taylor expansion of the reflection law for the twice-scattered emergent distribution, we encounter an integral that we observe is equal to the Laplace transform of the kernel,
    \begin{equation}
       \int_0^{\infty } \frac{\mu \left(1-\Lambda \left(\frac{i}{t}\right)\right)}{c \left(\pi 
   \left(1+t^2 \mu^2\right)\right)} \, dt = \mathcal{L}_{1/\mu} \left[ K(x) \right].
     \end{equation}
    The twice-scattered component of the reflection is then,
    \begin{equation}\label{eq:diffuselaw2}
        I(0,\mu;-\mu_\ell|2) = \frac{c^2}{2} \frac{\mathcal{L}_{1/\mu_\ell}\left[ K(x) \right]+\mathcal{L}_{1/\mu}\left[ K(x) \right]}{\mu_\ell + \mu}.
     \end{equation}
     We found the general form of the Laplace transform to reduce to
     \begin{equation}
         \mathcal{L}_{1/\mu}\left[ K(x) \right] =   \frac{1}{2} \left[\,
   _2F_1\left(\frac{1}{2},1;\frac{d}{2};\frac{1}{\mu^2}\right)-  \right. 
   \left. \frac{G(0)}{\mu (d-1)} \, _2F_1\left(1,1;\frac{1}{2}+\frac{d}{2};\frac{1}{\mu^2}\right)\right].
     \end{equation}
     Special cases include the known results for Flatland~\cite{EdEMMR18}
     \begin{equation}
        \mathcal{L}_{1/\mu}\left[ K(x) \right]  = \frac{\mu}{\pi} \frac{\text{sech}^{-1}(\mu)}{\sqrt{1 - \mu^2}},
     \end{equation}
     for 3D~\cite{sears75}
     \begin{equation}
        \mathcal{L}_{1/\mu}\left[ K(x) \right]  = \mu  \coth ^{-1}(1 + 2 \mu),
     \end{equation}
     and our new result for 4D,
     \begin{equation}
        \mathcal{L}_{1/\mu}\left[ K(x) \right]  =  \frac{\mu}{\pi} \left( 2 \sqrt{1-\mu ^2} \, \text{sech}^{-1}(\mu )+\pi  \mu -2 \right).
     \end{equation}
     Figure~\ref{fig-brdf4D} shows Monte Carlo validation of the emergent distributions for a 4D half space.

     The escape probability after exactly one scattering also involves the Laplace transform of the kernel,
     \begin{align}
        R(\mu_\ell|1) &= \int_0^1 \mu  I(0,\mu;-\mu_\ell|1) G(\mu) d\mu \\&= c \left( \frac{1}{2} - \mathcal{L}_{1/\mu_\ell}\left[ K(x) \right] \right).
     \end{align}
     An expression for $R(\mu_\ell|1)$ in terms of the characteristic function is also known~\cite{mullikin1968some}
     \begin{equation}
        R(\mu_\ell|1) = c \left[ \frac{1}{2} - \frac{\mu_\ell}{2} \int_0^1 \frac{G(\mu)}{\mu_\ell + \mu} d\mu \right].
     \end{equation}

    \subsubsection{Uniform diffuse illumination}

      We now consider the case of uniform diffuse illumination of the half space, with boundary condition given by Eq.(\ref{eq:boundarydiffuse}).  Integrating the diffuse reflection law (\ref{eq:diffuselaw}) against the source function
      \begin{equation}
        I(0,\mu) = \frac{\sqrt{\pi } \, \Gamma \left(\frac{d+1}{2}\right)}{\Gamma \left(\frac{d}{2}\right)} \int_0^1 \mu_\ell  I(0,\mu;-\mu_\ell) G(\mu_\ell) d \mu_\ell
      \end{equation}
      and applying Eq.(\ref{2ndH(mu)eq}), we find the emergent distribution
      \begin{equation}
        I(0,\mu) = \frac{\sqrt{\pi } \, \Gamma \left(\frac{d+1}{2}\right)}{\Gamma \left(\frac{d}{2}\right)} \left( 1 - \sqrt{1-c} H(\mu) \right)
      \end{equation}
      with albedo
      \begin{equation}\label{eq:diffusealbedo}
        R = \int_0^1 \mu \, I(0,\mu) G(\mu) \, d\mu.
      \end{equation}
      The emergent distribution $I(0,\mu)$ is proportional to Eq.(\ref{eq:albedo}), as it must be, by the optical reciprocity theorem.  For the conservative case, $c = 1$, $R$ reduces to Eq.(\ref{eq:Ghatnorm}), which is $1$ by construction.

      We find a number of analytic results for the single- and double-scattered components of the albedo for diffuse illumination, from a Taylor expansion about $c=0$ of Eq.(\ref{eq:diffusealbedo}).  
      For the probability of escape after one collision, in Flatland we find
      \begin{equation}
        R_1 = \frac{c}{16} \left(\frac{G_{3,3}^{3,2}\left(1\left|
        \begin{array}{c}
         -\frac{1}{2},0,\frac{3}{2} \\
         -\frac{1}{2},0,0 \\
        \end{array}
        \right.\right)}{\sqrt{\pi }}+12-3 \pi \right) \approx  0.214601836602552 c,
      \end{equation}
      for 3D\cite{hitchhiker}
      \begin{equation}
        R_1 = \frac{2}{3} c (1-\ln (2)),
      \end{equation}
      and for $d \ge 4$
      \begin{multline}
        R_1 =  \frac{c}{3}  \left(3 \,
       _3F_2\left(1,1,1-\frac{d}{2};\frac{3}{2},\frac{d}{2}+\frac{1}{2};1\right) \right. \\ \left. -\frac{4
       (d-2)}{d+1} \,
       _3F_2\left(2,2,2-\frac{d}{2};\frac{5}{2},\frac{d}{2}+\frac{3}{2};1\right)\right).
      \end{multline}
      It appears that for even dimensionalities $d \ge 4$, the escape probabilities are rational multiplies of $c$, including the remarkably simple single-scattering albedo for 4D, $R_1 = c / 5$.
      The double-scattering albedo for diffuse illumination in 4D is also a rational factor of $c^2$, $R_2 = 4 c^2 / 35$.

    \subsection{Coherent backscattering}

\begin{figure}
          \centering
          \includegraphics[width=.9\linewidth]{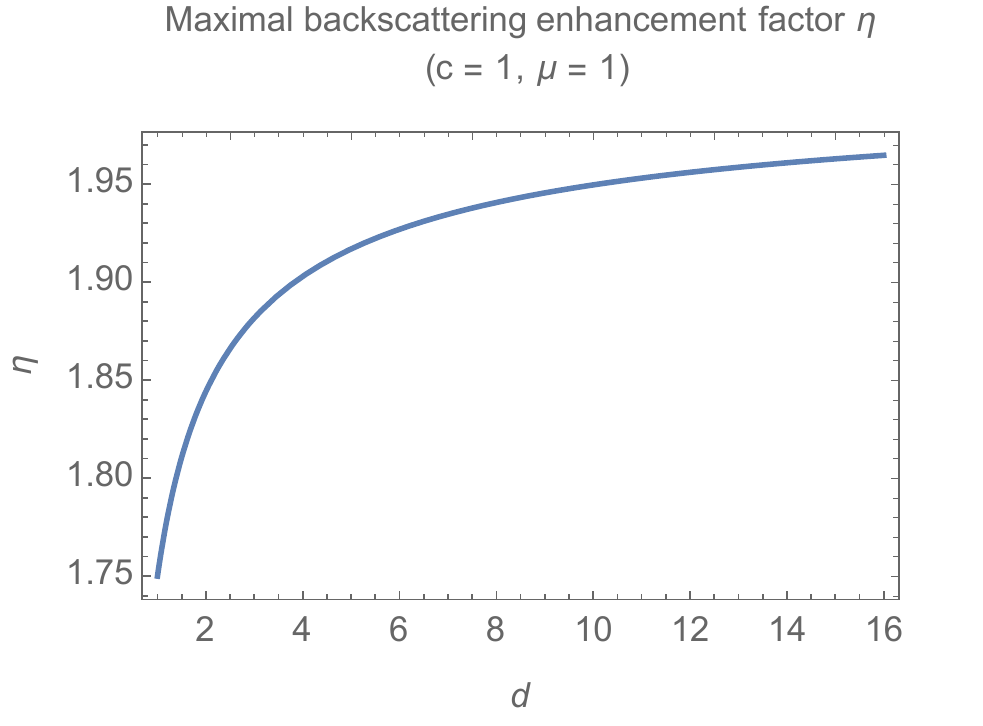}
          \caption{Maximal coherent backscattering enhancement factor $\eta$ under weak localization for an isotropically-scattering halfspace in $\mathbb{R}^d$ (normal incidence and conservative scattering).\label{fig:eta}} 
        \end{figure}

      In applications such as remote sensing it can be important to consider the effects of coherent backscattering.  Under the conditions of weak localization, the reflectance will be amplified in the backward direction by an enhancement factor $\eta$ that reaches a maximum for normal incidence and $c = 1$.  The maximal enhancement factors for 2D and 3D half spaces are~\cite{gorodnichev89,mishchenko1992multiple}
      \begin{equation}
        \eta = 2 - H(1)^{-2},
      \end{equation}
      which generalize to arbitrary dimension $d$ using Eq.(\ref{eq:HsolutionRd}), producing a monotonically increasing factor as dimension $d$ increases (Figure~\ref{fig:eta}), beginning at $\eta = 7/4$ for a 1D rod and approaching $\eta \rightarrow 2$.

\section{The albedo problem by Case's method}\label{sec:case}

  \subsection{Introduction}

    In this section we consider Case's approach~\cite{KMCPFZ67} to solving the $d$D albedo problem to illustrate its relationship to the Wiener-Hopf and resolvent approaches.  To solve Eq.(\ref{mu-transport}) with the classic Case-style eigenmode method\cite{KMCPFZ67}, we separate variables with the ansatz
    \begin{equation}
    I(x,\mu) = \phi(\nu,\mu)\exp(-x/\nu), \label{ansatz}
    \end{equation}
    for eigenvalues in the spectrum $\nu \in \sigma$, to obtain 
    \begin{equation}
    (\nu-\mu) \phi(\nu,\mu) = \frac{c\nu}{2}\int_{-1}^1 \phi(\nu,\mu) G(\mu) \, d\mu   \label{Gmuphinumu}.
    \end{equation}
    If we impose the normalization condition for the eigenmodes to be
    \begin{equation}
    \int_{-1}^1 \phi(\nu,\mu) G(\mu) \, d\mu = 1 \, , \quad \nu \in \sigma ,\label{Gmu normalization}
    \end{equation}
    then $\sigma = \{\nu \in [\,-1,1]\cup\pm \nu_0\}$ is the eigenvalue spectrum and the eigenmodes $\phi(\nu,\mu)$ satisfy the equation
    \begin{equation}
    (\nu - \mu)\phi(\nu,\mu) = c\nu/2,   \label{phinumu}
    \end{equation}
    identical to the form in a 3D domain (relative to our angular measure $G(\mu)$). From Eq.(\ref{phinumu}) the discrete eigenmodes, when they occur, satisfy
    \begin{equation}
    \phi(\pm \nu_0,\mu) = \frac{c\nu_0}{2(\nu_0 \mp \mu)} \label{discretemodes}
    \end{equation}
    with roots $\nu_0$ obtained from the dispersion function (\ref{eq:Lambdad}), which we have summarized in Section~\ref{sec:roots}.
    The eigenmodes for the continuum are
    \begin{equation}
    \phi(\nu,\mu) = \frac{c\nu}{2}\mathcal{P} \frac{1}{\nu-\mu} + \frac{\lambda(\nu)}{G(\nu)} \,\delta(\nu-\mu)\, ,\quad \nu \in [-1,1] \label{continuummodes}
    \end{equation}

    From Eqs.(\ref{GnDmu}) and (\ref{Gmu normalization}) it follows that the continuum eigenmodes also satisfy
    \begin{equation}
    \int_{-1}^1 \phi(\nu,\mu) G(\mu) d\mu = \ \!c\,\, _2F_1\left(\frac{1}{2},1;\frac{d}{2};\frac{1}{\nu^2} \right), \quad \nu \in \sigma.
    \end{equation}
    
    In this section, the collimated incident and diffuse illuminations for the albedo problem are selected to be
    \begin{align}
    	&I(0,\mu;\mu_\ell) = \delta(\mu-\mu_\ell), \quad  \mu, \mu_\ell \in [0,1]\label{deltabc} \\
    &I_k(0,\mu) = \mu^k , \quad \mu \in [0,1] \dots \, ,  \quad k = -1,\!
     0,\! 1 \label{kbc}
    \end{align}
    with $I(x,\mu;\mu_\ell)$ and $I_k(x,\mu)$ tending to $0$ as $x \rightarrow \infty$. The special case $k = -1$ corresponds to a unit current illumination. The notation $\psi(\mu)$ will be used henceforth to denote either surface illumination. 

    The constraint $I(x,\mu) \rightarrow 0$ as $x \rightarrow \infty$ forces the eigenmode expansion for the albedo problem to be written as
    \begin{multline}\label{half-range-expansion}
    I(x,\mu) = A(\nu_0)\phi(\nu_0,\mu)\exp(-x/\nu_0)  + \int_0^1 A(\nu)\phi(\nu,\mu)\, \exp(-x/\nu) \, d \nu \\
    \equiv \int_{\sigma \, +} A(\nu)\phi(\nu,\mu)\exp(-x/\nu) \, d\nu \ , \quad \mu \in [-1,1],
    \end{multline}
    where $\sigma  + = \{\nu \in [\,0,1] \, \cup \, \nu_0\} $ defines the half-spectrum subset of the full spectrum of eigenvalues $\sigma$. 
    From Eqs.(\ref{deltabc}), (\ref{kbc}), and 
    (\ref{half-range-expansion}) it follows that the expansion coefficients $A(\nu_0)$ and $A(\nu)$, $\nu \in [0,1]$, are to be determined from
    \begin{equation}
    I(0,\mu) \equiv \psi(\mu) = \int_{\sigma +} A(\nu) \phi(\nu,\mu) \,d\nu, \quad \mu \in [0,1] . \label{psi eq}
    \end{equation}

    We follow the approach in \cite{NMIK73} and \cite{NM15}\footnote{Tables 1, 3 and 4 of reference \cite{NM15} have numerical errors. The values for the column labeled $j_{ratio,1}(\mu_0)$ in Table 1 are all incorrect. The correct values for Table 3 with $c=0.7$ are $<\!\!x^2(\mu_0)\!\!> \,\,= 3.36647$ and $<\!\!x^3(\mu_0)\!\!>\,\,= 12.45400$ for $\mu_0 = 0.9$ and \\$<\!\!x^2(\mu_0)\!\!> \,\,= 3.83320$ and $<\!\!x^3(\mu_0)\!\!>\,\,= 14.86410$ for $\mu_0 = 1.0$.  The top right value of Table 4 should be $1.91257$.} to construct a Chandrasekhar $H(\mu)$ function by forcing the eigenfunctions $\phi(\nu,\mu)$ to obey a half-range-in-$\mu$ transport equation analogous to Eq.(\ref{Gmuphinumu}),
    \begin{equation}
    (\nu-\mu) \phi(\nu,\mu) = \frac{c\nu}{2}\int_{0}^1 \phi(\nu,\mu)  H(\mu) G(\mu) \, d\mu \, ,\quad \nu \in \sigma + . \label{Gmuphinumu+}
    \end{equation}
    Equation (\ref{Gmuphinumu}) then forces the constraint
    \begin{equation}
    \int_0^1 \phi(\nu,\mu)\, H(\mu) G(\mu) \, d\mu = 1 \, ,\quad \nu \in \sigma +.  \label{GH normalization}
    \end{equation}

    With the substitution of $\phi(\nu,\mu)$ from Eqs.(\ref{discretemodes}) and (\ref{continuummodes}) into Eq.(\ref{GH normalization}), the following equations are valid,
    \begin{equation}
    \frac{c\nu_0}{2}\int_0^1 \frac{H(\mu)G(\mu)}{\nu_0 - \mu}  \, d\mu = 1 \label{nuequation}
    \end{equation}
    and
    \begin{equation}
    \frac{c\nu}{2} \mathcal{P} \int_0^1 \frac{H(\mu)G(\mu)}{\nu - \mu} \, d\mu + \lambda (\nu)H(\nu) = 1 ,\quad \nu \in [0,1]\, .
    \end{equation}
    This suggests we construct $H(\mu)$ by considering in the complex plane the equation
    \begin{equation}
    \frac{cz}{2}\int_0^1 \frac{H(\mu)G(\mu)}{z-\mu} \, d\mu + \Lambda(z)H(z)=1\, , \quad z \notin [-1,1] \label{zequation}
    \end{equation}
    and examining the analyticity properties of $H(\mu)$, $0 \le \mu \le 1$. The factors $\Lambda(z)H(z)$ and $H(-z)$ are both continuous across $(-1,0)$; similarly, $\Lambda(z)H(-z)$ and $H(z)$ are continuous across $(0,1)$ so $\Lambda(z)H(z)H(-z)$ is analytic along $(-1,1)$. The points $z=\pm 1$ can be included so $\Lambda(z)H(z)H(-z)$ is analytic in the entire plane and from Liouville's theorem approaches a constant. With $H(0)=\Lambda(0) =1$ the Wiener-Hopf identity results,
    \begin{equation}
        H(z)H(-z) = 1/\Lambda(z), \label{WeinerHopf}
    \end{equation}
    with $H(z)$ satisfying Eq.(\ref{H(z)eq}) subject to the constraint \linebreak $1/H(-\nu_0) = 0$ imposed by Eq.(\ref{nuequation}). Thus, for $H(\mu)$, $0 \le \mu \le 1$, Eq.(\ref{1stH(mu)eq}) is recovered.  Equation (\ref{2ndH(mu)eq}) is needed, for example, to show that multiplication of Eq.(\ref{Gmuphinumu+}) by $G(\mu)H(\mu)$ and integration over $\mu \in [0,1]$ gives
    \begin{eqnarray}
    	\int_0^1 \phi(\nu,\mu)\,\mu\, H(\mu) G(\mu) \, d \mu = \nu (1-c)^{1/2} , \,  \nu \in \sigma + \label{GH+ normalization}
    \end{eqnarray}
    after use of Eq.(\ref{alpha0}). 

    The results of Eqs.(\ref{GH normalization}) and (\ref{GH+ normalization}) can be generalized, following a partial fraction analysis along with use of Eq.(\ref{alphadef}), to show that
    \begin{multline}
    	U_{k+1}(\nu) = \int_0^1 \mu^{k+1}\phi(\nu,\mu)\, H(\mu)G(\mu) \,d\mu =  \\
     \nu^{k+1}(1-c)^{1/2}- \frac{c}{2} \sum_{j=1}^k \nu^{k+1-j}\alpha_j,\quad k \ge 0,\ \nu \in \sigma + \label{Kk+1def}
    \end{multline}
    if the convention $\sum_{j=1}^0 X_j \equiv 0$ is understood here and elsewhere. For $k=-1$, Eq.(\ref{GH normalization}) gives $U_0(\nu) = 1$, $\nu \in \sigma +$.

\subsection{Orthogonality relations}

    Multiply Eq.(\ref{Gmuphinumu+}) by $\nu^{-1}\phi(\nu',\mu)$ and, in a second equation for $\nu'$, multiply by $\nu'^{-1} \phi(\nu,\mu)$ and then integrate both results over $\mu \in [0,1]$ and subtract to obtain
    \begin{equation}
        \int_0^1 \phi(\nu,\mu) \phi(\nu',\mu) \mu \,H(\mu) G(\mu) \,d\mu = 0, \quad \nu \ne \nu', \ \nu,\nu' \in {\sigma+} . \label{orthogonality}
    \end{equation}
    The corresponding normalization equations when $(d-3)/(d-2) < c< 1$ are
    \begin{align}
        &\int_0^1 \phi^2(\nu_0,\mu) \mu \, H(\mu)G(\mu) \,d\mu =  N(\nu_0)H(\nu_0)\label{discrete normalization} \\
        &\int_0^1 \phi(\nu,\mu)\phi(\nu',\mu) \mu \,H(\mu)G(\mu) \,d\mu = N(\nu) H(\nu)\,\delta(\nu - \nu'),  \label{continuum normalization}
    \end{align}
    for $\nu,\nu' \in (0,1)$. 

    The discrete normalization (\ref{discrete normalization}) can be derived by first differentiating Eq.(\ref{2ndH(mu)eq}), written for $-z$, and using Eq.(\ref{WeinerHopf}); then for $(d-3)/(d-2) < c< 1$ and with Eq.(\ref{discretemodes}) and $\Lambda(\nu_0) = 0$, it follows that
    {\small \begin{align}
        N(\nu_0) &= \int_{-1}^1 \phi^2(\nu_0,\mu)\,\mu \,G(\mu) \,d\mu 
        = \left. \frac{c\nu_0^2}{2} \frac{d \Lambda(z)}{dz} \right|_{z = \nu_0} \label{eq:Nnu01} \\
        &= c^2 \frac{ _2F_1\left(\frac{3}{2},2;\frac{d}{2}+1;\frac{1}{\nu _0^2}\right)}{d \, \nu _0} \\
        &= \frac{c \nu _0}{\nu
   _0^2-1} \left(1-(d-1)\frac{\nu _0^2}{2}   \left(1-c \,
   _2F_1\left(-\frac{1}{2},1;\frac{d}{2};\frac{1}{\nu _0^2}\right)\right)\right).  \label{Nnu0}
    \end{align}
    }
    For $d \in \{1,2,3,4\}$ we find, respectively,
    \begin{equation*}
      \left\{ \sqrt{1-c}, \frac{\sqrt{1-c^2}}{2 c}, \frac{c \nu _0}{2}  \left(\frac{c \nu _0^2}{\nu _0^2-1}-1\right) , \frac{\sqrt{(1-c) c}}{4 c-2}  \right\} \subset N(\nu_0).
    \end{equation*}
    Another convenient expression for $N(\nu_0)$, generalizing the derivation of Case and Zwiefel~\cite{KMCPFZ67} (p. 68), is found from writing the dispersion equation (\ref{eq:Lambdad})
    \begin{equation}\label{eq:dispersionF}
      \Lambda(\nu_0) = 1 - c F(\nu_0) = 0,
    \end{equation}
    where
    \begin{equation}
      F(z) = \, _2F_1\left(\frac{1}{2},1;\frac{d}{2};\frac{1}{z^2} \right).
    \end{equation}
    Differentiating (\ref{eq:dispersionF}) with respect to $c$, we find
    \begin{equation}
      c \frac{\partial \nu_0}{\partial c} F'(\nu_0)+F(\nu_0) = 0.
    \end{equation}
    From Eq.(\ref{eq:dispersionF}), we also have $F(\nu_0) = 1/c$.  Combining with Eq.(\ref{eq:Nnu01}), we find
    \begin{equation}
      \frac{1}{2 \, N(\nu_0)} = \frac{1}{\nu_0^2} \, \frac{\partial \nu_0}{\partial c} .
    \end{equation}

    The continuum normalization of
    \begin{equation}\label{N(nu)}
        N(\nu) = \nu  \Lambda^+(\nu)\Lambda^-(\nu)  / G(\nu), \quad \nu \in [-1,1]
    \end{equation}
    can be derived with the Poincar\'{e}-Bertrand formula and $\Lambda^{\pm}(\nu)$ from Eq.(\ref{Lambda pm}) yielding
    \begin{equation}
            N(\nu) = \frac{\nu}{G(\nu)} \left[ \lambda(\nu)^2 + \left(\frac{1}{2} c \pi \nu G(\nu) \right)^2 \right],
    \end{equation}
    where $\lambda(\nu)$ is given by Eq.(\ref{eq:lambdad}).
    Use of that formula enables an interchange of the order of integration from $\int_0^1 d\mu \int_{\sigma +} d\nu$ to $\int_{\sigma +} d\nu \int_0^1 d\mu$ when using the orthogonality relations.

    With the orthogonality relations we can determine the half-range expansion coefficients $A(\nu)$ in Eq.(\ref{psi eq}) from
    \begin{equation}
        A(\nu) = \frac{1}{N(\nu)H(\nu)}
        \int_0^1 \psi (\mu) \phi(\nu,\mu)\mu \,  H(\mu) G(\mu) \,d\mu, \nu \in \sigma +.  \label{A(nu)}
    \end{equation}
    Completeness of the eigenfunctions is assured by virtue of the closure relation
    \begin{equation}
        \delta(\mu-\mu_\ell) = \mu_\ell G(\mu_\ell)H(\mu_\ell) \int_{\sigma +} \frac{\phi(\nu,\mu)\phi(\nu,\mu_\ell)}{N(\nu)H(\nu)}  d\nu, \label{closure}
    \end{equation}
    for $\mu,\,\, \mu_\ell \in (0,1)$.  This last equation can be derived as in \cite{IKNMGS64,IKFS65}, again with the Poincar\'{e}-Bertrand formula, or easily confirmed by solving Eq.(\ref{A(nu)}) for $\psi(\mu) = \delta(\mu-\mu_\ell)$  and substituting the result into Eq.(\ref{psi eq}) to verify closure.

    Three identities directly follow from Eq.(\ref{closure}).  For the first, multiply by $1/\mu_\ell$ and integrate over $\mu_\ell \in [0,1]$; after interchanging orders of integration and using Eq.(\ref{GH normalization}) it follows that 
    \begin{subequations}
    \begin{equation}
        \int_{\sigma +} \frac{\phi(\nu,\mu)}{N(\nu)H(\nu)} \, d\nu = \frac{1}{\mu}\, . \label{identity1}
    \end{equation}
    Similarly, a direct integration of Eq.(\ref{closure}) over $\mu_\ell \in [0,1]$ gives, with the help of Eq.(\ref{GH+ normalization}), 
    \begin{equation}
        \int_{\sigma +} \frac{\nu \phi(\nu,\mu)}{N(\nu)H(\nu)} \, d\nu =  \frac{1}{(1-c)^{1/2}} \,.\label{identity2}
    \end{equation}
    Multiplication of this last equation by $G(\mu)H(\mu)$, integration over $\mu \in [0,1]$, and Eq.(\ref{alphadef}) gives
    \begin{equation}
        \int_{\sigma +} \frac{\nu}{N(\nu)H(\nu)} \, d\nu = \frac{\alpha_0}{(1-c)^{1/2}} \, . \label{identity3}
    \end{equation}
    \end{subequations}
    Yet another identity can be derived if Eq.(\ref{closure}) is multiplied by $\mu_\ell^n d \mu_\ell$ and $G(\mu)H(\mu)d\mu$ and the result integrated for both variables over $[0,1]$ to obtain
    \begin{equation}
    	\int_{\sigma +} \frac{U_{n+1}(\nu)}{N(\nu)H(\nu)} d\nu = \alpha_n, \, \, \, \, n \ge 0.
    \end{equation}
    For $ n = 0 $ the last result is consistent with Eq.(\ref{GH+ normalization}).

    Other equations can be derived by rewriting Eq.(\ref{2ndH(mu)eq}) for $-z$ and $-z'$ and subtracting the resulting equations to obtain
    \begin{multline}\label{eq:134}
        \int_0^1 \frac{cz}{2} \frac{1}{z-\mu'} \frac{cz'}{2} \frac{1}{z'-\mu'} \, \mu' \, H(\mu')G(\mu')\, d\mu' = \\  \frac{czz'}{2(z-z')} \left[ \frac{1}{H(-z)} - \frac{1}{H(-z')} \right ], \, \, \, \, \, z \neq z' 
    \end{multline}
    before specializing $z$ and $z'$ to variables $\nu$ and/or $\mu$.  After taking the appropriate limits as $z$ and $z'$ approach eigenvalues, all the orthogonality results can be condensed into the formula
    \begin{multline}
      \int_0^1 \phi(\nu,\mu) \phi(\nu',\mu) \mu G(\mu)H(\mu) d\mu  \\= [1-\Xi(\nu)]N(\nu)H(\nu)\delta(\nu - \nu')  
       -\Xi(\nu) \frac{\nu \phi(\nu',\nu)}{H(-\nu)} - \Xi(\nu') \frac{\nu'\phi(\nu,\nu')}{H(-\nu')}, \label{All_Orthog_Eqs}
    \end{multline}
    where $\Xi(\nu) = 0$ for $0 \le \nu \le 1$ and 1 otherwise.
     Thus, we find
    \begin{equation}
        \phi(\nu,-\mu) = \mu^{-1} H(\mu) \int_0^1 \phi(\nu,\mu') \phi(-\mu,\mu')\,\mu' \,H(\mu')G(\mu') \,d\mu' . \label{phinu-mu}
    \end{equation}

    Equation (\ref{phinu-mu}) leads to the reflection relation (or ``albedo operator'') that allows us to conveniently express the outgoing radiation from the surface in terms of the ingoing radiation.  To derive that equation, observe from Eq.(\ref{psi eq}) that
    \begin{equation}
        I(0,-\mu) = \int_{\sigma +} A(\nu) \phi(\nu,-\mu)\, d\nu \label{psi -mu}, \, \, \, \, \, \, \mu \in [0,1]
    \end{equation}
    so insert Eq.(\ref{phinu-mu}) into Eq.(\ref{psi -mu}), interchange the order of integrations, and use Eq.(\ref{psi eq}) to obtain 
    \begin{equation}
        I(0,-\mu) =  \mu^{-1}H(\mu) \int_0^1 \psi(\mu') \phi(-\mu,\mu')  \mu'  H(\mu') G(\mu') \,d\mu'\, . \label{reflection}
    \end{equation}

  \subsection{Albedo problem spatial moments}

    Substitution of Eqs.(\ref{deltabc}) and (\ref{kbc}) into Eq.(\ref{A(nu)}), followed by the use of Eq.(\ref{half-range-expansion}), yields general equations for the angular intensities for the collimated and diffuse illuminations, respectively, as
    \begin{subequations}
    \begin{equation}\label{a}
        I(x,\mu;\mu_\ell) =   \mu_\ell G(\mu)H(\mu_\ell)\int_{\sigma +} \frac{\phi(\nu,\mu)\phi(\nu,\mu_\ell)}{N(\nu)H(\nu)} \exp(-x/\nu) d\nu 
    \end{equation}
    and
    \begin{equation}\label{b}
        I_k(x,\mu) =   \int_{\sigma +} \frac{\exp(-x/\nu)}{N(\nu)H(\nu)} d\nu \int_0^1 \mu’^k \phi(\nu,\mu’) \mu’  H(\mu’) G(\mu’) d\mu’ .
    \end{equation}
    From Eq.(\ref{phinu-mu}), Eq.(\ref{b}) also can be written as
    \begin{equation}
        I_k(x,\mu) = \int_{\sigma +} \frac{U_{k+1}(\nu)}{N(\nu)H(\nu)} \exp(-x/\nu) d\nu . \label{c}
    \end{equation}
    \end{subequations}

    The objective here is to compute the $n$th spatial moments of the flux within the half-space, as given for $n = 0,\! 1,\! 2,$ etc., by
    \begin{align}
        &\rho_n(\mu_\ell) = \int_0^\infty x^n dx \int_{-1}^1 I(x,\mu;\mu_\ell)G(\mu) \, d\mu \, ,  \label{rhonmuelldef} \\
        &\rho_{n,k} = \int_0^\infty x^n dx \int_{-1}^1  I_k(x,\mu)G(\mu) \, d\mu\, , \ k = -1,\! 0,\! 1\! \dots \, . \label{rhonkdef}
    \end{align}
    Also of interest are the ratios of fluxes that give the $n$th mean distances of travel in $d$D before absorption or escape from the half space,
    \begin{subequations}
    \begin{eqnarray}
        \left< x^n(\mu_\ell) \right> &=& \rho_n(\mu_\ell)/\rho_0(\mu_\ell)  \label{xnmu} \\
        \left< x^n \right>_k &=& \rho_{n,k}/\rho_{0,k} \, .\label{xnk}
    \end{eqnarray}
    \end{subequations}

    Substitution of Eq.(\ref{half-range-expansion}) into either Eq.(\ref{rhonmuelldef}) or (\ref{rhonkdef}) gives a spatial moment
    \begin{equation}
        \rho_{\,n} = \int_{\sigma +} A(\nu)d\nu \int_{-1}^1 \phi(\nu,\mu) G(\mu) \,d\mu \int_0^\infty x^n\exp(-x/\nu) \,dx .
    \end{equation}
    After substitution of $A(\nu)$ from Eq.(\ref{A(nu)}), followed by integration over $x$ and $\mu$ and use of Eq.(\ref{Gmu normalization}), the result is
    \begin{equation}
        \rho_{\,n} = n!\int_{\sigma +} \frac{\nu^{n+1}}{N(\nu)H(\nu)}\,d\nu \int_0^1 \psi(\mu) \phi(\nu,\mu) \mu H(\mu) G(\mu) \,d\mu , \ n= 0,\! 1,\! 2 \dots   \label{rhon-full}
    \end{equation}
    that is the $d$D equivalent to the 3D Eq.(I42)\footnote{Eq.(XX) from \cite{NM15} will be cited as Eq.(IXX)} that has no factor $G(\mu)$ in the integral.

        Use of the Dirac delta $\psi(\mu)$ of Eq.(\ref{deltabc}) in Eq.(\ref{rhon-full}) gives
        \begin{equation}
            \rho_n(\mu_\ell) = n!\mu_\ell G(\mu_\ell)H(\mu_\ell)M_{n+1}(\mu_\ell) \label{rhosubnmuell}
        \end{equation}
        where 
        \begin{equation}
            M_{n+1}(\mu_\ell) = \int_{\sigma +} \frac{\nu^{n+1}\phi(\nu,\mu_\ell)}{N(\nu)H(\nu)}\, d\nu \, , \quad \mu_\ell \in [\,0,1] .\label{Mn+1def}
        \end{equation}
        Multiplication of closure relation (\ref{closure}) by $\mu^{n+1}H(\mu)G(\mu)d\mu$ and integration over $\mu$ gives
        \begin{equation}\label{eq:146}
            \mu^n_\ell = \int_{\sigma +} \frac{U_{n+1}(\nu)\,\phi(\nu,\mu_\ell)}{N(\nu)H(\nu)}\, d\nu.
        \end{equation}
        After substitution of Eq.(\ref{Kk+1def}) into (\ref{eq:146}) and a rearrangement of terms, it follows from Eq.(\ref{rhosubnmuell}) that
        \begin{equation}
        \rho_{\,n}(\mu_\ell) = \frac{n!}{(1-c)^{1/2}}\left[ \mu_\ell^{n+1} G(\mu_\ell)H(\mu_\ell) +  \frac{c}{2}\sum_{j=1}^n \frac{\alpha_j}{(n-j)!}\,\rho_{n-j}(\mu_\ell) \right], \ n \ge 0, \label{rhonrecursion}
        \end{equation} 
        with the understanding that $\sum_{j=1}^0 X_j \equiv 0$.

        Equations (\ref{xnmu}) and (\ref{rhonrecursion}) then give the recursion relation
        \begin{equation}
        \left<x^n(\mu_\ell)\right> = n!\left[\mu_\ell^n + \frac{c}{2(1-c)^{1/2}} \sum_{j=1}^n\frac{\alpha_j}{(n-j)!} \left<x^{n-j}(\mu_\ell) \right > \right ], \quad n \ge 1
        \end{equation}
        with $\left<x^0(\mu_\ell)\right> \equiv 1.$ The first two equations, for example, are
        \begin{align}
        &\left<x(\mu_\ell)\right> = \mu_\ell + \frac{c\alpha_1}{2(1-c)^{1/2}} \label{eq:meanx} \\
        &\left<x^2(\mu_\ell)\right> = 2\mu_\ell^2 + \frac{c}{(1-c)^{1/2}}\left[ \alpha_1\mu_\ell + \frac{c\alpha_1^2}{2(1-c)^{1/2}} + \alpha_2 \right] \, . \label{eq:meanx2}
        \end{align}
        The results for $\left< x^2 (\mu_\ell) \right>$ and $\left< x(\mu_\ell)\right>$ are useful because they can be combined to give the variance $V(x(\mu_\ell))$ of the spatial distribution,
        \begin{equation}
            V(x(\mu_\ell)) = \left< (x(\mu_\ell) - \left< x(\mu_\ell) \right>)^2 \right> = \left< x^2(\mu_\ell) \right > - \left< x(\mu_\ell) \right>^2  . \label{variance}
        \end{equation}
        \begin{figure*}
          \centering
          \subfigure[Fixed absorption $c = 0.7$]{\includegraphics[width=0.49 \linewidth]{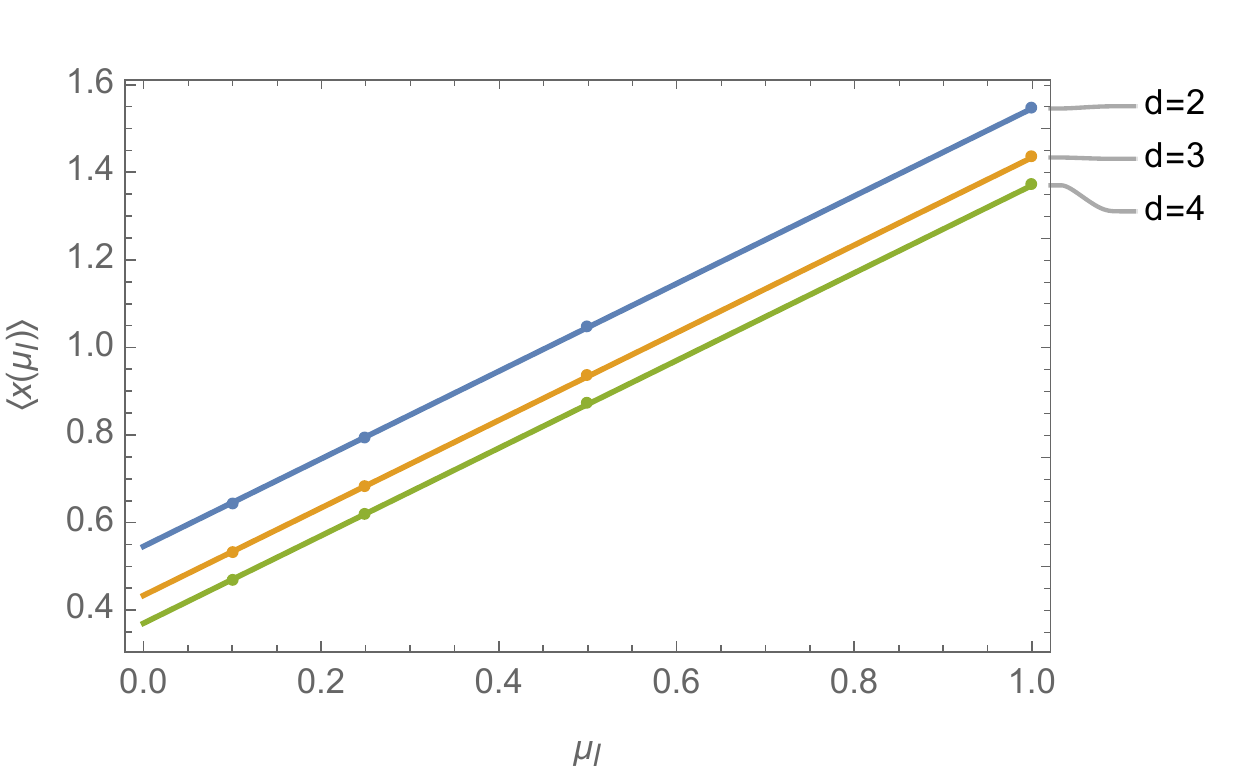}}
          \subfigure[Normally-incident illumination]{\includegraphics[width=0.49 \linewidth]{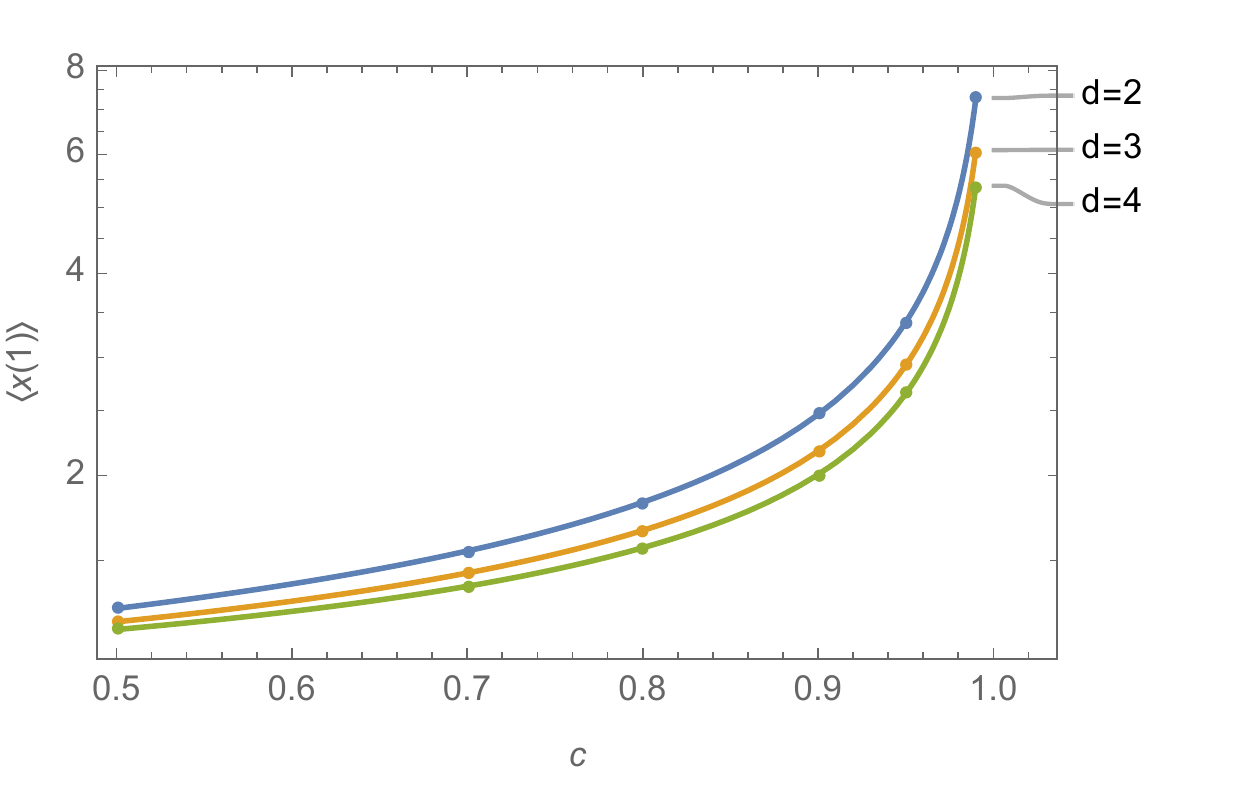}}
          \subfigure[Fixed absorption $c = 0.7$]{\includegraphics[width=0.49 \linewidth]{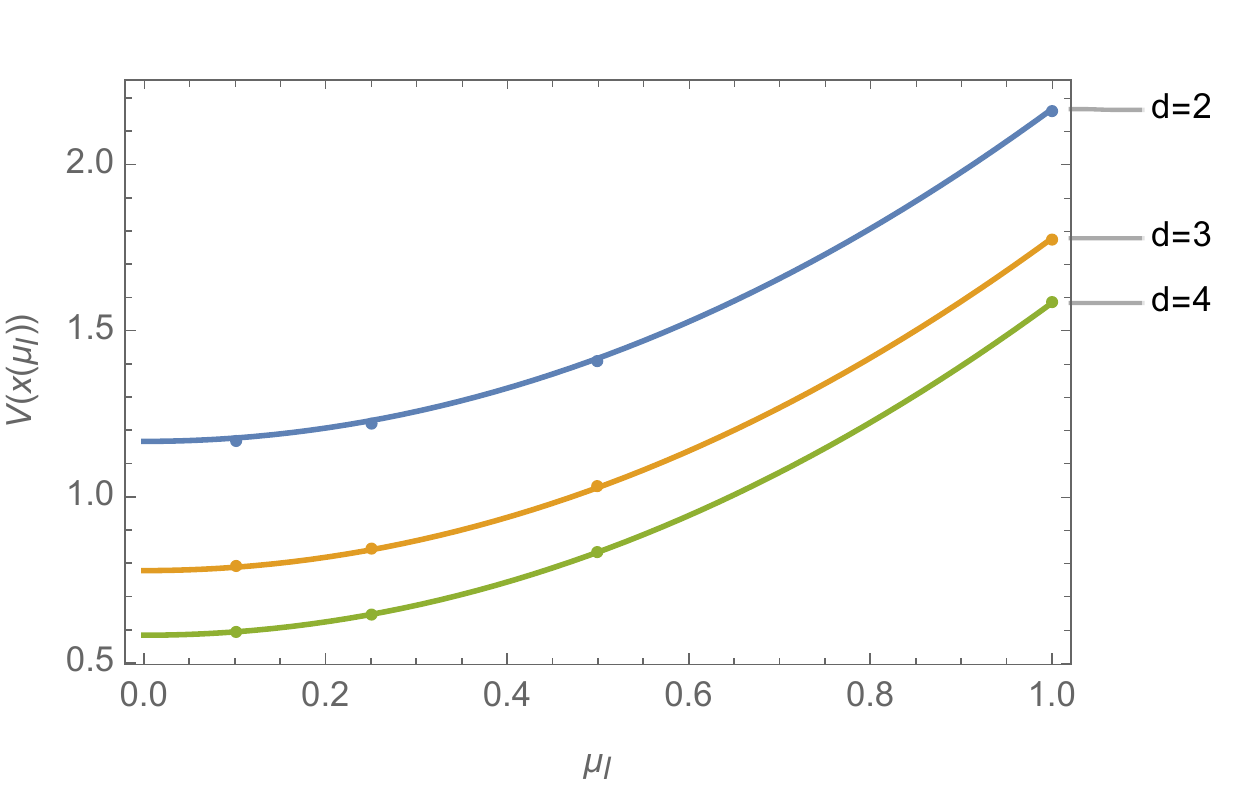}}
          \subfigure[Normally-incident illumination]{\includegraphics[width=0.49 \linewidth]{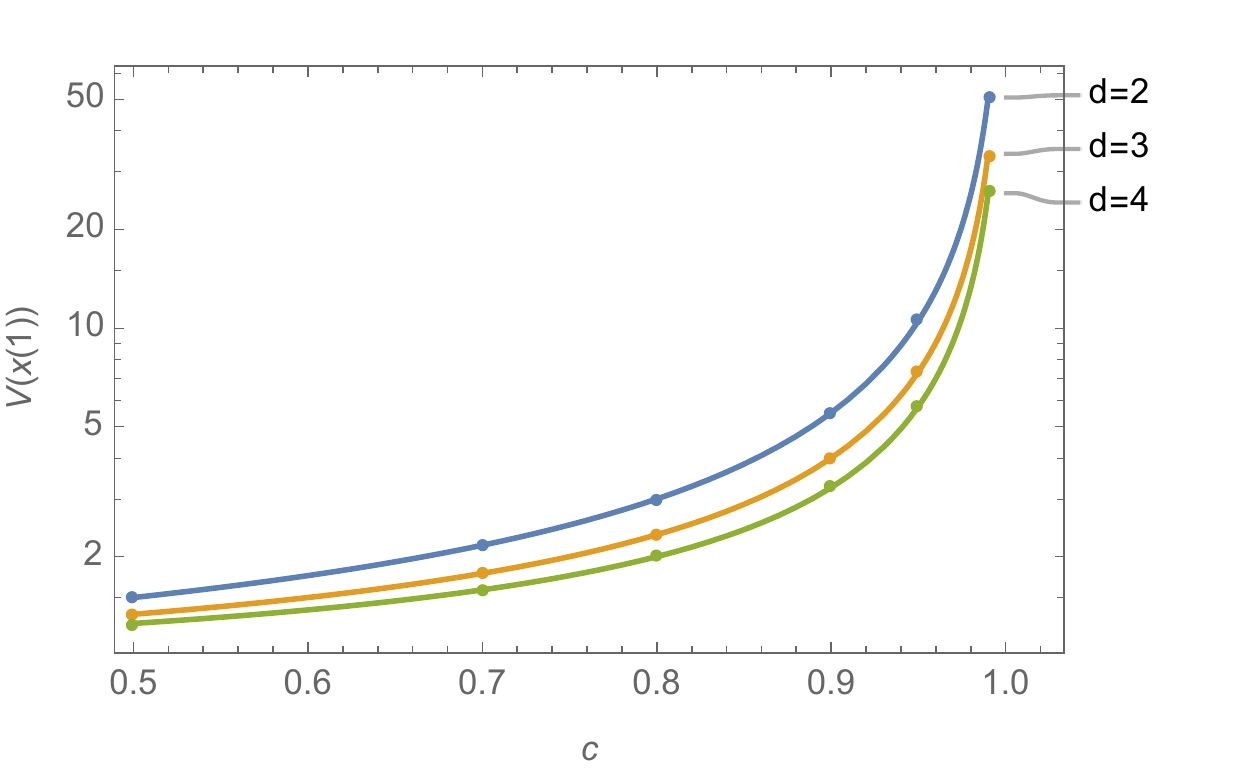}}
          \caption{Mean $\left< x (\mu_\ell) \right>$ and variance $V(x(\mu_\ell))$ of the optical depth of a photon undergoing isotropic scattering in a half space of dimension $d$ having arrived along cosine $\mu_\ell$, (Eqs.(\ref{eq:meanx}) and (\ref{variance})).  Monte Carlo reference solution shown as dots.}          
          \label{fig-delta-meanx} 
        \end{figure*}

        For a diffuse illumination, the spatial moments of Eq.(\ref{rhon-full}) for a diffuse illumination can be derived for $n \ge 1$  with Eq.(\ref{rhonrecursion}) used as a Green's function to obtain(I55)
        \begin{equation}
            \rho_{\,n,k} = \frac{n!}{(1-c)^{1/2}} \left[ \alpha_{n+k+1} + \frac{c}{2} \sum_{j=1}^n \frac{\alpha_j}{(n-j)!} \rho_{n-j,k} \right ], n \ge 0, \label{rhonk}
        \end{equation} 
        with the starting condition 
        \begin{equation}
            \rho_{\,0,k} = \frac{\alpha_{k+1}}{(1-c)^{1/2}} \, , \quad k = -1,\! 0,\! 1\dots.\label{rho0k}
        \end{equation}
        Equations (\ref{xnk}) and (\ref{rhonk}) then lead to 
        \begin{align}
            &\left<x\right>_k = \frac{\alpha_{k+2}}{\alpha_{k+1}} + \frac{c\alpha_1}{2(1-c)^{1/2}} \label{eq:meanxdiffuse}  \\
            &\left<x^2\right>_k = \frac{2\alpha_{k+3}}{\alpha_{k+1}} +  \frac{c\alpha_1 \alpha_{k+2}}{(1-c)^{1/2} \alpha_{k+1}}+ \frac{c^2\alpha_1^2 }{2(1-c)} + \frac{c\alpha_2}  {(1-c)^{1/2}} \label{x2k} 
        \end{align}
        for $k=-1,\ 0,\ 1$, etc.  We denote the variance of the optical depth of the photon under diffuse illumination by
        \begin{equation}\label{eq:Vk}
          V_k = \left<x^2\right>_k - \left<x\right>_k^2.
        \end{equation}

        The $d$D half-space results in Eqs.(\ref{rhonrecursion}) to (\ref{x2k}) are the same as for the 3D half-space except the integrals in $H(\mu)$ and $\alpha_n$ have the additional factor $G(\mu)$.  The equations for the mean and variance of the optical depth of a photon under unidirectional and diffuse illuminations are illustrated and compared to Monte Carlo simulation in Figures~\ref{fig-delta-meanx} and \ref{fig-diffuse-meanx}.  Monte Carlo sampling methods for $\mathbb{R}^d$ are described in \ref{appendix:MC}.
        \begin{figure*}
          \centering
          \subfigure[Mean optical depth]{\includegraphics[width=0.47 \linewidth]{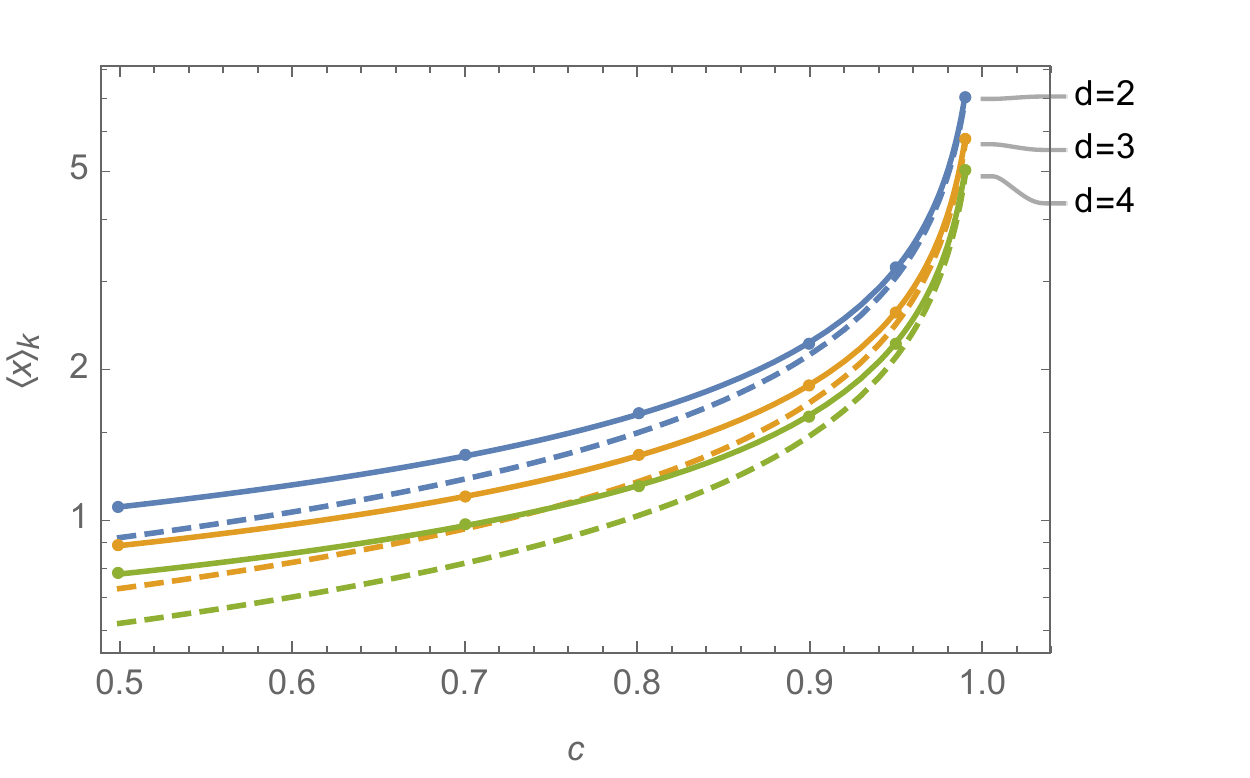}}
          \subfigure[Variance of optical depth]{\includegraphics[width=0.499 \linewidth]{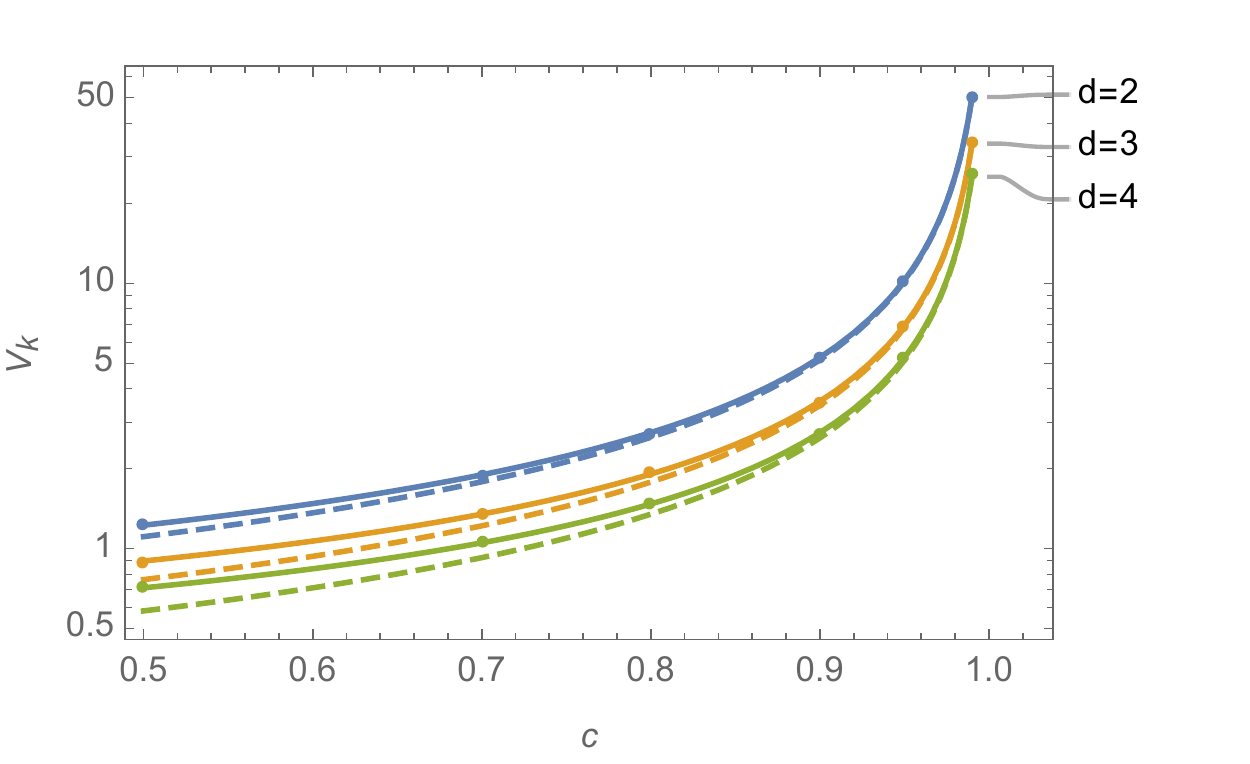}}
          \caption{Mean $\left<x\right>_k$ and variance $V_k$ of the optical depth of a photon undergoing isotropic scattering in a half space in dimension $d$ for diffuse illumination conditions, (Eqs.(\ref{eq:meanxdiffuse}) and (\ref{eq:Vk})).  Uniform diffuse illumination $k=0$ is shown as continuous, with the Monte Carlo reference solution shown as dots.  Isotropic illumination $k=-1$ is shown as dashed.}          
          \label{fig-diffuse-meanx} 
        \end{figure*}

  \subsection{Albedo problem directional moments} 
        Moments of the direction of photons emerging from the surface of the half-space also can be derived, in analogy to the emerging directional moments for a 3D half-space\cite{NM15}. 
        Equation (\ref{reflection}) for the collimated incident beam boundary condition of Eq.(\ref{deltabc}) immediately gives
        \begin{equation}
            I(0,-\mu;\mu_\ell) = \frac{c\mu_\ell G(\mu_\ell)H(\mu_\ell)H(\mu)}{2(\mu + \mu_\ell)}, \,  \label{I -mu mul}
        \end{equation}
        in agreement with Eq.(\ref{eq:diffuselaw}) once the differences between Eqs.(\ref{eq:eugboundary}) and (\ref{deltabc}) are accounted for.
        We choose to weight $I(0,-\mu;\mu_\ell)$ 
        with the powers $\mu^nG(\mu)$, $n = 0,\ 1,$ etc., so for the outward and inward moments of a collimated incident illumination we have
        \begin{align}
        j_{out,n}(\mu_\ell) &= \int_0^1 \mu^{n+1} I(0,-\mu;\mu_\ell) G(\mu) \, d\mu \, , \ n = 0,\! 1,\! 2\dots  \nonumber \\
        &= \frac{c\,\mu_\ell \,G(\mu_\ell)H(\mu_\ell)}{2}\int_0^1 
        \frac{\mu^{n+1}H(\mu)G(\mu)}{\mu + \mu_\ell} \, d\mu \label{joutnmuelldef} \\
        j_{\,in,n}(\mu_\ell) &= \int_0^1 \mu^{n+1}I(0,\mu;\mu_\ell) G(\mu) \, d\mu\, , \quad n = 0,\! 1,\! 2 \dots \nonumber \\
        &= \mu_\ell ^{n+1} G(\mu_\ell) \,. \label{jinnmuelldef}
        \end{align}  
        The ratios $R_n$ for $n = 0,\ 1,$ etc.\!\!\! , 
        \begin{subequations}
        \begin{eqnarray}
            R_n(\mu_\ell) &=& j_{\,out,n}(\mu_\ell)/j_{\,in,0}(\mu_\ell) \label{jratiomuell} \\
            R_{n,k} &=& j_{\,out,n,k}/j_{\,in,n,k} \, ,\label{jratiok}
        \end{eqnarray}
        \end{subequations}
        are the fractions of the incident current propagated in the $n$th outward directional moment, with $R_0$ the probability that an entering photon will escape the half space.  

\begin{figure*}
          \centering
          \subfigure[Fixed absorption $c = 0.99$]{\includegraphics[width=0.49 \linewidth]{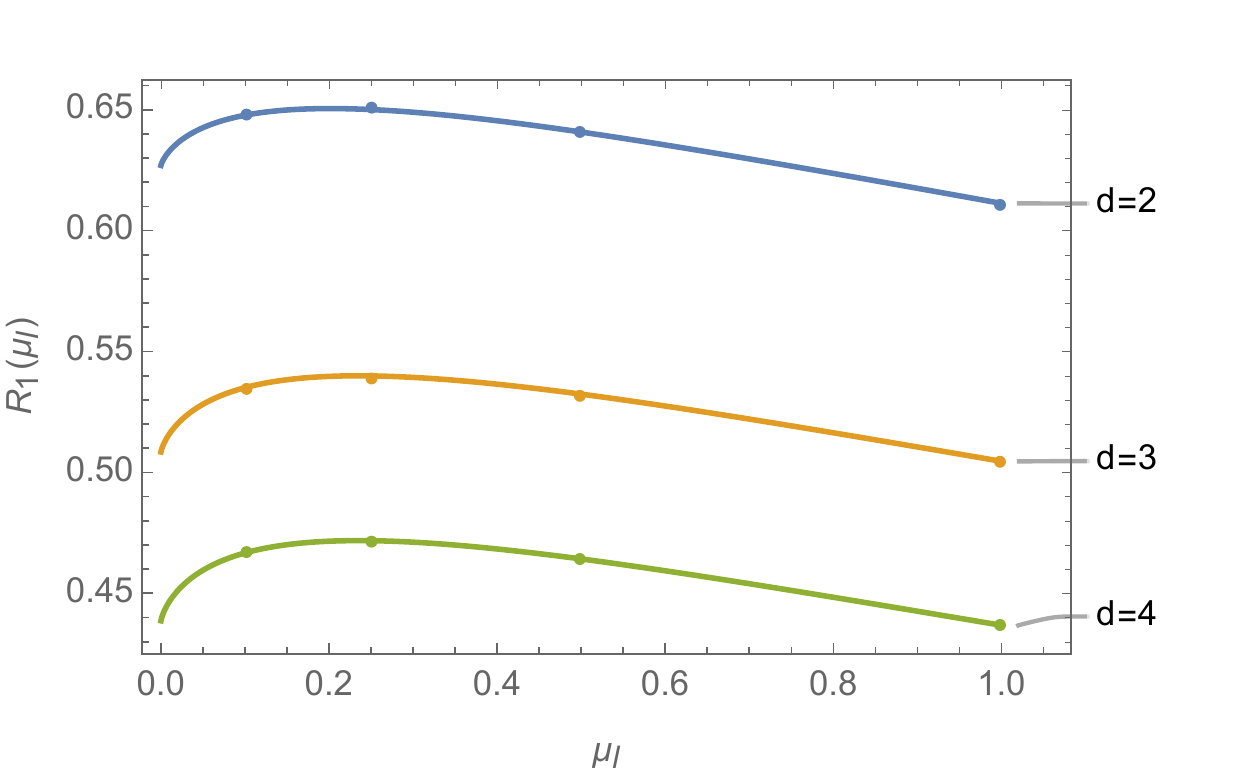}}
          \subfigure[Normally-incident illumination]{\includegraphics[width=0.49 \linewidth]{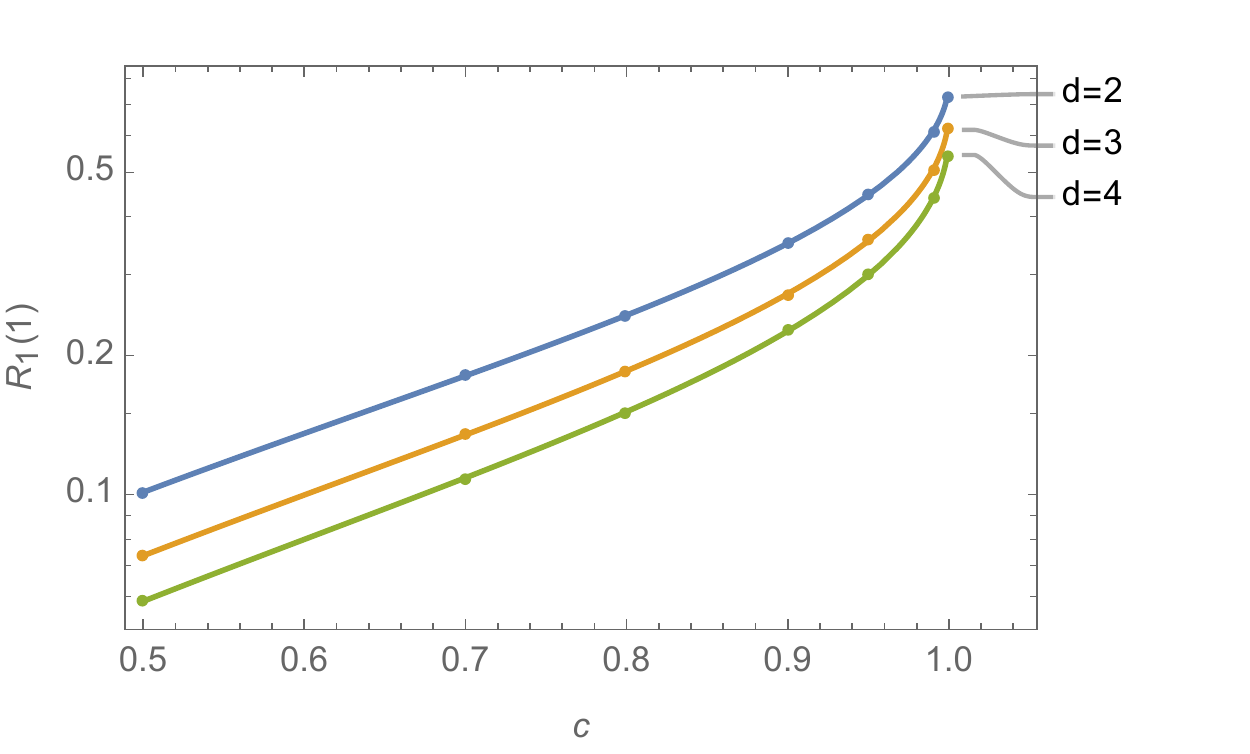}}
          \subfigure[Fixed absorption $c = 0.99$]{\includegraphics[width=0.49 \linewidth]{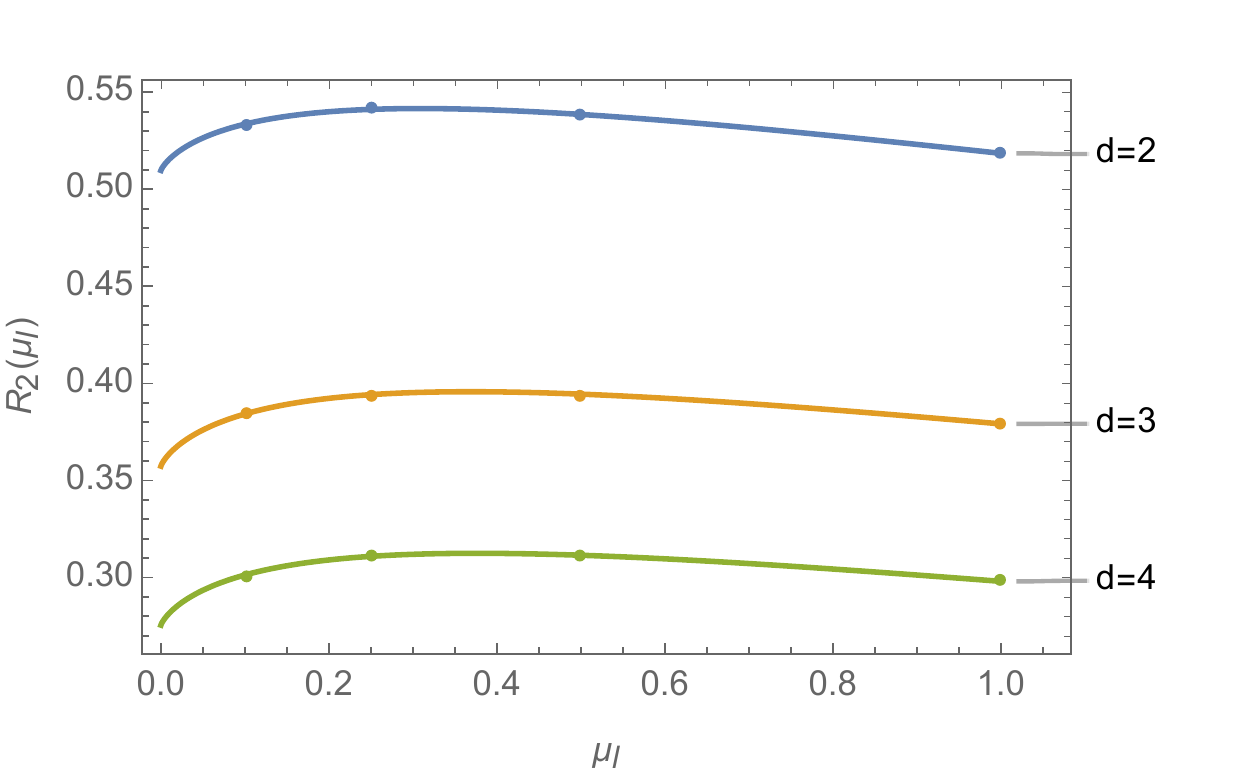}}
          \subfigure[Normally-incident illumination]{\includegraphics[width=0.49 \linewidth]{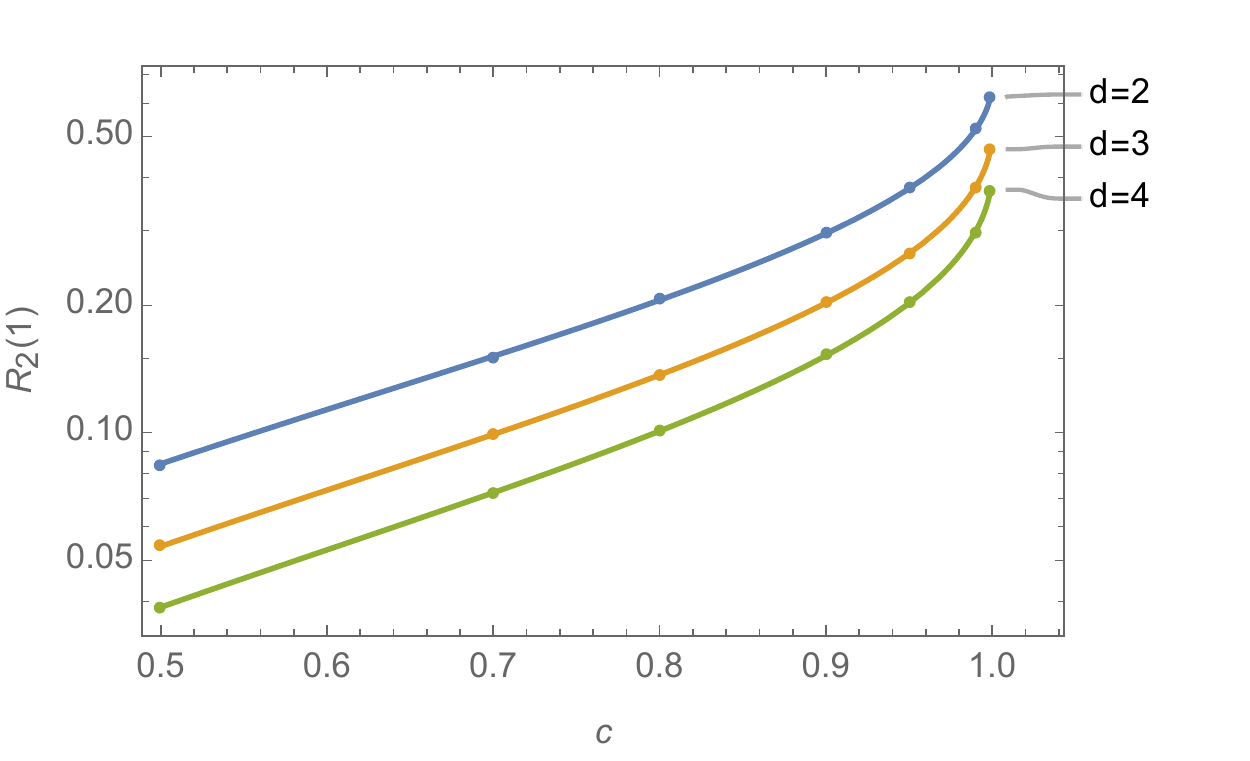}}
          \caption{Directional moments for the half space under unidirectional illumination (Eqs.(\ref{jratio1}) and (\ref{jratio2})).  Monte Carlo reference solution shown as dots.}          
          \label{fig-delta-meanu} 
        \end{figure*}

        With the partial fraction analysis of 
        \begin{equation}
            \frac{\mu^{n+1}}{\mu + \mu_\ell} = (-1)^n \frac{\mu\,\mu_\ell^n}{\mu + \mu_\ell} + \sum_{j=1}^n (-1)^{n+j} \mu^j \mu_\ell^{n-j}, \quad n \ge 0, \label{partialfraction}
        \end{equation}
        and Eqs.(\ref{2ndH(mu)eq}) and (\ref{joutnmuelldef}), the directional moments for a collimated surface illumination equal the 3D moments of Eq.(I37a) except for the extra factor $G(\mu_\ell)$, 
        \begin{equation}
            j_{\,out,n}(\mu_\ell) = (-1)^n G(\mu_\ell)\mu_\ell^{n+1} \left\{1 - H(\mu_\ell)\left[(1-c)^{1/2} \vphantom{\sum_{j=1}^n(-1)^j}  - \frac{c}{2} \sum_{j=1}^n(-1)^j\alpha_j\mu_\ell^{-j}\right]\right\}  ,
        \end{equation}
        so with $j_{\,in,n}(\mu_\ell) = \mu_\ell^{n+1} G(\mu_\ell)$ the ratios $R_n(\mu_\ell)$ can be computed for $n=0,\ 1,$ etc. Three examples are
        \begin{align}
            R_0(\mu_\ell) &= 1 - (1-c)^{1/2} H(\mu_\ell) \label{jratio0} \\
            R_1(\mu_\ell) &= H(\mu_\ell)[(1-c)^{1/2}\mu_\ell + (c/2)\alpha_1] - \mu_\ell \label{jratio1} \\
            R_2(\mu_\ell) &= \mu_\ell^2 \left(1-H\left(\mu_\ell\right) \left(\sqrt{1-c}-\frac{c}{2} \left(\frac{\alpha_2}{\mu_\ell^2}-\frac{\alpha _1}{\mu_\ell}\right)\right)\right) \label{jratio2}
        \end{align}
        with $R_0(\mu_\ell)$ giving the albedo of the halfspace, in agreement with Eq.(\ref{eq:albedo}). The ratios in Eqs.(\ref{jratio0}) and (\ref{jratio1}) are identical to the corresponding 3D ratios of Eqs.(I39a) and (I39b) except for the different definitions for $H(\mu_\ell)$ and $\alpha_n$.  We verify these in Figure~\ref{fig-delta-meanu}.

        For a diffuse illumination, Eqs.(\ref{kbc}) and (\ref{reflection}) initially give, for $k\ge 0$, 
        \begin{equation}
        I_k(0,-\mu)= H(\mu)\frac{c}{2}\int_0^1 \frac{(\mu')^{k+1}\,H(\mu')G(\mu')}{\mu + \mu'} d\,\mu'.
        \end{equation}
        With the help of Eqs.(\ref{2ndH(mu)eq}) and (\ref{partialfraction}), this becomes 
        \begin{equation}
        I_k(0,-\mu) = (-\mu)^k \left[1 - (1-c)^{1/2}\,H(\mu) + \frac{c}{2}H(\mu)\sum_{j=1}^k (-\mu)^j \alpha_j \right]\,.
        \end{equation}

\begin{figure*}
          \centering
          \subfigure[Mean exitant cosine]{\includegraphics[width=0.8 \linewidth]{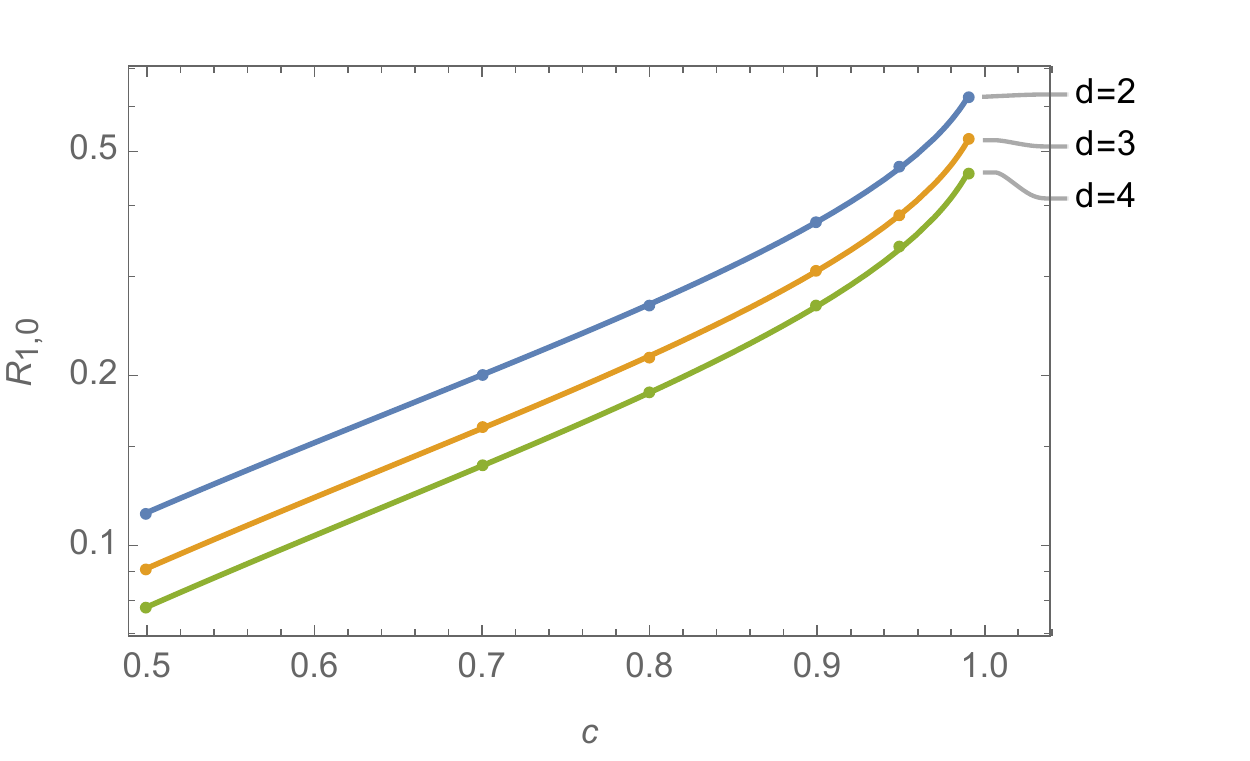}}
          \subfigure[Mean square exitant cosine]{\includegraphics[width=0.8 \linewidth]{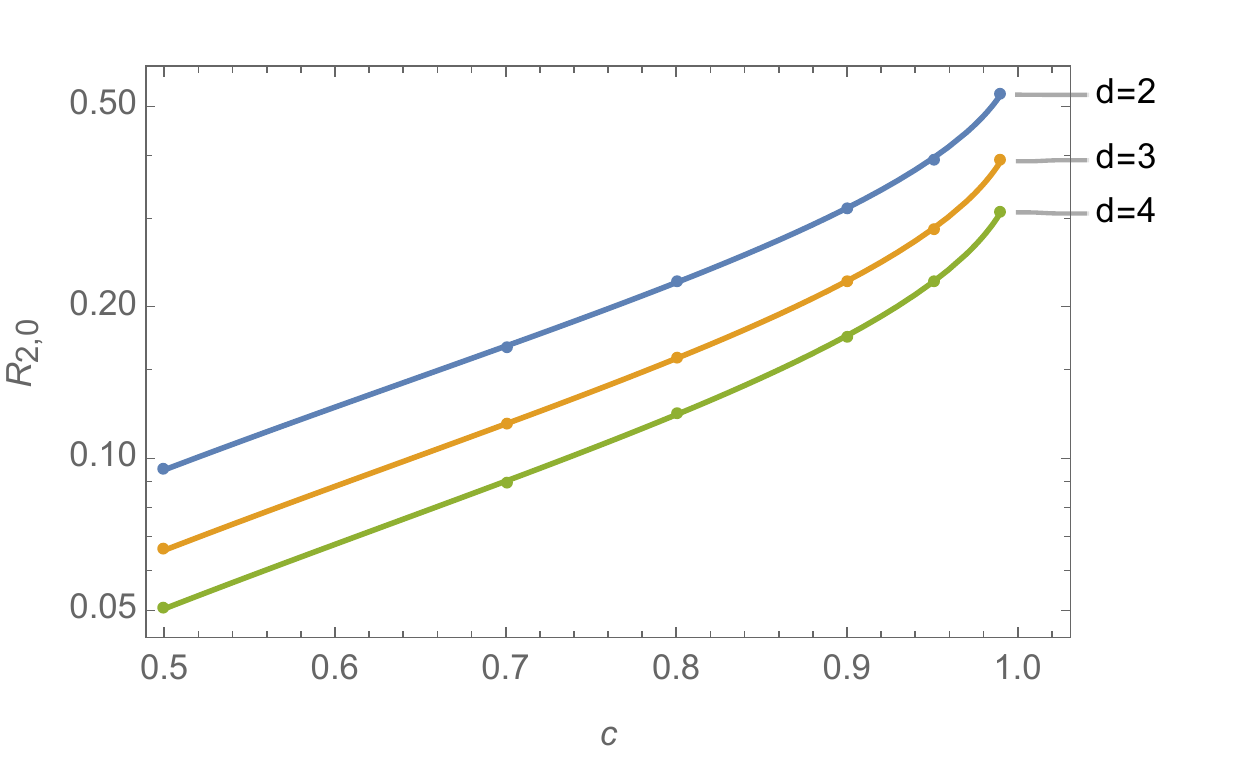}}
          \caption{Directional moments of the emergent intensity for a half space under uniform diffuse illumination ($k=0$), given by Eqs.(\ref{eq:R10}) and (\ref{eq:R20}).  Monte Carlo reference solution shown as dots.}          
          \label{fig-diffuse-meanu} 
        \end{figure*}

        With the definitions
        \begin{align}
        j_{\,out,n,k} &= \int_0^1 \mu^{n+1} I_k(0,-\mu) G(\mu) \, d\mu  \label{joutkdef}\\ 
        j_{\,in,n,k} &= \int_0^1 \mu^{n+1} I_k(0,\mu) G(\mu) \, d\mu  , \label{jinkdef}
        \end{align}
        it follows for $k \ge 0$ that
        \begin{subequations}
          \begin{eqnarray}
            j_{\,out,n,k} &=& (-1)^k \left[j_{in,n,k} -(1-c)^{1/2}\alpha_{n+k+1} \vphantom{\sum_{j=1}^k (-1)^j} \right.  \nonumber \\
            &\ & \quad \left. + \frac{c}{2} \sum_{j=1}^k (-1)^j \alpha_j\alpha_{n+k+1-j} \right] \label{joutk}\\
            j_{\,in,n,k} &=& \int_0^1 \mu^{n+k+1} G(\mu) d\,\mu \label{jink}\\
            \ \ &=& \frac{\Gamma \left(\frac{d}{2}\right) \Gamma \left(\frac{1}{2} (k+n+2)\right)}{\sqrt{\pi
       } \Gamma \left(\frac{1}{2} (d+k+n+1)\right)}. \nonumber
          \end{eqnarray}
        \end{subequations}
        We define the directional moments
        \begin{equation}\label{eq:diffusealbedocase}
          R_{n,k} = j_{\,out,n,k} / j_{\,in,0,k}
        \end{equation} 
        where again $R_{0,k}$ gives the albedo of the half-space.  For uniform diffuse illumination $k=0$, we find
        \begin{align}
          R_{1,0} &= \frac{\sqrt{\pi } \Gamma \left(\frac{d+1}{2}\right) }{\Gamma \left(\frac{d}{2}\right)} \left(\frac{1}{d}-\alpha _2
   \sqrt{1-c}\right) \label{eq:R10}\\
          R_{2,0} &= \frac{2}{d+1}-\frac{\sqrt{\pi }  \Gamma \left(\frac{d+1}{2}\right)}{\Gamma
   \left(\frac{d}{2}\right)} \alpha _3 \sqrt{1-c}, \label{eq:R20}
        \end{align}
        which we verify in Figure~\ref{fig-diffuse-meanu}.

        Turning now to the special case of $k=-1$ that corresponds to a unit incident current, Eqs.(\ref{1stH(mu)eq}) and (\ref{reflection}) give
        \begin{eqnarray}
            I_{-1}(0,-\mu) &=& \mu^{-1}H(\mu)\int_0^1 \phi(\mu,\mu') H(\mu') G(\mu') \, d\mu' \nonumber \\
            &=& \mu^{-1}[H(\mu)-1] \,.
        \end{eqnarray}
        The albedo of the half-space in this instance is
        \begin{equation}\label{eq:universalR}
          R_{0,-1} = \int_0^1 \mu I_{-1}(0,-\mu) G(\mu) d \mu = \frac{1}{c} \left( 2 - c - 2 \sqrt{1-c} \right)
        \end{equation}
        after use of Eq.(\ref{alpha0}).

        For other values of $k$, using Eqs.(\ref{joutkdef}) and (\ref{jinkdef}), we find
        \begin{subequations}
        \begin{equation}
            j_{\,out,n,-1} = \alpha_n -j_{in,n,-1}\,, \quad n = \,0,\ 1,\ 2, \dots
        \end{equation}
        and, for $d = 2$,
        \begin{eqnarray}
            j_{\,in,n,-1} &=& \int_0^1 \mu^{n+k+1} G(\mu) d\mu \label{jink}\\
            \ \ &=& \frac{n!!}{(n+1)!!}\,, \quad n = \ 0,2,4,\dots \nonumber \\
            \ \ &=& \frac{2}{\pi} \frac{n!!}{(n+1)!!}\,, \quad n=  1,3,5,
            \dots , \, \nonumber  
        \end{eqnarray}
        \end{subequations} 
        which differs from $j_{\,in,n,-1}$ for 3D.

    \section{The Milne problem}\label{sec:Milne}
    \subsection{Extrapolated endpoint for $c=1$}
      In this section we extend the derivation of Hopf (\cite{Hopf34} p. 85, \cite{Krein83} p.225) to find the distance $z_0$ such that the rigorous asymptotic portion of the scalar flux in the Milne problem with absorption $c = 1$, when extrapolated outside of the medium, reaches $\phi_{as}(-z_0) = 0$.  With Hopf's $\kappa(u) = \Lambda_{c=1}(1 / u)$ in our notation, we find
      \begin{equation}
        z_0(d) = \frac{1}{\pi} \int_0^\infty \frac{2}{t^2}-\frac{2 \, _2F_1\left(\frac{3}{2},2;\frac{d}{2}+1;-t^2\right)}{d\left(1- \,
       _2F_1\left(\frac{1}{2},1;\frac{d}{2};-t^2\right)\right)} dt,
      \end{equation}
      which we find to simplify further
      \begin{equation}\label{eq:z0simple}
        z_0(d) =  \frac{1}{\pi} \int_0^\infty  \left(\frac{1}{1+t^2}\right) \left( \frac{d}{t^2}+3-\frac{1}{1-\, _2F_1\left(\frac{1}{2},1;\frac{d}{2};-t^2\right)} \right) dt.
      \end{equation}
      The distance $z_0$ is also known to be related to the second moment of the $H$ function (\cite{Busbridge60} p. 56)
      \begin{equation}
        z_0(d) = \frac{\frac{1}{2}\alpha_2}{\sqrt{2 \psi_2}}.
      \end{equation}
      From the characteristic function we find
      \begin{equation}
        \psi_2 = \int_0^1 \Psi(\mu) \mu^2 d\mu = \frac{1}{2d}.
      \end{equation}
      From the last 2 equations, with $c = 1$, we find
      \begin{equation}
        z_0(d) = \frac{\sqrt{d}}{2}  \alpha_2.
      \end{equation}
      For the 1D rod we find $z_0(1) = 1$.  For Flatland we find
      \begin{equation}
        z_0(2) = \frac{1}{\pi} \int_0^{\infty } \frac{1}{1+t^2} \left(2-\frac{1}{1+\sqrt{1+t^2}}\right) \, dt = \frac{1}{2} + \frac{1}{\pi}
      \end{equation}
      in agreement with~\cite{EdEMMR18}.
      For 3D, Eq.(\ref{eq:z0simple}) matches Hopf's equation (forgiving the typesetting error) and is best evaluated using the procedure proposed by Plazcek and Seidel~\cite{plazcek47}, which we extend to higher odd dimensions.
      For 4D we find another closed form solution
      \begin{equation}
        z_0(4) = \frac{1}{\pi} \int_0^{\infty } \left(2-\frac{2}{\sqrt{1+t^2}}\right)\frac{1}{t^2} \, dt = \frac{2}{\pi}.
      \end{equation}
      In 5D, the integral
      \begin{equation}
        \int_0^{\infty } \frac{1}{1+t^2} \left(3+\frac{5}{t^2}-\frac{2 t^3}{t \left(3+2 t^2\right)-3 \left(1+t^2\right) \tan
       ^{-1}(t)}\right) \, dt
      \end{equation}
      can be evaluating using Plazcek and Seidel's approach giving
      \begin{align}
        z_0(5) &= \frac{10}{\pi^2} + \frac{1}{\pi} \int_0^{\infty } \left(\frac{12 (-1+x \cot (x))}{5+\cos (2 x)-6 x \cot
       (x)}+\frac{5}{x^2}\right) \, dx \nonumber \\ &\approx 0.5819457611.
      \end{align}
      With d = 6,
      \begin{align}
        z_0(6) &= \frac{1}{\pi } \int_0^{\infty } \frac{6}{4+3 t^2+4 \sqrt{1+t^2}} \, dt \nonumber \\ 
        &= -\frac{3}{4 \sqrt{2}}+\frac{3}{\pi }+\frac{3 \cot ^{-1}\left(2 \sqrt{2}\right)}{2 \sqrt{2} \pi }.
      \end{align}
      Mathematica is able to determine a bulky but closed form solution for $z_0(8)$, which contains an isolated $4 / \pi$, which is interesting to compare to the other even-dimension solutions.  Numerical constants for the first $8$ values of $d$ are summarized in Table~\ref{tab:z0}.
        \begin{table}[]
            \centering
            \begin{tabular}{|l|l|}
            \hline
            $d$ & $z_0(d)$        \\
            \hline
            $1$   & $1.0000000000$ \\
            $2$   & $0.8183098862$ \\
            $3$   & $0.7104460896$ \\
            $4$   & $0.6366197724$ \\
            $5$   & $0.5819457611$ \\
            $6$   & $0.5393348406$ \\
            $7$   & $0.5049086837$ \\
            $8$   & $0.4763395251$ \\
            \hline         
            \end{tabular}
            \caption{(Milne) extrapolation distance $z_0(d)$ with ($c = 1$) in dimension $d$.\label{tab:z0}}
        \end{table}

      \subsection{Extrapolated endpoint for $c < 1$}
        With absorption, we consider Case's approach to the Milne problem with boundary conditions
        \begin{subequations} 
          \begin{eqnarray}
          I(0,\mu) &=& 0, \quad \mu \in [\,0,1], \\
          I(x,\mu) &=& \phi(-\nu_0,\mu) \exp(x/\nu_0), \quad x \rightarrow \infty ,
          \end{eqnarray}
        \end{subequations}
        where the asymptotic portion of the intensity can be written as
        \begin{equation}
          I_{as}(x,\mu) = A(\nu_0)\phi(\nu_0,\mu)\exp(-x/\nu_0) + \phi(-\nu_0,\mu) \exp(x/\nu_0).
        \end{equation}
        The extrapolation distance $z_0(c,d)$ now depends on absorption and dimension, and the asymptotic flux satisfies $\int_{-1}^1 I_{as}(-z_0(c,d),\mu) d\mu = 0$, leading to 
        \begin{equation}
          A(\nu_0) = -\exp[-2z_0(c,d)/\nu_0]. \label{Anu0Milne}
        \end{equation}
        The expansion coefficient $A(\nu_0)$ is determined from Eq.(\ref{A(nu)}) 
        \begin{equation}
          A(\nu_0) = \frac{1}{N(\nu_0)H(\nu_0)}\int_0^1 \psi(\mu) \phi(\nu_0,\mu)\,\mu \, H(\mu) G(\mu) \, d\mu 
        \end{equation}
        with $\psi(\mu) = -\phi(-\nu_0,\mu)$.
        But from Eq.(\ref{All_Orthog_Eqs}), it follows for $\nu' \in \sigma_+$ that
        \begin{equation}
          \int_0^1 \phi(-\nu_0,\mu)\phi(\nu',\mu)\mu\,  H(\mu)G(\mu)\, d\mu = \frac{1}{4} \frac{c\, \nu_0}{H(\nu_0)}, \label{MilneIntegral}
        \end{equation}
        and so the combination of Eqs.(\ref{Anu0Milne}) and (\ref{MilneIntegral}) gives
        \begin{equation}
          z_0(c,d) = \frac{\nu_0}{2} \ln \left[ \frac{4N(\nu_0)H^2(\nu_0)}{c\nu_0} \right],  \label{z0(c,d)}
        \end{equation}
        in agreement with Busbridge (\cite{Busbridge60}, Eq.(23.13)).  

        We found numerical agreement with Eq.(\ref{z0(c,d)}) and a previous derivation for Flatland and 3D~\cite{EdEMMR18}
        \begin{equation}\label{eq:z0cd}
          z_0(c,d) = \frac{\nu_0}{2} \ln \left[ \frac{\nu_0 + 1}{\nu_0 - 1} \right] - \frac{1}{\pi} \int_0^1 \frac{\theta(t)}{1-t^2/\nu_0^2}  dt,
        \end{equation}
        which generalizes to general dimension when $0 \le \theta(t) \le \pi$ is determined from Eq.(\ref{eq:theta}).

\begin{figure*}
          \centering
          \includegraphics[width=.99\linewidth]{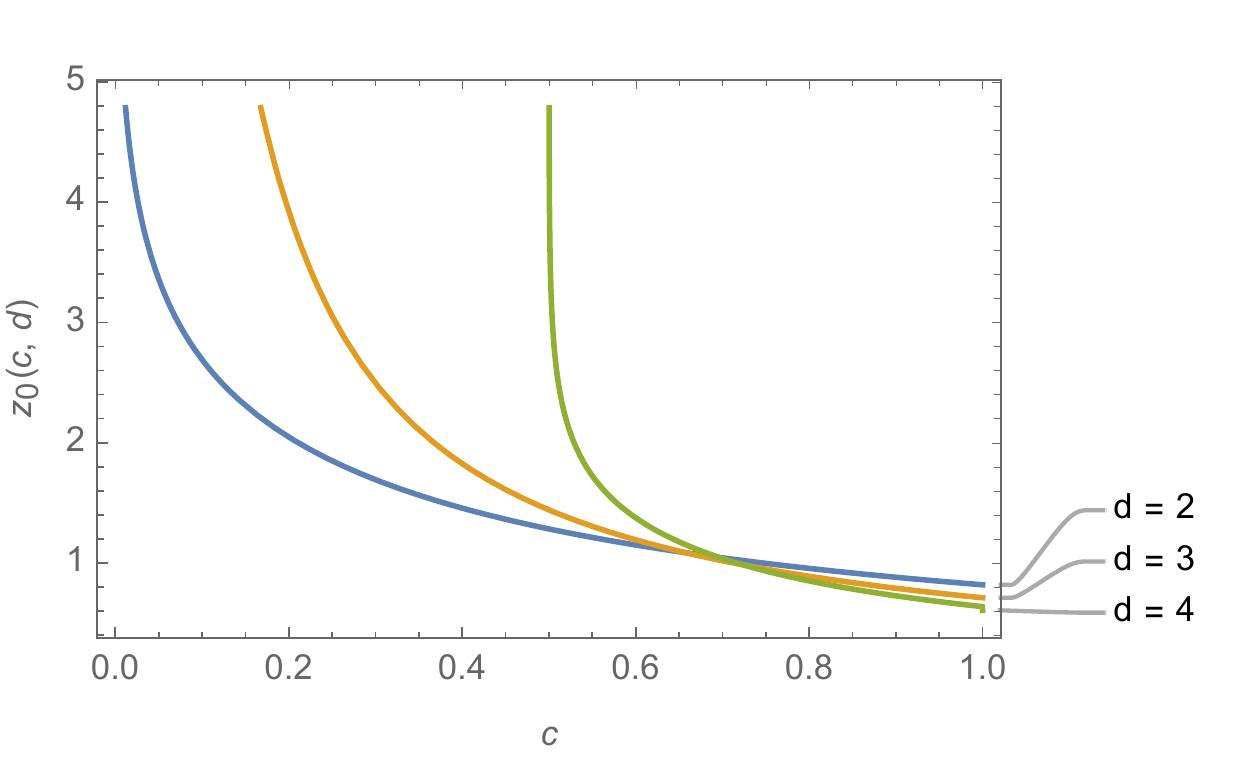}
          \includegraphics[width=.99\linewidth]{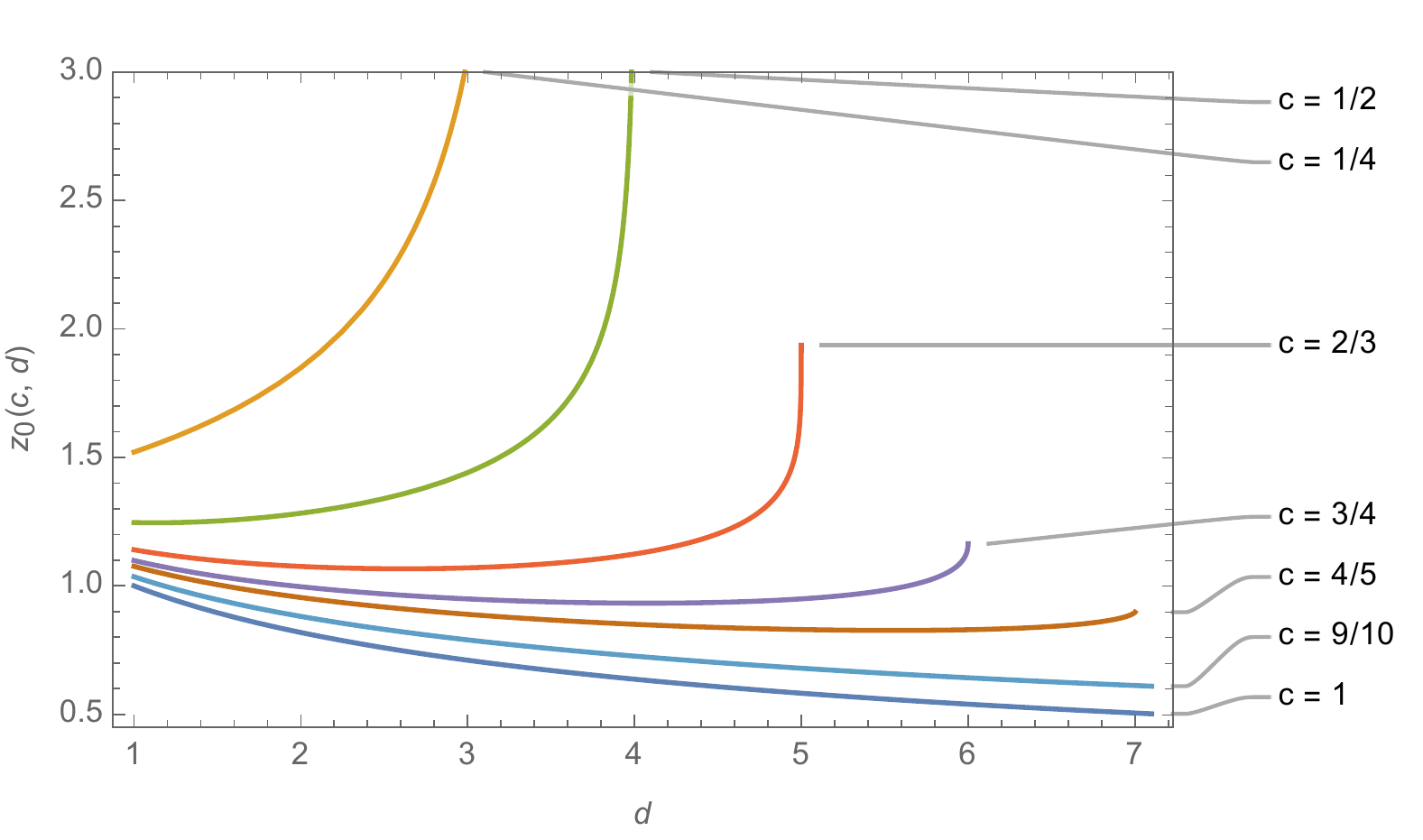}
          \caption{\label{fig-z0}  Milne extrapolation distance for the half space in $\mathbb{R}^d$ for varying $c$ and $d$ (Eq.(\ref{z0(c,d)})).}          
     \end{figure*}

        For monoenergetic isotropic scattering and Fresnel-\newline matched boundaries in $d \le 3$, the extrapolation distance always exists, but an extrapolation distance is not universal for $d > 3$, and disappears when $c < (d-3)/(d-2)$, as seen in Figure~\ref{fig-z0}.

     For the Milne problem emerging intensity, we find from Eqs.(\ref{reflection}) and (\ref{MilneIntegral}) that
      \begin{equation}
        I(0,-\mu) = \frac{\phi(\nu_0,\mu) H(\mu)}{H(\nu_0)}.
      \end{equation}
      Because $I(0,\mu) = 0$ and $G(-\mu) = G(\mu)$ for $\mu \in [0,1]$, the directional moments
      \begin{equation}
        \Phi_n(0) = \int_{-1}^1 \mu^n I(0,\mu) G(\mu) d\mu, \quad n=0,\ 1
      \end{equation}
      are, from Eqs.(\ref{GH normalization}) and (\ref{GH+ normalization}),
      \begin{equation}
        \Phi_0(0) = 1/H(\nu_0) {\mathrm{\ \ and\ \ }} \Phi_1(0) = -\nu_0\sqrt{1-c}/H(\nu_0).
      \end{equation}
    \subsection{Extrapolated endpoint for adjacent half-spaces}
      Problems of two adjacent half-spaces with isotropic scattering also can be analyzed\cite{kuvsvcer1964orthogonality,mendelson1964one,leuthauser68,mccormick1969neutron,NMIK73,smedleystevenson12}, with different absorption and scattering properties denoted by the single scattering albedo values $c_k = c_1$ for $x > 0$ and $c_k = c_2$ for $x < 0$. The characteristic functions now are $\Psi_k(\mu) = (c_k/2) G(\mu)$, for $k=1,\ 2$, where $G(\mu)$ again is given for $d$D in Eq.(\ref{GnDmu}). Completeness of the discrete and continuum eigenfunctions, 
      \begin{align}
        \phi_k(\pm\nu_{0k},\mu) &= \frac{c_k \,\nu_{0k}}{2} \frac{1}{\nu_{0k} \mp \mu} \nonumber \\
        \phi_k(\nu,\mu) &= \frac{c_k\,\nu}{2} \mathcal{P} \frac{1}{\nu-\mu} + \frac{\lambda_k(\nu)}{G(\mu)}\delta(\nu-\mu), \quad [-1,1], \nonumber
      \end{align}
      leads to the expansions
      \begin{equation}
      I(x,\mu) =
      \begin{cases}
      \int_{\sigma_1 +}A_1(\nu)\phi_1(\nu,\mu) \exp(-x/\nu) d\nu + f_1(x,\mu), &  x > 0 \\
       -\int_{\sigma_2 -}A_2(\nu)\phi_2(\nu,\mu)\exp(-x/\nu) d\nu + f_2(x,\mu) ,& x < 0
       \end{cases}
      \end{equation}
    with 
    \begin{subequations}
    \begin{eqnarray}
    \sigma_1 + &=& \{\nu \in [\,0,1] \cup \nu_{01}\} \\
    \sigma_2 - &=& \{\nu \in [-1,0\,] \cup -\nu_{02}\}.
    \end{eqnarray}
    \end{subequations}

    For continuity at the interface at $x=0$, 
    \begin{equation}
    \mathcal{I}(\mu) = \int_{\sigma_1 +}A_1(\nu)\phi_1(\nu,\mu) d\nu + \int_{\sigma_2 -}A_2(\nu)\phi_2(\nu,\mu) d\nu, \label{LongInterface}
    \end{equation}
    where 
    \begin{equation}
    \mathcal{I}(\mu) = I(0^+,\mu) - I(0^-,\mu) - f_1(0^+,\mu) + f_2(0^-,\mu). \label{x=0 condition}
    \end{equation}
    If we denote
    {\small
    \begin{subequations}
    \begin{eqnarray*}
    c(\nu) &=& c_1,\ \phi(\nu,\mu) = \phi_1(\nu,\mu),\nonumber \\ &\quad& \mathrm{and} \ A(\nu) = A_1(\nu) \ \mathrm{for} \ \nu \in (0,1)\ \mathrm{or}\ \nu= \pm \nu_{01} \\ 
    c(\nu) &=& c_2,\ \phi(\nu,\mu) = \phi_2(\nu,\mu),\nonumber \\
    &\quad& \mathrm{and} \ A(\nu) = A_2(\nu)\ \mathrm{for} \ \nu \in (-1,0)\ \mathrm{or} \ \nu = \pm \nu_{02}, 
    \end{eqnarray*}
    \end{subequations}
    }
    then Eq.(\ref{LongInterface}) can be written as
    \begin{equation}
    \mathcal{I}(\mu) = \int_{\sigma_1 + \, \cup \, \sigma_2 -} A(\nu)\Phi(\nu,\mu) d\nu. \label{ShortInterface}
    \end{equation}

    The weight function $\mu W(\mu)$ for two-media orthogonality relations can be expressed in terms of the $H$-functions for each medium\cite{kuvsvcer1964orthogonality} with
    \begin{equation}
    W(\nu) = c(\nu) \left[\frac{H_1(|\nu|)}{H_2(|\nu|)} \right]^{\mathrm{sign(\nu)}}, \quad \nu \in \{(-1,1)\cup  \pm \nu_{01}\cup  \pm \nu_{02}\}.
    \end{equation}
    The adjacent half-space orthogonality relation is
    \begin{equation}
    \int_{-1}^1 \phi(\nu,\mu)\phi(\nu',\mu) \mu W(\mu) G(\mu) d \mu = 0, \quad \nu \ne \nu' 
    \end{equation}
    and the associated normalization integrals are
    {\small
    \begin{eqnarray*}
    \int_{-1}^1 \phi^2(\nu_{01},\mu) \mu W(\mu) G(\mu) d\mu &=& N(\nu_{01})W(\nu_{01}) \\
    \int_{-1}^1 \phi^2(-\nu_{02},\mu) \mu W(\mu) G(\mu) d\mu &=& N(-\nu_{02}) W(-\nu_{02}) \\
    \int_{-1}^1 \phi(\nu,\mu)\phi(\nu',\mu) \,\mu W(\mu)G(\mu) d\mu &=& N(\nu) W(\nu) \delta(\nu-\nu')
    \end{eqnarray*}}
    In a similar manner\cite{NMIK73}, other adjacent-media equations follow, as in Eq.(\ref{All_Orthog_Eqs}).

\begin{figure}
        \centering
        \subfigure[2D]{\includegraphics[width=0.55 \linewidth]{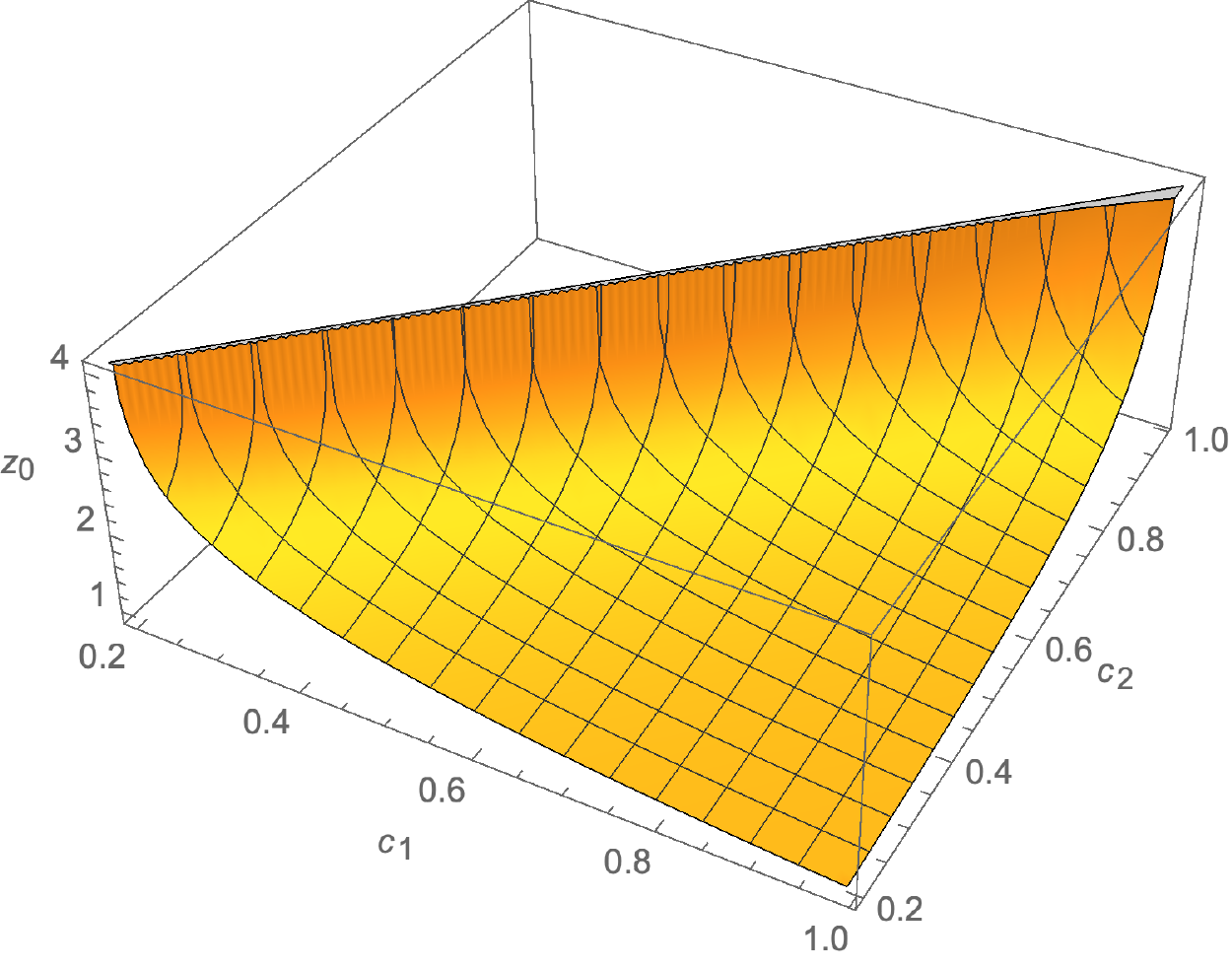}}
        \subfigure[3D]{\includegraphics[width=0.55 \linewidth]{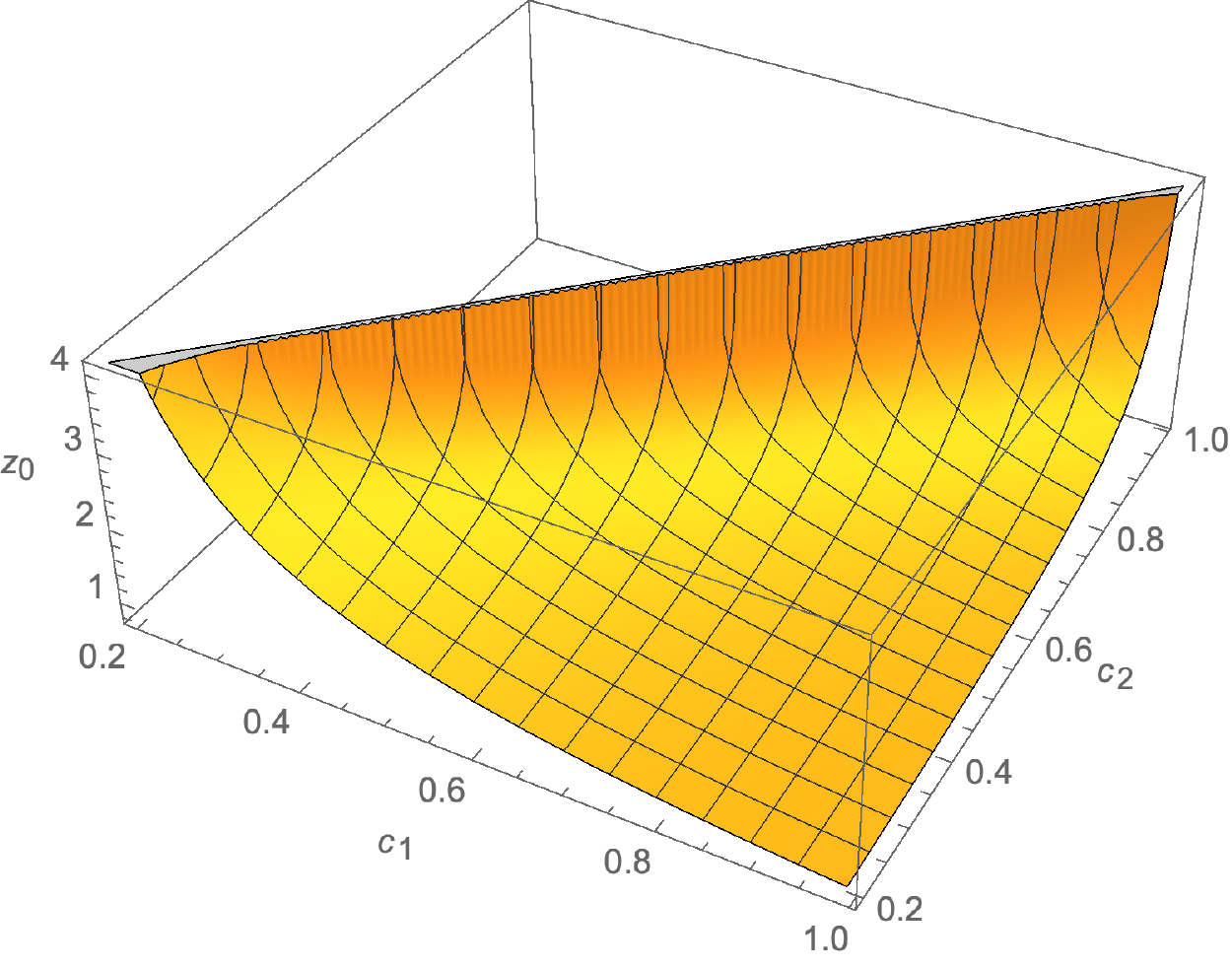}}
        \subfigure[4D]{\includegraphics[width=0.55 \linewidth]{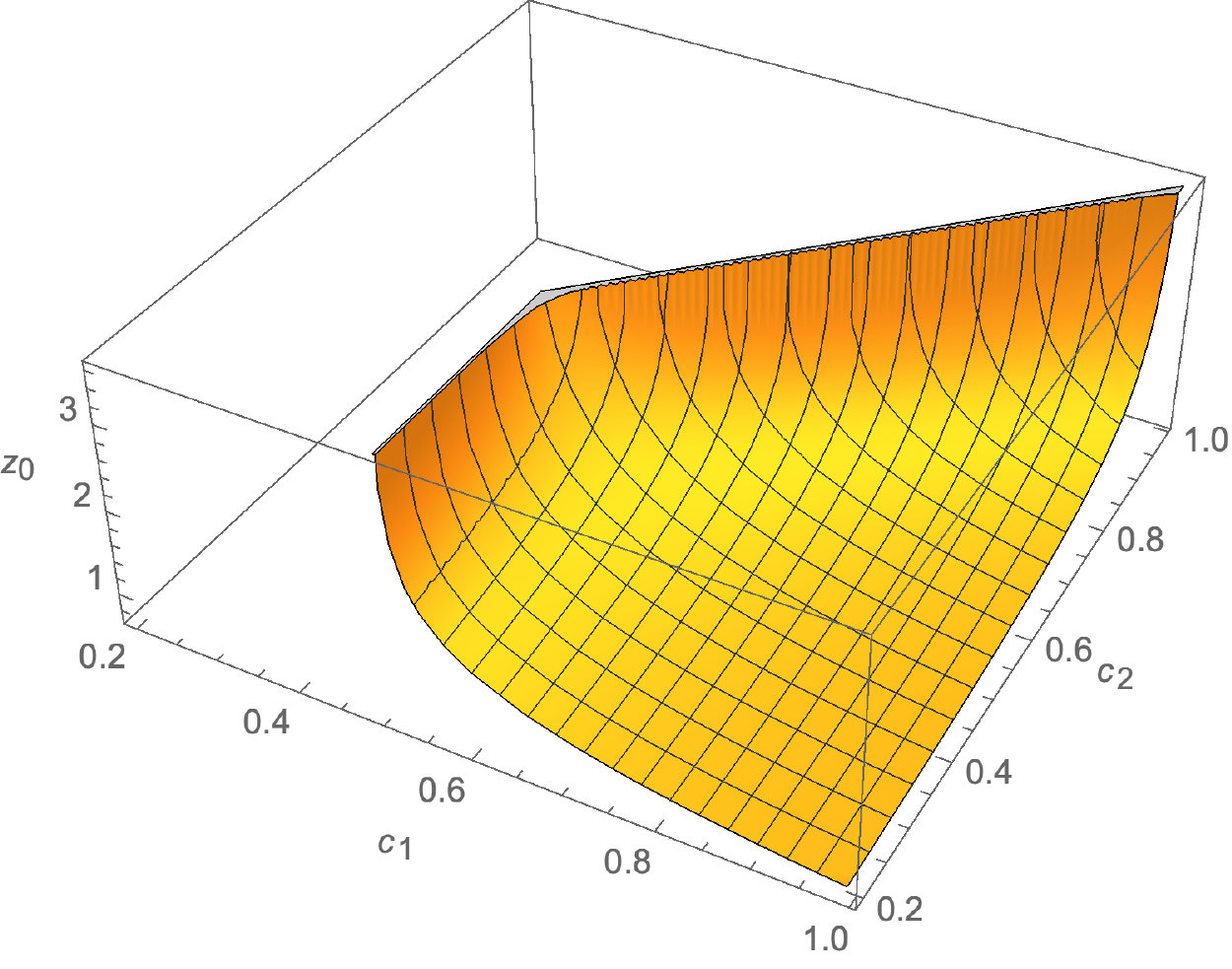}}
        \caption{Extrapolation distance $z_0(c_1,c_2,d)$ for adjacent half spaces in various dimensions.}          
        \label{fig:z0adjacent} 
      \end{figure}

    With the exponentially growing source in $x>0$, $\mathcal{I}(\mu) = -\phi_1(-\nu_{01},\mu)$ and the adjacent-medium extrapolation distance $z_0(c_1,c_2,d)$ becomes
    \begin{equation}
    z_0(c_1,c_2,d) = \frac{\nu_{01}}{2} \ln \left [\frac{4N_1(\nu_{01}) H_1^2(\nu_{01})H_2(-\nu_{01})}{c_1\nu_{01} H_2(\nu_{01})} \right] . \label{z0for12}
    \end{equation} 
    Tabulated results for 3D are available\cite{leuthauser68,NMIK73} and Figure \ref{fig:z0adjacent} and Tables~\ref{tab:z0adjacent2D}, \ref{tab:z0adjacent3D}, and \ref{tab:z0adjacent4D} illustrate results for different values of $c$ and $d$.

      \begin{table}[]
            \centering
            \begin{tabular}{lcccc}
            \hline
            & & $c_2$        \\
            \cline{2-5}
            $c_1$ & 0.6 & 0.8 & 0.9 & 0.95 \\
            \hline
            0.8  & 1.5882436 &  &  &  \\
            0.9  & 1.3584815 & 2.1003544 &  &  \\
            0.99 & 1.2131854 & 1.6762401 & 2.3696245 &  \\
            1.01 & 1.1858938 & 1.6127114 & 2.2060217 & 5.5553 \\
            1.1  & 1.0791665 & 1.3925977 & 1.748799 & 2.7656641 \\
            1.2  & 0.9837306 & 1.222884 & 1.4652013 & 2.0342982 \\
            1.4  & 0.83958884 & 0.99755241 & 1.1399741 & 1.4286662 \\
            \hline         
            \end{tabular}
            \caption{Two-media extrapolation distances $z_0(c_1,c_2,2)$ for 2D.\label{tab:z0adjacent2D}}
        \end{table}

      \begin{table}[]
            \centering
            \begin{tabular}{lcccc}
            \hline
            & & $c_2$        \\
            \cline{2-5}
            $c_1$ & 0.6 & 0.8 & 0.9 & 0.95 \\
            \hline
            0.8  & 1.4127134  \\
            0.9  & 1.175452 & 1.7835171 \\
            0.99 & 1.0265018 & 1.3985129 & 1.9595943 \\
            1.01 & 0.9987603 & 1.3402434 & 1.8181755 & 4.5360605 \\
            1.1  & 0.89132113 & 1.1377555 & 1.4192792 & 2.2264309 \\
            1.2  & 0.79703336 & 0.98183071 & 1.1697516 & 1.6118864 \\
            1.4  & 0.65879753 & 0.77729562 & 0.88422937 & 1.1004687 \\
            \hline         
            \end{tabular}
            \caption{Two-media extrapolation distances $z_0(c_1,c_2,3)$ for 3D.\label{tab:z0adjacent3D}}
        \end{table}

        \begin{table}[]
            \centering
            \begin{tabular}{lcccc}
            \hline
            & & $c_2$        \\
            \cline{2-5}
            $c_1$ & 0.6 & 0.8 & 0.9 & 0.95 \\
            \hline
            0.8  & 1.3093552 & \text{} & \text{} & \text{} \\
            0.9  &  1.0594888 & 1.5882436 & \text{} & \text{} \\
            0.99 & 0.90809624 & 1.2274169 & 1.7108654 & \text{} \\
            1.01 &  0.8804248 & 1.1727963 & 1.5834373 & 3.9282009 \\
            1.1  &  0.77489064 & 0.9837306 & 1.222884 & 1.9098907 \\
            1.2  &  0.68446724 & 0.83958884 & 0.99755241 & 1.3693007 \\
            1.4  &  0.55567088 & 0.65367435 & 0.74209513 & 0.92048675 \\
            \hline         
            \end{tabular}
            \caption{Two-media extrapolation distances $z_0(c_1,c_2,4)$ for 4D.\label{tab:z0adjacent4D}}
        \end{table}

\section{Half-space and adjacent half-space properties invariant to $d$}\label{sec:universal}

    A well known property of sums of independent identically distributed (i.i.d.) symmetric random numbers is that the mean first-passage time (number of terms in the sum) for the sum to go negative is universal and does not depend on the distribution of the individual terms.  Early proofs of this theorem were given by Sparre-Andersen and Pollaczek in the 1950s~\cite{frisch95,majumdar2010universal}.  

    The depth coordinate of a photon undergoing isotropic scattering in a half space corresponds precisely to such a stochastic process, with depth displacements between collisions drawn from the transport kernel $K(x)$, which varies with $d$.  Escape occurs the first time the sum of displacements goes negative.  For internal sources, unidirectional illumination, and uniform diffuse illumination, we have seen that the mean number of collisions and escape probabilities for a half space depend on $d$.  However, when the first displacement into the medium is also drawn from $K(x)$, by placing an isotropic point or plane source at the boundary interface, then the i.i.d. property is satisfied, and we find mean collision rates and escape probabilities that are universal and that are given by very simple equations.

    \subsection{Half-space invariants}

    Frankel and Nelson~\cite{frankel53} considered a half space in 3D with an isotropic plane source at the boundary, and gave a derivation of the albedo, including the generating function for escape over number of collisions.  Their derivation does not rely on the exponential free-path distribution between collisions, nor that the medium is three dimensional, and so, in some sense, constitutes one of the earliest proofs of this universal property of sums of random variables.  Their derivation finds the universal albedo for the one-sided isotropic surface source (only emitting into the medium) to be
    \begin{equation}\label{eq:rodR}
        \mathcal{R}(c) = 1 - (1-c) \sum_{j=0}^\infty  2 (-1)^{j+1} \binom{-\frac{1}{2}}{j+1} c^j = \frac{2-c-2 \sqrt{1-c}}{c}.
    \end{equation}
    This result agrees with our prior derivation for a one-sided isotropic plane source at the boundary, Eq.(\ref{eq:universalR}).  Equivalently, if the first displacement is $0$, which is the case under grazing illumination $\mu_\ell = 0$, then we also find a universal albedo, directly from Eq.(\ref{eq:albedo}), given that $H(0) = 1$,
    \begin{equation}
      R(\mu_\ell = 0) = 1 - \sqrt{1-c}.
    \end{equation}

    A closely-related universal property is the collision rate density at the boundary $C(0)$ due to a unit uniform isotropic source inside the medium.  Davison~\cite{davison57} noted the relationship of this problem, by reciprocity, to that of the total collision rate inside the medium due to an isotropic plane source at the boundary.  This deviates slightly from the earlier example by now considering a two-sided source at the boundary, with half of the photons escaping directly outward causing no collisions.  The mean number of collisions inside the medium (and by reciprocity, $C(0)$ for the constant source problem), is now half of the number created by the one-sided isotropic source,
    \begin{equation}
        C(0) = \frac{1}{2} \sum_{j=0}^\infty  2 (-1)^{j+1} \binom{-\frac{1}{2}}{j+1} c^j = \frac{1}{c} \left( \frac{1}{\sqrt{1-c}} -1 \right)
    \end{equation}
    agreeing with prior derivations for 3D~\cite{davison57} and 2D~\cite{EdEMMR18}.

    \subsection{Adjacent half-space invariants}

    Let us now consider another related scenario: two adjacent half spaces with different single-scattering albedos, $c_1 \neq c_2$, with an isotropic point or plane source at the boundary.  With emission and scattering being isotropic, it suffices to consider only a 1D rod, with displacements given by a normalized continuous symmetric distribution $K(x)$.  We can compute the mean number of collisions using Stokes' ``glass plates'' analysis of interreflections between the two half rods.    Let us define a reflection operator $\mathcal{R}(c)$ for a half rod for light arriving at the interface, using Eq.(\ref{eq:rodR})
    and a related absorption operator $\mathcal{A}(c) = 1 - \mathcal{R}(c)$.  The total absorption $\mathcal{A}_{1}$ in rod $1$ can be found by considering that the initial intensity entering rod $1$ is $1/2$ from the photons moving towards rod $1$ initially, and $(1/2)R(c_2)$ from the photons that reflect off of rod $2$ before arriving at rod $1$.  Combining these intensities, $((1/2)+(1/2)R(c_2))\mathcal{A}(c_1)$ will be absorbed in rod $1$ with no further collisions.  But some portion of this intensity will reflect off of rod $1$, and off of rod $2$, arriving back at rod $1$, etc.  Completing the geometric series, we find,
    \begin{equation}
        \mathcal{A}_1 = \left( \frac{1}{2} + \frac{\mathcal{R}(c_2)}{2} \right) \mathcal{A}(c_1) \left( \frac{1}{1 - \mathcal{R}(c_2) \mathcal{R}(c_1) } \right). 
    \end{equation}
    The total number of collisions in rod $1$ is then
    \begin{equation}
        C_1 = \frac{\mathcal{A}_1}{1-c_1} = \frac{1}{\sqrt{1-c_1} \sqrt{1-c_2}-c_1+1}.
    \end{equation}
    This agrees with known results for the collision rate density $C(0)$ at the boundary between two half spaces due to a uniform unit source in the half space with absorption $c_1$~\cite{mendelson1964one}, which again follows from the reciprocity theorem.

    The total number of collisions $C_2$ in rod 2 is $C_1$ with $c_1$ and $c_2$ swapped, and the total number of collisions in the entire system given by
    \begin{equation}\label{eq:C1C2}
        \int_{-\infty}^\infty C(x) dx = C_1 + C_2 = \frac{1}{\sqrt{1-c_1}\sqrt{1-c_2}}.
    \end{equation}
    To the best of our knowledge, Eq.(\ref{eq:C1C2}) is new and contains within it two classical universal properties of linear transport.  When both absorption albedos match, $c = c_1 = c_2$, we form a homogeneous infinite medium with the desired mean number of collisions $C_1 + C_2 = 1 / (1-c)$.  Also, when $c_2 = 0$, we recover the $\sqrt{\epsilon}$ law~\cite{ivanov94} for the half space. 

    We found Monte Carlo simulation to validate Eq.(\ref{eq:C1C2}) for a variety of dimensions and free-path distributions between collisions.

\section{Conclusion}\label{sec:conclusion}

  We have solved the half-space albedo and Milne problems of linear transport theory in a generalized setting of Euclidean $\mathbb{R}^d$ space.  The familiar solution of the 3D problem is given new mathematical context as a member of a new family of pseudo problems whose general solutions are given exactly in terms of hypergeometric functions.  The 3D problem is a unique member of this family, the only dimension where the characteristic $\Psi(1) = c/2$ is not $0$ and not $\infty$ at $\mu = 1$, and the largest dimension that includes a discrete diffusion component for all absorption levels $c > 0$.

  In Section 89 of his book\cite{SC60}, Chandrasekhar concludes ``$\dots$ a full discussion of the relation of the `pseudo-problems' $\dots$ will throw some light on the basic structure of the theory of radiative transfer'' and we have found this to be the case.  The new family of pseudo problems includes a two dimensional wave propagation problem that corresponds exactly to isotropic scattering in Flatland half spaces, and an old solution of that problem has led to an exact analytic $H$ function for Flatland, which should be useful in practical settings.  

  That this family also includes $H$ functions arising for 3D anisotropic scattering, as given in~\ref{appendix:connections}, was unexpected.  These, and the strange, near correspondences between $H$ function moments in various dimension, seem worthy of future attention.

\section{Acknowledgments}
Thanks to B. D. Ganapol for pointing out the numerical errors in \cite{NM15} and for tracking down a twice-scattered BRDF solution for 3D~\cite{sears75}.

\appendix

  \section{Analytic expression for discrete eigenvalues}\label{appendix:eigen}

  From Eqs.(\ref{eq:HthetaZero}) and (\ref{WeinerHopf}) with $z = 0$ we follow Siewert~\cite{siewert1980computing} to find a closed-form expression for the discrete eigenvalues in any real dimensionality $d > 1$,
    \begin{equation}
      \nu_0 = \pm \frac{1}{\sqrt{1-c}} \exp \left( -\frac{1}{\pi} \int_0^1 \frac{\theta(t)}{t} dt \right),
    \end{equation}
    provided $(d-3)/(d-2) < c < 1$.  The term $\theta(t)$ is known analytically for all $d \ge 1$ by combining Eqs.(\ref{eq:theta}) and (\ref{eq:lambdad}) and must be restricted to be positive.

  \section{Exact analytic forms of $H$}\label{sec:Heval}

  In our initial study of the family of $H$ functions for isotropic scattering in $\mathbb{R}^d$ we found analytic forms for several special cases, as well as simple relationships between different dimensionalities at different levels of absorption. 

  \subsection{The $H$ function for the 1D rod}

  For the 1D rod, the $H$ function reduces to a simple expression,
  \begin{align}
  	H_{1D}(\mu) &= \exp \left( \frac{-\mu}{\pi} \int_0^\infty \frac{1}{1+\mu^2 t^2} \ln \left(1 -  \frac{c}{1+t^2} \right) dt \right) \nonumber \\ &= \frac{1+\mu}{1+\mu \, \sqrt{1-c}}.
  \end{align}

  \subsection{The Flatland $H$ function}

  In 2D Flatland, analytic forms have been reported~\cite{EdEMMR18} for normal incidence ($\mu = 1$)
  \begin{equation}\label{eq:H2Dmu1}
  	H_{2D}(1) = \sqrt{\frac{2}{1+\sqrt{1-c^2}}} \exp \left(\frac{c \,
   _3F_2\left(\frac{1}{2},1,1;\frac{3}{2},\frac{3}{2};c^2\right)}{\pi }\right)
  \end{equation}
  involving a hypergeometric function $_3F_2$ with derivative
  \begin{equation}
		\frac{\partial}{\partial x} \left( x \, _3F_2\left(\frac{1}{2},1,1;\frac{3}{2},\frac{3}{2};x^2\right) \right) = \frac{\sin ^{-1}(x)}{x \sqrt{1-x^2}}, \, \, \, \, \, \, 0 < x < 1.
  \end{equation}
  The same $_3F_2$ function appears in the case of conservative scattering in Flatland~\cite{EdEMMR18},
  \begin{multline}\label{eq:H2Dc1}
  	H_{2D}(\mu,c=1) = \sqrt{1+\mu} \left| \exp \left(\frac{
   _3F_2\left(\frac{1}{2},1,1;\frac{3}{2},\frac{3}{2};\frac{1}{\mu^2}\right)}{\pi \, \mu }\right) \right| \\
   = \sqrt{1+\mu} \exp \left(\frac{
   _3F_2\left(\frac{1}{2},1,1;\frac{3}{2},\frac{3}{2};\frac{1}{\mu^2}\right)}{\pi \, \mu } + \frac{1}{2}\cosh ^{-1}(\mu)\right).
  \end{multline}
  We now show that Flatland $H$ can be expressed analytically for all $0 < \mu < 1, 0 < c < 1$ by exploiting an integral form presented by Fock~\cite{fock44} for a diffraction problem with the MacDonald kernel of Eq.(\ref{eq:MacDonaldK}).  Fock found (see also \cite{Krein62}),
  \begin{equation}
  	H_{2D}(\mu) =  \sqrt{ \frac{1 + \mu}{1 + \mu \sqrt{1-c^2}} } \exp \left[ \frac{1}{2\pi} \int_{\cos^{-1}(1/\mu)-\sin^{-1}(c)}^{\cos^{-1}(1/\mu)+\sin^{-1}(c)} \frac{u}{\sin u} du \right].
  \end{equation}
  If we define
  \begin{equation}
  	\chi(x) \equiv \int \frac{x}{\sin x} dx  = -2 i \text{Li}_2\left(e^{i x}\right)+\frac{1}{2} i \text{Li}_2\left(e^{2 i x}\right)-2 x \tanh
   ^{-1}\left(e^{i x}\right),
  \end{equation}
  where $\text{Li}_2$ is the dilogarithm function,
  \begin{equation}
  	\text{Li}_2(z) = \sum_{k=1}^\infty \frac{z^k}{k^2} = \int_{z}^0 \frac{\ln( 1 - t)}{t}dt,
  \end{equation}  
  then we can write the analytic form of Flatland $H$,
  \begin{multline}\label{eq:FlatlandHanalytic}
  	H_{2D}(\mu) =  \sqrt{ \frac{1 + \mu}{1 + \mu \sqrt{1-c^2}} }  \exp \left[ \frac{1}{2\pi} \left( \chi(\cos^{-1}(1/\mu)+\sin^{-1}(c)) \right. \right. \\ \left. \left. - \chi(\cos^{-1}(1/\mu)-\sin^{-1}(c)) \right ) \vphantom{\frac{1}{2\pi}} \right] \\
  	= \sqrt{ \frac{1 + \mu}{1 + \mu \sqrt{1-c^2}} } \left| \exp \left[ \frac{1}{\pi} \chi \left[ \cos^{-1}(1/\mu)+\sin^{-1}(c) \right]  \right] \right|.
  \end{multline}
  We also observe that
  \begin{equation}\label{eq:hyperH}
  	\chi(x) = \text{Re} \left(\sin (x) \, _3F_2\left(\frac{1}{2},1,1;\frac{3}{2},\frac{3}{2};\sin
   ^2(x)\right)\right),
  \end{equation}
  which coincides with Eqs.(\ref{eq:H2Dmu1}) and (\ref{eq:H2Dc1}) by noting that 
  \begin{align}
    &\sin\left( \cos^{-1}(1/\mu)+\sin^{-1}(c) \right)_{c \rightarrow 1} = \frac{1}{\mu}; \\ &\sin\left( \cos^{-1}(1/\mu)+\sin^{-1}(c) \right)_{\mu \rightarrow 1} = c.
  \end{align}

  Similar expressions involving $\text{Li}_2(z)$ and $_3F_2$ also arise in the case of a Cauchy random flight in a 1D rod~\cite{kulczycki2010spectral} and path integrals over half spaces~\cite{inayat1987new}.  These analytic forms have advantages relative to the numerical evaluation of integrals using quadrature and related methods.  Software such as Mathematica can easily produce trusted benchmark values of any desired precision when given such analytic inputs. Using Eq.(\ref{eq:FlatlandHanalytic}), we found complete agreement with the benchmark values for Flatland $H$ reported in a previous paper~\cite{EdEMMR18}.  We caution that the above expressions are restricted to $0 < \mu < 1$, and do not apply to terms such as $H(\nu_0)$, which appear in the Milne extrapolation distance equations.

  The new analytic form of $H$ also permits additional exact simplified benchmark values involving Catalan's constant $C = 0.9159655942...$.  To add to the previously reported result~\cite{mishchenko1992multiple,EdEMMR18}
  \begin{equation}
    H_{2D}(1,c=1) = \sqrt{2} e^{\frac{2 C}{\pi }}
  \end{equation}
  we observe that
  \begin{equation}
  	H_{2D}(1,c=1/2) = \frac{2 e^{\frac{4 C}{3 \pi }}}{\left(2+\sqrt{3}\right)^{2/3}}.
  \end{equation}

  \subsection{The 4D $H$-Function}
  With a similar approach to the derivation in Flatland~\cite{EdEMMR18} we also found a closely-related analytic form of $H$ in 4D with conservative scattering,
  \begin{equation}
  	H_{4D}(\mu,c=1) = \frac{\left( H_{2D}(\mu,c=1) \right)^2}{1+\mu}.
  \end{equation}

  \subsection{Interdimensional correspondences}
  We also noticed an interesting relationship between dimensionalities two apart for specific values of absorption.  Let us write Eq.(\ref{eq:HsolutionRd}) for $H$ in a new notation, where $f(t)$ is a general input and the argument to the logarithm,
  \begin{equation}
  	H(\mu,f(t)) = \exp  \left( \frac{-\mu}{\pi} \int_0^\infty \frac{1}{1+t^2\mu^2} \ln f(t) dt \right).
  \end{equation}
  We have, by linearity of the integral and properties of the log and exp,
  \begin{equation}\label{eq:Hfactoring}
  	\frac{H\left(\mu,f_1(t)\right)}{ H\left(\mu,f_2(t)\right)} = \exp  \left( \frac{-\mu}{\pi} \int_0^\infty \frac{1}{1+t^2\mu^2} \ln \frac{f_1(t)}{ f_2(t)} dt \right)  = H\left(\mu,\frac{f_1(t)}{ f_2(t)} \right).
  \end{equation}
  If we now write Eq.(\ref{eq:Lambdad}) as a function with three arguments,
  \begin{equation}
  	\Lambda(z,c,d) = 1 - c _2F_1\left(\frac{1}{2},1;\frac{d}{2};\frac{1}{z^2} \right),
  \end{equation}
  we conjecture
  \begin{equation}
  	\frac{\Lambda \left(\frac{i}{t},1,d-2\right)}{\Lambda
   \left(\frac{i}{t},\frac{d-3}{d-2},d\right)} = \frac{t^2}{1+t^2}
  \end{equation}
  for $d \ge 3$ (which we verified for $d$ up to $30$).  Given that
  \begin{equation}
  	H\left(\mu,\frac{t^2}{1+t^2}\right) = \exp  \left( \frac{-\mu}{\pi} \int_0^\infty \frac{1}{1+t^2\mu^2} \ln \frac{t^2}{1+t^2} dt \right) = 1 + \mu
  \end{equation}
  we therefore conjecture that
  \begin{equation}
  	H_{(d-2)D}(\mu,c=1) = (1+\mu) H_{dD}\left(\mu,c=\frac{d-3}{d-2}\right).
  \end{equation}
  By this notation we mean
  \begin{equation}
    H_{dD}(\mu,c=x) \equiv H(\mu,\Lambda(i/t,x,d)).
  \end{equation}
  This relates the $H$ function for conservative scattering at some dimensionality $d$ to the $H$ function in two dimensions higher where absorption is set to $c = \frac{d-3}{d-2}$, which is the absorption level where the discrete eigenspectra disappears.  For example,
  \begin{subequations}
  \begin{align}
  	&H_{1D}(\mu,c=1) = (1+\mu) H_{3D}(\mu,c=0) \\
  	&H_{2D}(\mu,c=1) = (1+\mu) H_{4D}(\mu,c=1/2) \\
  	&H_{3D}(\mu,c=1) = (1+\mu) H_{5D}(\mu,c=2/3) \\
  	&H_{4D}(\mu,c=1) = (1+\mu) H_{6D}(\mu,c=3/4)
  \end{align}
  \end{subequations}
  and so on.  These are, to the best of our knowledge, the only known analytic expressions for (or relationships between) $H$-functions that correspond to non-trivial multiple-scattering processes.  

  Similar unexpected correspondences have been noted in infinite $dD$ transport processes, such as Brownian motion, where a particle leaving the origin spends the same total time in a sphere in dimension $d+2$ as the first time to escape the sphere of the same radius in dimension $d$, for any $d > 1$ and any radius~\cite{ciesielski1962first}.

\section{Monte-Carlo sampling in $\mathbb{R}^d$}\label{appendix:MC}

  \subsection{Isotropic direction sampling}

    Uniform random sampling of directions on the $d$D sphere $S^{d-1}$ are generated by normalizing a vector of $d$ independent normally-distributed random numbers with a mean of $0$, which we generate using Box-Muller transforms.

  \subsection{Unit uniform diffuse illumination sampling}
    For uniform diffuse illumination, we can invert the cumulative distribution function
    \begin{equation}
        \frac{\sqrt{\pi } \, \Gamma \left(\frac{d+1}{2}\right)}{\Gamma \left(\frac{d}{2}\right)} \int_0^\mu  \mu' G(\mu') d\mu' = \xi = 1-\left(1-\mu^2\right)^{\frac{d-1}{2}}
    \end{equation}
    and solve for $\mu$ to find
    \begin{equation}\label{eq:diffuseMC}
    	\mu = \sqrt{1-(1-\xi)^{\frac{2}{d-1}}}.
    \end{equation}
    This gives a Monte-Carlo sampling procedure for uniform diffuse illumination of the half space by selecting incoming cosines $\mu$ from Eq.(\ref{eq:diffuseMC}) where $0 < \xi < 1$ is a uniform random number.

\section{Connections to anisotropic scattering and nonclassical transport}\label{appendix:connections}

  In studying the variation of the transport kernel and related eigenspectra as a function of the dimensionality $d$ of the half space with isotropic scattering, we have come across transport kernels and related $H$ functions that exactly describe transport in 3D domains under anisotropic scattering or with correlation between the scattering centers in the medium.  We briefly identify these correspondences here.

  \subsection{Anisotropic scattering}
        In the exact law of diffuse reflection for a 3D half space with linearly-anisotropic scattering (\cite{SC60} p. 140) the solution involves an $H$ function with characteristic
        \begin{equation}
        	\Psi(\mu) = \frac{1}{4} b c (1 - \mu^2),
        \end{equation}
        where the phase function $(1 + b \cos \theta)$ has anisotropy parameter $-1 < b < 1$.
        This maps precisely to the $H$ function for isotropic scattering in 5D
        \begin{equation}
        	\Psi(\mu) = \frac{3}{4} c_{5} (1 - \mu^2),
        \end{equation}
        with a reduced absorption level
        \begin{equation}
        	c_{5} = \frac{1}{3} b c.
        \end{equation}
        Similarly, in the case of Rayleigh scattering in 3D, one of the three $H$ functions has a characteristic
        \begin{equation}
        	\Psi(\mu) = \frac{3}{32} (1 - \mu^2)^2,
        \end{equation}
        which corresponds to isotropic scattering in 7D with single-scattering albedo $c_7 = 1 / 10$.  Results such as Eqs.(\ref{eq:HthetaZero}) and (\ref{eq:HthetaNoZero}) may offer previously unconsidered ways to compute transport solutions for these 3D problems.

  \subsection{Nonclassical transport}\label{A:nonexp}

        In the case of a random medium with spatially-correlated scatterers~\cite{audic93,larsen11}, the exponential distribution of path-lengths between collisions in Eq.(\ref{eq:WHK}) becomes a general, non-negative, normalized distribution $p_c(s)$ and the plane-parallel generalized Schwarzschild-Milne kernel for the collision rate density becomes~\cite{deon18ii}
        \begin{equation}
            K(x) = \frac{1}{2} \int_0^1   p_c\left(|x|/\mu\right) \frac{1}{\mu} G(\mu) \, d\mu.
        \end{equation}
        In infinite spherical geometries it was shown~\cite{deon13} that the diffusion asymptotics of isotropic scattering in a classical medium of dimensionality $d$ can be exhibited by a non-exponential transport process in dimensionality $d'$ by appropriate choice of $p_c(s)$.  The same procedure can translate plane-parallel collision kernels $K(x)$ from one dimensionality to another.  Of particular interest are the Picard and MacDonald kernels, whose related $H$ functions have analytic expressions.  Either kernel can be exhibited in $\mathbb{R}^d$ under isotropic scattering by selecting the free-paths between collisions from a family of distributions involving Bessel-$K$ functions,
        \begin{equation}\label{eq:pcdiffusionpower}
            p_c(s) = \frac{d \left(\frac{s}{2}\right)^{-1+\frac{d}{2}+p} K_{\frac{1}{2} (d-2 p)}(s)}{\Gamma
        \left(\frac{d}{2}+1\right) \Gamma (p)}, \; \; \; \; d \ge 1, p \ge \frac{1}{2}
        \end{equation}
        whose Fourier-transformed propagator in $\mathbb{R}^d$ is a diffusion mode to a power $p$~\cite{deon18ii},
        \begin{equation}
          \tilde{K}(t) = \left( \frac{1}{1+t^2} \right)^p.
        \end{equation}
        The Picard kernel in Eq.(\ref{eq:PicardK}) arises when $p=1$, which happens in Flatland~\cite{deon13} via
        \begin{equation}
            p_c(s) = s K_0(s)
        \end{equation}
        and in 3D via~\cite{deon13}
        \begin{equation}
            p_c(s) = \e^{-s} s.
        \end{equation}
        The MacDonald kernel in Eq.(\ref{eq:MacDonaldK}) arises when $p=1/2$, which happens in 3D via
        \begin{equation}
            p_c(s) = \frac{2 s K_1(s)}{\pi }.
        \end{equation}
        The $H$ function for $p = 2$ can also be written analytically by combining the 1D $H$ Function and Eq.(\ref{eq:Hfactoring}) to yield
        \begin{equation}
            H\left(\mu,1-c \, \left(\frac{1}{1+t^2}\right)^2\right)  = \frac{(\mu +1)^2}{\left(\sqrt{1-\sqrt{c}} \mu +1\right) \left(\sqrt{\sqrt{c}+1} \mu +1\right)}.
        \end{equation}

        While the same kernel can describe two different scattering processes in half spaces of differing dimensionalities, the solutions of the respective albedo problems are different because the rate density of initial collisions $C_0(z)$ is non-exponential in the case of correlation, and the relationship between the scalar collision rate and emerging distribution is less directly expressed via Laplace transforms.  Complete details of solving half space albedo problems with correlation will be considered in a future paper~\cite{deon18iv}.

\bibliographystyle{elsarticle-num-names}
\bibliography{halfspaceDd}

\begin{thebibliography}{99}
\providecommand{\natexlab}[1]{#1}
\providecommand{\url}[1]{\texttt{#1}}
\providecommand{\urlprefix}{URL }
\expandafter\ifx\csname urlstyle\endcsname\relax
  \providecommand{\doi}[1]{doi:\discretionary{}{}{}#1}\else
  \providecommand{\doi}[1]{doi:\discretionary{}{}{}\begingroup
  \urlstyle{rm}\url{#1}\endgroup}\fi
\providecommand{\bibinfo}[2]{#2}

\bibitem[{Chandrasekhar(1960)}]{SC60}
\bibinfo{author}{S.~Chandrasekhar}, \bibinfo{title}{Radiative Transfer},
  \bibinfo{publisher}{Dover}, \bibinfo{year}{1960}.

\bibitem[{Davison(1957)}]{davison57}
\bibinfo{author}{B.~Davison}, \bibinfo{title}{{Neutron Transport Theory}},
  \bibinfo{publisher}{Oxford University Press}, \bibinfo{year}{1957}.

\bibitem[{van Rossum and Nieuwenhuizen(1999)}]{vanrossum99}
\bibinfo{author}{M.~C.~W. van Rossum}, \bibinfo{author}{T.~M. Nieuwenhuizen},
  \bibinfo{title}{{Multiple scattering of classical waves: microscopy,
  mesoscopy, and diffusion}}, \bibinfo{journal}{Reviews of Modern Physics}
  \bibinfo{volume}{71}~(\bibinfo{number}{1}) (\bibinfo{year}{1999})
  \bibinfo{pages}{313--371}, ISSN \bibinfo{issn}{1539-0756},
  \urlprefix\url{https://doi.org/10.1103/RevModPhys.71.313}.

\bibitem[{Grosjean(1953)}]{grosjean53}
\bibinfo{author}{C.~C. Grosjean}, \bibinfo{title}{{Solution of the
  non-isotropic random flight problem in the $k$-dimensional space}},
  \bibinfo{journal}{Physica} \bibinfo{volume}{19}~(\bibinfo{number}{1-12})
  (\bibinfo{year}{1953}) \bibinfo{pages}{29--45}, ISSN
  \bibinfo{issn}{0031-8914},
  \urlprefix\url{https://doi.org/10.1016/S0031-8914(53)80004-2}.

\bibitem[{Wing(1962)}]{wing62}
\bibinfo{author}{G.~M. Wing}, \bibinfo{title}{{An Introduction to Transport
  Theory}}, \bibinfo{publisher}{Wiley}, \bibinfo{year}{1962}.

\bibitem[{Kohler and Papanicolaou(1973)}]{kohler1973power}
\bibinfo{author}{W.~Kohler}, \bibinfo{author}{G.~C. Papanicolaou},
  \bibinfo{title}{Power statistics for wave propagation in one dimension and
  comparison with radiative transport theory}, \bibinfo{journal}{Journal of
  Mathematical Physics} \bibinfo{volume}{14}~(\bibinfo{number}{12})
  (\bibinfo{year}{1973}) \bibinfo{pages}{1733--1745},
  \urlprefix\url{https://doi.org/10.1063/1.1666247}.

\bibitem[{Hoogenboom(2008)}]{hoogenboom08}
\bibinfo{author}{J.~E. Hoogenboom}, \bibinfo{title}{{The two-direction
  neutral-particle transport model: A useful tool for research and education}},
  \bibinfo{journal}{Transport Theory and Statistical Physics}
  \bibinfo{volume}{37}~(\bibinfo{number}{1}) (\bibinfo{year}{2008})
  \bibinfo{pages}{65--108},
  \urlprefix\url{https://doi.org/10.1080/00411450802271791}.

\bibitem[{Davis and Xu(2014)}]{davis14}
\bibinfo{author}{A.~B. Davis}, \bibinfo{author}{F.~Xu}, \bibinfo{title}{A
  generalized linear transport model for spatially correlated stochastic
  media}, \bibinfo{journal}{Journal of Computational and Theoretical Transport}
  \bibinfo{volume}{43}~(\bibinfo{number}{1-7}) (\bibinfo{year}{2014})
  \bibinfo{pages}{474--514},
  \urlprefix\url{https://doi.org/10.1080/23324309.2014.978083}.

\bibitem[{Jarosz et~al.(2012)Jarosz, Sch{\"o}nefeld, Kobbelt, and
  Jensen}]{JSKJ12}
\bibinfo{author}{W.~Jarosz}, \bibinfo{author}{V.~Sch{\"o}nefeld},
  \bibinfo{author}{L.~Kobbelt}, \bibinfo{author}{H.~W. Jensen},
  \bibinfo{title}{Theory, analysis and applications of 2D global illumination},
  \bibinfo{journal}{ACM Transactions on Graphics (TOG)}
  \bibinfo{volume}{31}~(\bibinfo{number}{5}) (\bibinfo{year}{2012})
  \bibinfo{pages}{125},
  \urlprefix\url{https://doi.org/10.1145/2231816.2231823}.

\bibitem[{Bitterli(2015)}]{tantalum}
\bibinfo{author}{B.~Bitterli}, \bibinfo{title}{The Secret Life of Photons:
  Simulating 2D Light Transport}
  \urlprefix\url{https://benedikt-bitterli.me/tantalum/}, \bibinfo{note}{(last
  accessed Dec 2018)}.

\bibitem[{Schuster(1905)}]{schuster05}
\bibinfo{author}{A.~Schuster}, \bibinfo{title}{Radiation through a foggy
  atmosphere}, \bibinfo{journal}{The Astrophysical Journal}
  \bibinfo{volume}{21} (\bibinfo{year}{1905}) \bibinfo{pages}{1},
  \urlprefix\url{https://doi.org/10.1086/141186}.

\bibitem[{Kubelka(1931)}]{kubelka31}
\bibinfo{author}{P.~Kubelka}, \bibinfo{title}{{Ein Beitrag zur Optik der
  Farbanstriche (Contribution to the optic of paint)}},
  \bibinfo{journal}{Zeitschrift fur technische Physik} \bibinfo{volume}{12}
  (\bibinfo{year}{1931}) \bibinfo{pages}{593--601}.

\bibitem[{Meador and Weaver(1980)}]{meador80}
\bibinfo{author}{W.~E. Meador}, \bibinfo{author}{W.~R. Weaver},
  \bibinfo{title}{{Two-stream approximations to radiative transfer in planetary
  atmospheres- A unified description of existing methods and a new
  improvement}}, \bibinfo{journal}{Journal of the Atmospheric Sciences}
  \bibinfo{volume}{37}~(\bibinfo{number}{3}) (\bibinfo{year}{1980})
  \bibinfo{pages}{630--643}, ISSN \bibinfo{issn}{1520-0469},
  \urlprefix\url{https://doi.org/10.1175/1520-0469(1980)037<0630:TSATRT>2.0.CO;2}.

\bibitem[{Jakeman and Pusey(1976)}]{jakeman1976model}
\bibinfo{author}{E.~Jakeman}, \bibinfo{author}{P.~N. Pusey}, \bibinfo{title}{A
  model for non-Rayleigh sea echo}, \bibinfo{journal}{IEEE Transactions on
  Antennas and Propagation} \bibinfo{volume}{24}~(\bibinfo{number}{6})
  (\bibinfo{year}{1976}) \bibinfo{pages}{806--814},
  \urlprefix\url{https://doi.org/10.1109/TAP.1976.1141451}.

\bibitem[{Sato et~al.(2012)Sato, Fehler, and Maeda}]{sato}
\bibinfo{author}{H.~Sato}, \bibinfo{author}{M.~C. Fehler},
  \bibinfo{author}{T.~Maeda}, \bibinfo{title}{Seismic Wave Propagation and
  Scattering in the Heterogeneous Earth}, vol. \bibinfo{volume}{496},
  \bibinfo{publisher}{Springer},
  \urlprefix\url{https://doi.org/10.1007/978-3-540-89623-4},
  \bibinfo{year}{2012}.

\bibitem[{Jeanson et~al.(2005)Jeanson, Rivault, Deneubourg, Blanco, Fournier,
  Jost, and Theraulaz}]{jeanson05}
\bibinfo{author}{R.~Jeanson}, \bibinfo{author}{C.~Rivault},
  \bibinfo{author}{J.-L. Deneubourg}, \bibinfo{author}{S.~Blanco},
  \bibinfo{author}{R.~Fournier}, \bibinfo{author}{C.~Jost},
  \bibinfo{author}{G.~Theraulaz}, \bibinfo{title}{Self-organized aggregation in
  cockroaches}, \bibinfo{journal}{Animal Behaviour}
  \bibinfo{volume}{69}~(\bibinfo{number}{1}) (\bibinfo{year}{2005})
  \bibinfo{pages}{169--180},
  \urlprefix\url{https://doi.org/10.1016/j.anbehav.2004.02.009}.

\bibitem[{Yang and Deb(2009)}]{XYSD09}
\bibinfo{author}{X.-S. Yang}, \bibinfo{author}{S.~Deb}, \bibinfo{title}{Cuckoo
  search via L{\'e}vy flights}, in: \bibinfo{booktitle}{Nature \& Biologically
  Inspired Computing, 2009. NaBIC 2009. World Congress on},
  \bibinfo{organization}{IEEE}, \bibinfo{pages}{210--214},
  \urlprefix\url{https://doi.org/10.1109/NABIC.2009.5393690},
  \bibinfo{year}{2009}.

\bibitem[{Fock(1944)}]{fock44}
\bibinfo{author}{V.~Fock}, \bibinfo{title}{Some integral equations of
  mathematical physics}, in: \bibinfo{booktitle}{Doklady AN SSSR},
  vol.~\bibinfo{volume}{26}, \bibinfo{pages}{147--151},
  \urlprefix\url{http://mi.mathnet.ru/eng/msb6183}, \bibinfo{year}{1944}.

\bibitem[{Bal et~al.(2000)Bal, Freilikher, Papanicolaou, and Ryzhik}]{Bal00}
\bibinfo{author}{G.~Bal}, \bibinfo{author}{V.~Freilikher},
  \bibinfo{author}{G.~Papanicolaou}, \bibinfo{author}{L.~Ryzhik},
  \bibinfo{title}{Wave transport along surfaces with random impedance},
  \bibinfo{journal}{Physical Review B}
  \bibinfo{volume}{62}~(\bibinfo{number}{10}) (\bibinfo{year}{2000})
  \bibinfo{pages}{6228},
  \urlprefix\url{https://doi.org/10.1103/PhysRevB.62.6228}.

\bibitem[{Meylan and Masson(2006)}]{MHDM06}
\bibinfo{author}{M.~H. Meylan}, \bibinfo{author}{D.~Masson}, \bibinfo{title}{{A
  linear Boltzmann equation to model wave scattering in the marginal ice
  zone}}, \bibinfo{journal}{Ocean Modelling}
  \bibinfo{volume}{11}~(\bibinfo{number}{3-4}) (\bibinfo{year}{2006})
  \bibinfo{pages}{417--427},
  \urlprefix\url{https://doi.org/10.1016/j.ocemod.2004.12.008}.

\bibitem[{Vynck et~al.(2012)Vynck, Burresi, Riboli, and Wiersma}]{KVDSW12}
\bibinfo{author}{K.~Vynck}, \bibinfo{author}{M.~Burresi},
  \bibinfo{author}{F.~Riboli}, \bibinfo{author}{D.~S. Wiersma},
  \bibinfo{title}{Photon management in two-dimensional disordered media},
  \bibinfo{journal}{Nature Materials}
  \bibinfo{volume}{11}~(\bibinfo{number}{12}) (\bibinfo{year}{2012})
  \bibinfo{pages}{1017}, \urlprefix\url{https://doi.org/10.1038/nmat3442}.

\bibitem[{Gorodnichev et~al.(1989)Gorodnichev, Dudarev, and
  Rogozkin}]{gorodnichev89}
\bibinfo{author}{E.~E. Gorodnichev}, \bibinfo{author}{S.~L. Dudarev},
  \bibinfo{author}{D.~B. Rogozkin}, \bibinfo{title}{{Coherent backscattering
  enhancement under conditions of weak wave localization in disordered 3D and
  2D systems}}, \bibinfo{journal}{Soviet Physics JETP}
  \bibinfo{volume}{69}~(\bibinfo{number}{3}) (\bibinfo{year}{1989})
  \bibinfo{pages}{481--490}.

\bibitem[{Gorodnichev et~al.(1990)Gorodnichev, Dudarev, and
  Rogozkin}]{gorodnichev90}
\bibinfo{author}{E.~E. Gorodnichev}, \bibinfo{author}{S.~L. Dudarev},
  \bibinfo{author}{D.~B. Rogozkin}, \bibinfo{title}{{Coherent wave
  backscattering by random medium. Exact solution of the albedo problem}},
  \bibinfo{journal}{Physics Letters A}
  \bibinfo{volume}{144}~(\bibinfo{number}{1}) (\bibinfo{year}{1990})
  \bibinfo{pages}{48--54},
  \urlprefix\url{https://doi.org/10.1016/0375-9601(90)90047-R}.

\bibitem[{Mishchenko et~al.(1992)Mishchenko, Dlugach, and
  Yanovitskij}]{mishchenko1992multiple}
\bibinfo{author}{M.~I. Mishchenko}, \bibinfo{author}{J.~M. Dlugach},
  \bibinfo{author}{E.~G. Yanovitskij}, \bibinfo{title}{Multiple light
  scattering by polydispersions of randomly distributed, perfectly-aligned,
  infinite Mie cylinders illuminated perpendicularly to their axes},
  \bibinfo{journal}{Journal of Quantitative Spectroscopy and Radiative
  Transfer} \bibinfo{volume}{47}~(\bibinfo{number}{5}) (\bibinfo{year}{1992})
  \bibinfo{pages}{401--410},
  \urlprefix\url{https://doi.org/10.1016/0022-4073(92)90041-2}.

\bibitem[{Grzesik(2018)}]{grzesik2018radiative}
\bibinfo{author}{J.~A. Grzesik}, \bibinfo{title}{Radiative albedo from a
  linearly fibered half-space}, \bibinfo{journal}{The European Physical Journal
  Plus} \bibinfo{volume}{133}~(\bibinfo{number}{5}) (\bibinfo{year}{2018})
  \bibinfo{pages}{178},
  \urlprefix\url{https://doi.org/10.1140/epjp/i2018-12004-4}.

\bibitem[{Marschner et~al.(2003)Marschner, Jensen, Cammarano, Worley, and
  Hanrahan}]{marschner03}
\bibinfo{author}{S.~R. Marschner}, \bibinfo{author}{H.~W. Jensen},
  \bibinfo{author}{M.~Cammarano}, \bibinfo{author}{S.~Worley},
  \bibinfo{author}{P.~Hanrahan}, \bibinfo{title}{Light scattering from human
  hair fibers}, in: \bibinfo{booktitle}{ACM Transactions on Graphics (TOG)},
  vol.~\bibinfo{volume}{22}, \bibinfo{organization}{ACM},
  \bibinfo{pages}{780--791},
  \urlprefix\url{https://doi.org/10.1145/1201775.882345}, \bibinfo{year}{2003}.

\bibitem[{Kelley and Larsen(2015)}]{BWKEWL2016}
\bibinfo{author}{B.~W. Kelley}, \bibinfo{author}{E.~W. Larsen},
  \bibinfo{title}{A consistent 2D/1D approximation to the 3D neutron transport
  equation}, \bibinfo{journal}{Nuclear Engineering and Design}
  \bibinfo{volume}{295} (\bibinfo{year}{2015}) \bibinfo{pages}{598--614},
  \urlprefix\url{https://doi.org/10.1016/j.nucengdes.2015.07.026}.

\bibitem[{Paasschens(1997)}]{paasschens97}
\bibinfo{author}{J.~C.~J. Paasschens}, \bibinfo{title}{{Solution of the
  time-dependent Boltzmann equation}}, \bibinfo{journal}{Physical Review E}
  \bibinfo{volume}{56}~(\bibinfo{number}{1}) (\bibinfo{year}{1997})
  \bibinfo{pages}{1135--1141},
  \urlprefix\url{https://doi.org/10.1103/PhysRevE.56.1135}.

\bibitem[{Martelli et~al.(2007)Martelli, Sassaroli, Pifferi, Torricelli,
  Spinelli, and Zaccanti}]{martelli07}
\bibinfo{author}{F.~Martelli}, \bibinfo{author}{A.~Sassaroli},
  \bibinfo{author}{A.~Pifferi}, \bibinfo{author}{A.~Torricelli},
  \bibinfo{author}{L.~Spinelli}, \bibinfo{author}{G.~Zaccanti},
  \bibinfo{title}{{Heuristic Green's function of the time dependent radiative
  transfer equation for a semi-infinite medium}}, \bibinfo{journal}{Optics
  Express} \bibinfo{volume}{15}~(\bibinfo{number}{26}) (\bibinfo{year}{2007})
  \bibinfo{pages}{18168--18175},
  \urlprefix\url{https://doi.org/10.1364/OE.15.018168}.

\bibitem[{Reimberg and Abramo(2015)}]{reimberg2015random}
\bibinfo{author}{P.~H.~F. Reimberg}, \bibinfo{author}{L.~R. Abramo},
  \bibinfo{title}{Random flights through spaces of different dimensions},
  \bibinfo{journal}{Journal of Mathematical Physics}
  \bibinfo{volume}{56}~(\bibinfo{number}{1}) (\bibinfo{year}{2015})
  \bibinfo{pages}{013512}, \urlprefix\url{https://doi.org/10.1063/1.4906808}.

\bibitem[{Ciesielski and Taylor(1962)}]{ciesielski1962first}
\bibinfo{author}{Z.~Ciesielski}, \bibinfo{author}{S.~J. Taylor},
  \bibinfo{title}{{First passage times and sojourn times for Brownian motion in
  space and the exact Hausdorff measure of the sample path}},
  \bibinfo{journal}{Transactions of the American Mathematical Society}
  \bibinfo{volume}{103}~(\bibinfo{number}{3}) (\bibinfo{year}{1962})
  \bibinfo{pages}{434--450}, \urlprefix\url{https://doi.org/10.2307/1993838}.

\bibitem[{Majumdar et~al.(2006)Majumdar, Comtet, and Ziff}]{majumdar06}
\bibinfo{author}{S.~N. Majumdar}, \bibinfo{author}{A.~Comtet},
  \bibinfo{author}{R.~M. Ziff}, \bibinfo{title}{Unified solution of the
  expected maximum of a discrete time random walk and the discrete flux to a
  spherical trap}, \bibinfo{journal}{Journal of Statistical Physics}
  \bibinfo{volume}{122}~(\bibinfo{number}{5}) (\bibinfo{year}{2006})
  \bibinfo{pages}{833--856},
  \urlprefix\url{https://doi.org/10.1007/s10955-005-9002-x}.

\bibitem[{Comtet and Tourigny(2011)}]{comtet2011excursions}
\bibinfo{author}{A.~Comtet}, \bibinfo{author}{Y.~Tourigny},
  \bibinfo{title}{Excursions of diffusion processes and continued fractions},
  in: \bibinfo{booktitle}{Annales de l'Institut Henri Poincar{\'e},
  Probabilit{\'e}s et Statistiques}, vol.~\bibinfo{volume}{47},
  \bibinfo{organization}{Institut Henri Poincar{\'e}},
  \bibinfo{pages}{850--874},
  \urlprefix\url{https://doi.org/10.1214/10-AIHP390}, \bibinfo{year}{2011}.

\bibitem[{Chandrasekhar(1943)}]{chandrasekhar43}
\bibinfo{author}{S.~Chandrasekhar}, \bibinfo{title}{Stochastic problems in
  physics and astronomy}, \bibinfo{journal}{Reviews of Modern Physics}
  \bibinfo{volume}{15}~(\bibinfo{number}{1}) (\bibinfo{year}{1943})
  \bibinfo{pages}{1--89},
  \urlprefix\url{https://doi.org/10.1103/RevModPhys.15.1}.

\bibitem[{Watanabe and Watanabe(1970)}]{watanabe1970convergence}
\bibinfo{author}{S.~Watanabe}, \bibinfo{author}{T.~Watanabe},
  \bibinfo{title}{{Convergence of isotropic scattering transport process to
  Brownian motion}}, \bibinfo{journal}{Nagoya Mathematical Journal}
  \bibinfo{volume}{40} (\bibinfo{year}{1970}) \bibinfo{pages}{161--171},
  \urlprefix\url{https://doi.org/10.1017/S0027763000013933}.

\bibitem[{Fournier and Frisch(1978)}]{fournier78}
\bibinfo{author}{J.-D. Fournier}, \bibinfo{author}{U.~Frisch},
  \bibinfo{title}{$d$-Dimensional turbulence}, \bibinfo{journal}{Physical
  Review A} \bibinfo{volume}{17}~(\bibinfo{number}{2}) (\bibinfo{year}{1978})
  \bibinfo{pages}{747},
  \urlprefix\url{https://doi.org/10.1103/PhysRevA.17.747}.

\bibitem[{Dutka(1985)}]{dutka1985problem}
\bibinfo{author}{J.~Dutka}, \bibinfo{title}{On the problem of random flights},
  \bibinfo{journal}{Archive for History of Exact Sciences}
  \bibinfo{volume}{32}~(\bibinfo{number}{3-4}) (\bibinfo{year}{1985})
  \bibinfo{pages}{351--375}.

\bibitem[{Zoia et~al.(2011)Zoia, Dumonteil, and Mazzolo}]{zoia11}
\bibinfo{author}{A.~Zoia}, \bibinfo{author}{E.~Dumonteil},
  \bibinfo{author}{A.~Mazzolo}, \bibinfo{title}{Collision densities and mean
  residence times for d-dimensional exponential flights},
  \bibinfo{journal}{Physical Review E}
  \bibinfo{volume}{83}~(\bibinfo{number}{4}) (\bibinfo{year}{2011})
  \bibinfo{pages}{041137},
  \urlprefix\url{https://doi.org/10.1103/PhysRevE.83.041137}.

\bibitem[{Kingman(1963)}]{kingman1963random}
\bibinfo{author}{J.~F.~C. Kingman}, \bibinfo{title}{Random walks with spherical
  symmetry}, \bibinfo{journal}{Acta Mathematica}
  \bibinfo{volume}{109}~(\bibinfo{number}{1}) (\bibinfo{year}{1963})
  \bibinfo{pages}{11--53}, \urlprefix\url{https://doi.org/10.1007/BF02391808}.

\bibitem[{Abramowitz and Stegun(1972)}]{abramowitz1965handbook}
\bibinfo{editor}{M.~Abramowitz}, \bibinfo{editor}{I.~A. Stegun} (Eds.),
  \bibinfo{title}{Handbook of Mathematical Functions: with Formulas, Graphs,
  and Mathematical Tables}, \bibinfo{publisher}{Dover}, \bibinfo{edition}{9}
  edn., \bibinfo{year}{1972}.

\bibitem[{Kolesnik and Orsingher(2005)}]{kolesnik2005planar}
\bibinfo{author}{A.~D. Kolesnik}, \bibinfo{author}{E.~Orsingher},
  \bibinfo{title}{A planar random motion with an infinite number of directions
  controlled by the damped wave equation}, \bibinfo{journal}{Journal of Applied
  Probability} \bibinfo{volume}{42}~(\bibinfo{number}{4})
  (\bibinfo{year}{2005}) \bibinfo{pages}{1168--1182},
  \urlprefix\url{https://doi.org/10.1239/jap/1134587824}.

\bibitem[{Kolesnik(2008)}]{kolesnik2008random}
\bibinfo{author}{A.~D. Kolesnik}, \bibinfo{title}{Random motions at finite
  speed in higher dimensions}, \bibinfo{journal}{Journal of Statistical
  Physics} \bibinfo{volume}{131}~(\bibinfo{number}{6}) (\bibinfo{year}{2008})
  \bibinfo{pages}{1039--1065},
  \urlprefix\url{https://doi.org/10.1007/s10955-008-9532-0}.

\bibitem[{d'Eon(2013)}]{deon13}
\bibinfo{author}{E.~d'Eon}, \bibinfo{title}{{Rigorous asymptotic and
  moment-preserving diffusion approximations for generalized linear Boltzmann
  transport in arbitrary dimension}}, \bibinfo{journal}{Transport Theory and
  Statistical Physics} \bibinfo{volume}{42}~(\bibinfo{number}{6-7})
  (\bibinfo{year}{2013}) \bibinfo{pages}{237--297},
  \urlprefix\url{https://doi.org/10.1080/00411450.2014.910231}.

\bibitem[{d'Eon(2019{\natexlab{a}})}]{deon18ii}
\bibinfo{author}{E.~d'Eon}, \bibinfo{title}{{A reciprocal formulation of
  nonexponential radiative transfer. 2: Monte Carlo estimation and diffusion
  approximation}}, \bibinfo{journal}{Journal of Computational and Theoretical
  Transport (submitted)} .

\bibitem[{Liemert and Kienle(2011)}]{liemert11}
\bibinfo{author}{A.~Liemert}, \bibinfo{author}{A.~Kienle},
  \bibinfo{title}{Radiative transfer in two-dimensional infinitely extended
  scattering media}, \bibinfo{journal}{Journal of Physics A: Mathematical and
  Theoretical} \bibinfo{volume}{44} (\bibinfo{year}{2011})
  \bibinfo{pages}{505206},
  \urlprefix\url{https://doi.org/10.1088/1751-8113/44/50/505206}.

\bibitem[{Machida(2016)}]{MM16}
\bibinfo{author}{M.~Machida}, \bibinfo{title}{The radiative transport equation
  in flatland with separation of variables}, \bibinfo{journal}{Journal of
  Mathematical Physics} \bibinfo{volume}{57}~(\bibinfo{number}{7})
  (\bibinfo{year}{2016}) \bibinfo{pages}{073301},
  \urlprefix\url{https://doi.org/10.1063/1.4958976}.

\bibitem[{Rossetto(2017)}]{rossetto17}
\bibinfo{author}{V.~Rossetto}, \bibinfo{title}{Space--time domain velocity
  distributions in isotropic radiative transfer in two dimensions},
  \bibinfo{journal}{Journal of Physics A: Mathematical and Theoretical}
  \bibinfo{volume}{50}~(\bibinfo{number}{16}) (\bibinfo{year}{2017})
  \bibinfo{pages}{165001},
  \urlprefix\url{https://doi.org/10.1088/1751-8121/aa5f66}.

\bibitem[{Birkhoff and Abu-Shumays(1970)}]{birkhoff70}
\bibinfo{author}{G.~Birkhoff}, \bibinfo{author}{I.~K. Abu-Shumays},
  \bibinfo{title}{Exact analytic solutions of transport equations},
  \bibinfo{journal}{Journal of Mathematical Analysis and Applications}
  \bibinfo{volume}{32}~(\bibinfo{number}{3}) (\bibinfo{year}{1970})
  \bibinfo{pages}{468--481},
  \urlprefix\url{https://doi.org/10.1016/0022-247X(70)90271-4}.

\bibitem[{d'Eon and Williams(2018)}]{EdEMMR18}
\bibinfo{author}{E.~d'Eon}, \bibinfo{author}{M.~M.~R. Williams},
  \bibinfo{title}{{Isotropic scattering in a Flatland Half-Space}},
  \bibinfo{journal}{Journal of Computational and Theoretical Transport}
  \bibinfo{volume}{47}~(\bibinfo{number}{1-3}) (\bibinfo{year}{2018})
  \bibinfo{pages}{226--245},
  \urlprefix\url{https://doi.org/10.1080/23324309.2018.1544566}.

\bibitem[{Liemert and Kienle(2012)}]{liemert12a}
\bibinfo{author}{A.~Liemert}, \bibinfo{author}{A.~Kienle},
  \bibinfo{title}{Analytical approach for solving the radiative transfer
  equation in two-dimensional layered media}, \bibinfo{journal}{Journal of
  Quantitative Spectroscopy and Radiative Transfer}
  \bibinfo{volume}{113}~(\bibinfo{number}{7}) (\bibinfo{year}{2012})
  \bibinfo{pages}{559--564},
  \urlprefix\url{https://doi.org/10.1016/j.jqsrt.2012.01.013}.

\bibitem[{McDowall et~al.(2010)McDowall, Stefanov, and
  Tamasan}]{mcdowall2009stability}
\bibinfo{author}{S.~McDowall}, \bibinfo{author}{P.~Stefanov},
  \bibinfo{author}{A.~Tamasan}, \bibinfo{title}{Stability of the gauge
  equivalent classes in inverse stationary transport},
  \bibinfo{journal}{Inverse Problems}
  \bibinfo{volume}{26}~(\bibinfo{number}{2}),
  \urlprefix\url{https://doi.org/10.1088/0266-5611/26/2/025006}.

\bibitem[{Abu-Shumays(1967)}]{abushumays67}
\bibinfo{author}{I.~K. Abu-Shumays}, \bibinfo{title}{Generating functions and
  reflection and transmission functions}, \bibinfo{journal}{Journal of
  Mathematical Analysis and Applications}
  \bibinfo{volume}{18}~(\bibinfo{number}{3}) (\bibinfo{year}{1967})
  \bibinfo{pages}{453--471},
  \urlprefix\url{https://doi.org/10.1016/0022-247X(67)90038-8}.

\bibitem[{McCormick(2015)}]{NM15}
\bibinfo{author}{N.~J. McCormick}, \bibinfo{title}{Angular and spatial moments
  for half-space albedo transport problems}, \bibinfo{journal}{Annals of
  Nuclear Energy} \bibinfo{volume}{86} (\bibinfo{year}{2015})
  \bibinfo{pages}{72--79},
  \urlprefix\url{https://doi.org/10.1016/j.anucene.2014.12.021}.

\bibitem[{d'Eon(2014)}]{deon14dual}
\bibinfo{author}{E.~d'Eon}, \bibinfo{title}{{A dual-beam 3D searchlight
  BSSRDF}}, \bibinfo{journal}{ACM SIGGRAPH 2014 Talks}
  \bibinfo{volume}{65}~(\bibinfo{number}{1}) (\bibinfo{year}{2014})
  \bibinfo{pages}{1},
  \urlprefix\url{http://doi.acm.org/10.1145/2614106.2614140}.

\bibitem[{K\v{r}iv\'{a}nek and d'Eon(2014)}]{krivanek14zero}
\bibinfo{author}{J.~K\v{r}iv\'{a}nek}, \bibinfo{author}{E.~d'Eon},
  \bibinfo{title}{{A Zero-variance-based sampling scheme for Monte Carlo
  subsurface scattering}}, \bibinfo{journal}{ACM SIGGRAPH 2014 Talks}
  \bibinfo{volume}{66}~(\bibinfo{number}{1}) (\bibinfo{year}{2014})
  \bibinfo{pages}{1},
  \urlprefix\url{http://doi.acm.org/10.1145/2614106.2614138}.

\bibitem[{Monin(1956)}]{monin1956statistical}
\bibinfo{author}{A.~S. Monin}, \bibinfo{title}{A statistical interpretation of
  the scattering of microscopic particles}, \bibinfo{journal}{Theory of
  Probability \& Its Applications} \bibinfo{volume}{1}~(\bibinfo{number}{3})
  (\bibinfo{year}{1956}) \bibinfo{pages}{298--311},
  \urlprefix\url{https://doi.org/10.1137/1101024}, \bibinfo{note}{(English
  translation)}.

\bibitem[{Birkhoff and Abu-Shumays(1969)}]{birkhoff69}
\bibinfo{author}{G.~Birkhoff}, \bibinfo{author}{I.~Abu-Shumays},
  \bibinfo{title}{Harmonic solutions of transport equations},
  \bibinfo{journal}{Journal of Mathematical Analysis and Applications}
  \bibinfo{volume}{28}~(\bibinfo{number}{1}) (\bibinfo{year}{1969})
  \bibinfo{pages}{211--221},
  \urlprefix\url{https://doi.org/10.1016/0022-247X(69)90123-1}.

\bibitem[{Hanrahan and Krueger(1993)}]{hanrahan}
\bibinfo{author}{P.~Hanrahan}, \bibinfo{author}{W.~Krueger},
  \bibinfo{title}{Reflection from layered surfaces due to subsurface
  scattering}, in: \bibinfo{booktitle}{Proceedings of ACM SIGGRAPH 1993},
  \bibinfo{pages}{164--174},
  \urlprefix\url{https://doi.org/10.1145/166117.166139}, \bibinfo{year}{1993}.

\bibitem[{Busbridge(1960)}]{Busbridge60}
\bibinfo{author}{I.~W. Busbridge}, \bibinfo{title}{The Mathematics of Radiative
  Transfer}, \bibinfo{publisher}{Cambridge University Press},
  \bibinfo{year}{1960}.

\bibitem[{Chwolson(1889)}]{chwolson1889grundzuge}
\bibinfo{author}{O.~D. Chwolson}, \bibinfo{title}{{Grundz{\"u}ge einer
  mathematischen Theorie der inneren Diffusion des Lichtes}},
  \bibinfo{journal}{Bull. Acad. Imp. Sci. St. Petersburg} \bibinfo{volume}{33}
  (\bibinfo{year}{1889}) \bibinfo{pages}{221--256}.

\bibitem[{Jensen et~al.(2001)Jensen, Marschner, Levoy, and Hanrahan}]{jensen01}
\bibinfo{author}{H.~W. Jensen}, \bibinfo{author}{S.~R. Marschner},
  \bibinfo{author}{M.~Levoy}, \bibinfo{author}{P.~Hanrahan}, \bibinfo{title}{A
  practical model for subsurface light transport}, in:
  \bibinfo{booktitle}{Proceedings of the 28th Annual Conference on Computer
  Graphics and Interactive Techniques}, \bibinfo{organization}{ACM},
  \bibinfo{pages}{511--518}, \bibinfo{year}{2001}.

\bibitem[{Bouwkamp(1954)}]{bouwkamp1954diffraction}
\bibinfo{author}{C.~J. Bouwkamp}, \bibinfo{title}{Diffraction theory},
  \bibinfo{journal}{Reports on Progress in Physics}
  \bibinfo{volume}{17}~(\bibinfo{number}{1}) (\bibinfo{year}{1954})
  \bibinfo{pages}{35},
  \urlprefix\url{https://doi.org/10.1088/0034-4885/17/1/302}.

\bibitem[{Daniele and Zich(2014)}]{daniele2014wiener}
\bibinfo{author}{V.~G. Daniele}, \bibinfo{author}{R.~Zich}, \bibinfo{title}{The
  Wiener-Hopf method in electromagnetics}, \bibinfo{publisher}{Scitech
  Publishing}, \bibinfo{year}{2014}.

\bibitem[{Case(1957)}]{case57}
\bibinfo{author}{K.~M. Case}, \bibinfo{title}{On Wiener-Hopf equations},
  \bibinfo{journal}{Annals of Physics}
  \bibinfo{volume}{2}~(\bibinfo{number}{4}) (\bibinfo{year}{1957})
  \bibinfo{pages}{384--405},
  \urlprefix\url{https://doi.org/10.1016/0003-4916(57)90027-1}.

\bibitem[{Krein(1962)}]{Krein62}
\bibinfo{author}{M.~G. Krein}, \bibinfo{title}{Integral equations on a
  half-line with kernel depending upon the difference of the arguments},
  \bibinfo{journal}{Amer. Math. Soc. Transl.(2)} \bibinfo{volume}{22}
  (\bibinfo{year}{1962}) \bibinfo{pages}{163--288}, \bibinfo{note}{(English
  translation)}.

\bibitem[{Ivanov(1994)}]{ivanov94}
\bibinfo{author}{V.~Ivanov}, \bibinfo{title}{Resolvent method: Exact solutions
  of half-space transport problems by elementary means},
  \bibinfo{journal}{Astronomy and Astrophysics} \bibinfo{volume}{286}
  (\bibinfo{year}{1994}) \bibinfo{pages}{328--337}.

\bibitem[{Williams(2005)}]{williams05}
\bibinfo{author}{M.~M.~R. Williams}, \bibinfo{title}{{The Milne problem with
  Fresnel reflection}}, \bibinfo{journal}{Journal of Physics A: Mathematical
  and General} \bibinfo{volume}{38} (\bibinfo{year}{2005})
  \bibinfo{pages}{3841},
  \urlprefix\url{https://doi.org/10.1088/0305-4470/38/17/009}.

\bibitem[{Stewart et~al.(1966)Stewart, Ku{\v{s}}{\v{c}}er, and
  McCormick}]{JSIKNM66}
\bibinfo{author}{J.~C. Stewart}, \bibinfo{author}{I.~Ku{\v{s}}{\v{c}}er},
  \bibinfo{author}{N.~J. McCormick}, \bibinfo{title}{Equivalence of special
  models in energy-dependent neutron transport and nongrey radiative transfer},
  \bibinfo{journal}{Annals of Physics}
  \bibinfo{volume}{40}~(\bibinfo{number}{2}) (\bibinfo{year}{1966})
  \bibinfo{pages}{321--333},
  \urlprefix\url{https://doi.org/10.1016/0003-4916(66)90030-3}.

\bibitem[{Ivanov(1973)}]{ivanov73}
\bibinfo{author}{V.~V. Ivanov}, \bibinfo{title}{Transfer of Radiation in
  Spectral Lines}, vol. \bibinfo{volume}{385}, \bibinfo{publisher}{US
  Government Printing Office}, \bibinfo{year}{1973}.

\bibitem[{Frisch(1988)}]{frisch88}
\bibinfo{author}{H.~Frisch}, \bibinfo{title}{{A Cauchy integral equation method
  for analytic solutions of half-space convolution equations}},
  \bibinfo{journal}{Journal of Quantitative Spectroscopy and Radiative
  Transfer} \bibinfo{volume}{39}~(\bibinfo{number}{2}) (\bibinfo{year}{1988})
  \bibinfo{pages}{149--162},
  \urlprefix\url{https://doi.org/10.1016/0022-4073(88)90082-9}.

\bibitem[{Carlstedt and Mullikin(1966)}]{carlstedt66}
\bibinfo{author}{J.~L. Carlstedt}, \bibinfo{author}{T.~W. Mullikin},
  \bibinfo{title}{{Chandrasekhar's $X$-and $Y$-functions}},
  \bibinfo{journal}{The Astrophysical Journal Supplement Series}
  \bibinfo{volume}{12} (\bibinfo{year}{1966}) \bibinfo{pages}{449},
  \urlprefix\url{https://doi.org/10.1086/190133}.

\bibitem[{Mullikin(1968)}]{mullikin1968some}
\bibinfo{author}{T.~W. Mullikin}, \bibinfo{title}{Some probability
  distributions for neutron transport in a half-space},
  \bibinfo{journal}{Journal of Applied Probability}
  \bibinfo{volume}{5}~(\bibinfo{number}{2}) (\bibinfo{year}{1968})
  \bibinfo{pages}{357--374}, \urlprefix\url{https://doi.org/10.2307/3212258}.

\bibitem[{Muskhelishvili(1958)}]{muskhelishvili2008singular}
\bibinfo{author}{N.~I. Muskhelishvili}, \bibinfo{title}{Singular Integral
  Equations: Boundary Problems of Function Theory and their Application to
  Mathematical Physics}, \bibinfo{publisher}{Wolters-Noordhoff},
  \urlprefix\url{https://doi.org/10.1007/978-94-009-9994-7},
  \bibinfo{year}{1958}.

\bibitem[{Busbridge(1957)}]{Busbridge57}
\bibinfo{author}{I.~W. Busbridge}, \bibinfo{title}{{On the $H$-functions of S.
  Chandrasekhar}}, \bibinfo{journal}{The Quarterly Journal of Mathematics}
  \bibinfo{volume}{8}~(\bibinfo{number}{1}) (\bibinfo{year}{1957})
  \bibinfo{pages}{133--140},
  \urlprefix\url{https://doi.org/10.1093/qmath/8.1.133}.

\bibitem[{Fox(1961)}]{fox61}
\bibinfo{author}{C.~Fox}, \bibinfo{title}{{A solution of Chandrasekhar's
  integral equation}}, \bibinfo{journal}{Transactions of the American
  Mathematical Society} \bibinfo{volume}{99}~(\bibinfo{number}{2})
  (\bibinfo{year}{1961}) \bibinfo{pages}{285--291},
  \urlprefix\url{https://doi.org/10.2307/1993398}.

\bibitem[{Williams(2012)}]{williams12}
\bibinfo{author}{M.~M.~R. Williams}, \bibinfo{title}{{A simple method for
  calculating the moments of the Chandrasekhar $H$ function}},
  \bibinfo{journal}{Annals of Nuclear Energy} \bibinfo{volume}{46}
  (\bibinfo{year}{2012}) \bibinfo{pages}{232--233},
  \urlprefix\url{https://doi.org/10.1016/j.anucene.2012.03.030}.

\bibitem[{Sears(1975)}]{sears75}
\bibinfo{author}{V.~F. Sears}, \bibinfo{title}{Slow-neutron multiple
  scattering}, \bibinfo{journal}{Advances in Physics}
  \bibinfo{volume}{24}~(\bibinfo{number}{1}) (\bibinfo{year}{1975})
  \bibinfo{pages}{1--45},
  \urlprefix\url{https://doi.org/10.1080/00018737500101361}.

\bibitem[{d'Eon(2016)}]{hitchhiker}
\bibinfo{author}{E.~d'Eon}, \bibinfo{title}{A hitchhiker's guide to multiple
  scattering, v0.1.3}, \bibinfo{publisher}{(self published)},
  \urlprefix\url{http://www.eugenedeon.com/hitchhikers}, \bibinfo{year}{2016}.

\bibitem[{Case and Zweifel(1967)}]{KMCPFZ67}
\bibinfo{author}{K.~M. Case}, \bibinfo{author}{P.~F. Zweifel},
  \bibinfo{title}{Linear Transport Theory},
  \bibinfo{publisher}{Addison-Wesley}, \bibinfo{year}{1967}.

\bibitem[{McCormick and Ku{\v{s}}{\v{c}}er(1973)}]{NMIK73}
\bibinfo{author}{N.~J. McCormick}, \bibinfo{author}{I.~Ku{\v{s}}{\v{c}}er},
  \bibinfo{title}{Singular eigenfunction expansions in neutron transport
  theory}, in: \bibinfo{editor}{E.~Henley}, \bibinfo{editor}{J.~Lewins} (Eds.),
  \bibinfo{booktitle}{Advances in Nuclear Science and Technology},
  vol.~\bibinfo{volume}{7}, \bibinfo{publisher}{Academic Press},
  \bibinfo{pages}{181--282},
  \urlprefix\url{https://doi.org/10.1016/B978-0-12-029307-0.50010-X},
  \bibinfo{year}{1973}.

\bibitem[{Ku{\v{s}}{\v{c}}er et~al.(1964{\natexlab{a}})Ku{\v{s}}{\v{c}}er,
  McCormick, and Summerfield}]{IKNMGS64}
\bibinfo{author}{I.~Ku{\v{s}}{\v{c}}er}, \bibinfo{author}{N.~J. McCormick},
  \bibinfo{author}{G.~C. Summerfield}, \bibinfo{title}{{Orthogonality of Case's
  eigenfunctions in one-speed transport theory}}, \bibinfo{journal}{Annals of
  Physics} \bibinfo{volume}{30}~(\bibinfo{number}{3})
  (\bibinfo{year}{1964}{\natexlab{a}}) \bibinfo{pages}{411--421},
  \urlprefix\url{https://doi.org/10.1016/0003-4916(64)90127-7}.

\bibitem[{Ku{\v{s}}{\v{c}}er and Shure(1967)}]{IKFS65}
\bibinfo{author}{I.~Ku{\v{s}}{\v{c}}er}, \bibinfo{author}{F.~Shure},
  \bibinfo{title}{Closure relations for the eigenfunctions of the one-speed
  transport equation}, \bibinfo{journal}{Journal of Mathematical Physics}
  \bibinfo{volume}{8}~(\bibinfo{number}{4}) (\bibinfo{year}{1967})
  \bibinfo{pages}{823--826}, \urlprefix\url{https://doi.org/10.1063/1.1705284}.

\bibitem[{Hopf(1934)}]{Hopf34}
\bibinfo{author}{E.~Hopf}, \bibinfo{title}{Mathematical Problems of Radiative
  Equilibrium}, \bibinfo{publisher}{Cambridge University Press},
  \bibinfo{year}{1934}.

\bibitem[{Krein(1983)}]{Krein83}
\bibinfo{author}{M.~G. Krein}, \bibinfo{title}{{On nonlinear integral equations
  which play a role in the theory of Wiener-Hopf equations. I, II}}, in:
  \bibinfo{booktitle}{Topics in Differential and Integral Equations and
  Operator Theory}, \bibinfo{publisher}{Springer}, \bibinfo{pages}{173--242},
  \urlprefix\url{https://doi.org/10.1007/978-3-0348-5416-0\_3},
  \bibinfo{year}{1983}.

\bibitem[{Placzek and Seidel(1947)}]{plazcek47}
\bibinfo{author}{G.~Placzek}, \bibinfo{author}{W.~Seidel},
  \bibinfo{title}{Milne's problem in transport theory},
  \bibinfo{journal}{Physical Review} \bibinfo{volume}{72}~(\bibinfo{number}{7})
  (\bibinfo{year}{1947}) \bibinfo{pages}{550--555},
  \urlprefix\url{https://doi.org/10.1103/PhysRev.72.550}.

\bibitem[{Ku{\v{s}}{\v{c}}er et~al.(1964{\natexlab{b}})Ku{\v{s}}{\v{c}}er,
  McCormick, and Summerfield}]{kuvsvcer1964orthogonality}
\bibinfo{author}{I.~Ku{\v{s}}{\v{c}}er}, \bibinfo{author}{N.~J. McCormick},
  \bibinfo{author}{G.~Summerfield}, \bibinfo{title}{{Orthogonality of Case's
  eigenfunctions in one-speed transport theory}}, \bibinfo{journal}{Annals of
  Physics} \bibinfo{volume}{30}~(\bibinfo{number}{3})
  (\bibinfo{year}{1964}{\natexlab{b}}) \bibinfo{pages}{411--421},
  \urlprefix\url{https://doi.org/10.1016/0003-4916(64)90127-7}.

\bibitem[{Mendelson and Summerfield(1964)}]{mendelson1964one}
\bibinfo{author}{M.~R. Mendelson}, \bibinfo{author}{G.~C. Summerfield},
  \bibinfo{title}{One-speed neutron transport in two adjacent half-spaces},
  \bibinfo{journal}{Journal of Mathematical Physics}
  \bibinfo{volume}{5}~(\bibinfo{number}{5}) (\bibinfo{year}{1964})
  \bibinfo{pages}{668--674}, \urlprefix\url{https://doi.org/10.1063/1.1704161}.

\bibitem[{Leuth{\"a}user(1968)}]{leuthauser68}
\bibinfo{author}{K.~D. Leuth{\"a}user}, \bibinfo{title}{Der extrapolierte
  Endpunkt benachbarter Halbr{\"a}ume}, \bibinfo{journal}{Atomkernenergie}
  \bibinfo{volume}{13}~(\bibinfo{number}{6}) (\bibinfo{year}{1968})
  \bibinfo{pages}{385--386}.

\bibitem[{McCormick(1969)}]{mccormick1969neutron}
\bibinfo{author}{N.~J. McCormick}, \bibinfo{title}{{Neutron transport for
  anisotropic scattering in adjacent half-spaces. I. Theory}},
  \bibinfo{journal}{Nuclear Science and Engineering}
  \bibinfo{volume}{37}~(\bibinfo{number}{2}) (\bibinfo{year}{1969})
  \bibinfo{pages}{243--251},
  \urlprefix\url{https://doi.org/10.13182/NSE69-A20684}.

\bibitem[{Smedley-Stevenson(2012)}]{smedleystevenson12}
\bibinfo{author}{R.~Smedley-Stevenson}, \bibinfo{title}{A new analytic solution
  of the one-speed neutron transport equation for adjacent half-spaces with
  isotropic scattering}, \bibinfo{journal}{Annals of Nuclear Energy}
  \bibinfo{volume}{46} (\bibinfo{year}{2012}) \bibinfo{pages}{218--231},
  \urlprefix\url{https://doi.org/10.1016/j.anucene.2012.03.034}.

\bibitem[{Frisch and Frisch(1995)}]{frisch95}
\bibinfo{author}{U.~Frisch}, \bibinfo{author}{H.~Frisch},
  \bibinfo{title}{Universality of escape from a half-space for symmetrical
  random walks}, in: \bibinfo{booktitle}{L{\'e}vy Flights and Related Topics in
  Physics}, \bibinfo{publisher}{Springer}, \bibinfo{pages}{262--268},
  \urlprefix\url{https://doi.org/10.1007/3-540-59222-9\_39},
  \bibinfo{year}{1995}.

\bibitem[{Majumdar(2010)}]{majumdar2010universal}
\bibinfo{author}{S.~N. Majumdar}, \bibinfo{title}{Universal first-passage
  properties of discrete-time random walks and L{\'e}vy flights on a line:
  Statistics of the global maximum and records}, \bibinfo{journal}{Physica A:
  Statistical Mechanics and its Applications}
  \bibinfo{volume}{389}~(\bibinfo{number}{20}) (\bibinfo{year}{2010})
  \bibinfo{pages}{4299--4316},
  \urlprefix\url{https://doi.org/10.1016/j.physa.2010.01.021}.

\bibitem[{Frankel and Nelson(1953)}]{frankel53}
\bibinfo{author}{S.~Frankel}, \bibinfo{author}{E.~Nelson},
  \bibinfo{title}{Methods of treatment of displacement integral equations},
  \bibinfo{type}{Tech. Rep.} \bibinfo{number}{AECD-3497; LADC-79},
  \bibinfo{institution}{Los Alamos Scientific Lab.},
  \urlprefix\url{https://doi.org/10.2172/4371404}, \bibinfo{year}{1953}.

\bibitem[{Siewert(1980)}]{siewert1980computing}
\bibinfo{author}{C.~E. Siewert}, \bibinfo{title}{On computing eigenvalues in
  radiative transfer}, \bibinfo{journal}{Journal of Mathematical Physics}
  \bibinfo{volume}{21}~(\bibinfo{number}{9}) (\bibinfo{year}{1980})
  \bibinfo{pages}{2468--2470},
  \urlprefix\url{https://doi.org/10.1063/1.524684}.

\bibitem[{Kulczycki et~al.(2010)Kulczycki, Kwa{\'s}nicki, Ma{\l}ecki, and
  Stos}]{kulczycki2010spectral}
\bibinfo{author}{T.~Kulczycki}, \bibinfo{author}{M.~Kwa{\'s}nicki},
  \bibinfo{author}{J.~Ma{\l}ecki}, \bibinfo{author}{A.~Stos},
  \bibinfo{title}{{Spectral properties of the Cauchy process on half-line and
  interval}}, \bibinfo{journal}{Proceedings of the London Mathematical Society}
  \bibinfo{volume}{101}~(\bibinfo{number}{2}) (\bibinfo{year}{2010})
  \bibinfo{pages}{589--622},
  \urlprefix\url{https://doi.org/10.1112/plms/pdq010}.

\bibitem[{Inayat-Hussain(1987)}]{inayat1987new}
\bibinfo{author}{A.~A. Inayat-Hussain}, \bibinfo{title}{{New properties of
  hypergeometric series derivable from Feynman integrals II. A generalisation
  of the H function}}, \bibinfo{journal}{Journal of Physics A: Mathematical and
  General} \bibinfo{volume}{20}~(\bibinfo{number}{13}) (\bibinfo{year}{1987})
  \bibinfo{pages}{4119},
  \urlprefix\url{https://doi.org/10.1088/0305-4470/20/13/020}.

\bibitem[{Audic and Frisch(1993)}]{audic93}
\bibinfo{author}{S.~Audic}, \bibinfo{author}{H.~Frisch},
  \bibinfo{title}{{Monte-Carlo simulation of a radiative transfer problem in a
  random medium: Application to a binary mixture}}, \bibinfo{journal}{Journal
  of Quantitative Spectroscopy and Radiative Transfer}
  \bibinfo{volume}{50}~(\bibinfo{number}{2}) (\bibinfo{year}{1993})
  \bibinfo{pages}{127--147},
  \urlprefix\url{https://doi.org/10.1016/0022-4073(93)90113-V}.

\bibitem[{Larsen and Vasques(2011)}]{larsen11}
\bibinfo{author}{E.~W. Larsen}, \bibinfo{author}{R.~Vasques},
  \bibinfo{title}{{A generalized linear Boltzmann equation for non-classical
  particle transport}}, \bibinfo{journal}{Journal of Quantitative Spectroscopy
  and Radiative Transfer} \bibinfo{volume}{112}~(\bibinfo{number}{4})
  (\bibinfo{year}{2011}) \bibinfo{pages}{619--631},
  \urlprefix\url{https://doi.org/10.1016/j.jqsrt.2010.07.003}.

\bibitem[{d'Eon(2019{\natexlab{b}})}]{deon18iv}
\bibinfo{author}{E.~d'Eon}, \bibinfo{title}{{A reciprocal formulation of
  nonexponential radiative transfer. 4: Half spaces}}, \bibinfo{journal}{(in
  preparation)} .

\end{thebibliography}

\end{document}